\documentclass[pre,twocolumn,notitlepage]{revtex4-1}
\usepackage{amsmath}
\usepackage{graphicx,xcolor}
\usepackage{lipsum}
\usepackage{longtable}
\usepackage{array}
\usepackage{xcolor}
\newcolumntype{L}{>{$}l<{$}}
\newcommand{\gammar}{\Gamma_R}
\newcommand{\reps}{r^{-\epsilon}}

\begin{document}

%-----------------------------------------------------------------	
\title{Critical dynamics of \\ 
  non-conserved strongly anisotropic permutation symmetric three-vector model}
	
\author{Rajiv G. Pereira}
\affiliation{School of Physics, IISER Thiruvananthapuram, Vithura, Kerala 695551, India}

%-----------------------------------------------------------------	
\begin{abstract}

We explore, employing the renormalization-group theory, the critical scaling behavior of the permutation symmetric three-vector model that obeys \textit{non-conserving} dynamics and has a relevant anisotropic perturbation which drives the system into a non-equilibrium steady state. \textcolor{black}{We explicitly find the independent critical exponents with corrections up to two loops. They include the static exponents $\nu$ and $\eta$, the off equilibrium exponent $\widetilde{\eta}$, the dynamic exponent $z$ and the strong anisotropy exponent $\Delta$. We also express the other anisotropy exponents in terms of these.}  

\end{abstract}

%------------------------------------------------------------------

\maketitle
%-----------------------------------------------------------------

\section{Introduction}

Universality exhibited by systems out of equilibrium has been a prominent object of study in statistical physics, especially since the formulation of the renormalization-group (\textit{RG}) theory \cite{GezaOdor2004, tauber2014critical, antonov2020, young2020nonequilibrium}. A variety of genuine non-equilibrium (\textit{NE}) universality classes have been identified and well studied in the past few decades. The driven diffusive systems \cite{schmittmann1995, tauber1999, tauber2014critical} and the percolation models \cite{Kinzel1983, janssen1981} are a few examples. However, in comparison to equilibrium and near-equilibrium classes \cite{ma2018modern, hohenberg1977}, genuine non-equilibrium ones remain far less explored.  

Driven-diffusive models constitute an important category that violates the detailed-balance condition.  They have been widely used to describe physical systems, such as fast ionic conductors \cite{katz1983,katz1984} and traffic jams \cite{helbing2001traffic, chowdhury329statistical}, in order to investigate physics far from equilibrium. A variety of such models have been explored in the past \cite{schmittmann1995}, and they continue to appear in recent studies, for instance, Bose condensation transition \cite{sieberer2014nonequilibrium, tauber2014} and systems coupled to mutually interacting Langmuir kinetics \cite{vuijk2015driven}. These are essentially Ising-like models with anisotropic forces, and many of them exhibit universality distinct from that of any equilibrium class \cite{janssen1986, schmittmann1996, bassler1994b}. However, those models with spatially biased forces that violate detailed-balance even at the long-distance and large-time limit are mostly the ones that follow conserving dynamics.

Non-conserved Ising-like systems with relevant anisotropic perturbations are rare, and their critical properties are far less explored. One such exception can be found in Ref. \citep{DuttaPark2011}, wherein a cyclic permutation symmetric three-vector model with non-conserving dynamics and anisotropic perturbations was introduced. It was shown that for this model, below the critical dimension $d_c=4$, there exists an infrared stable fixed point at which one of the anisotropic perturbations is relevant, thus identifying a new genuine non-equilibrium universality class.

Though the anisotropic \textit{NE} fixed point was identified, the critical behavior of this class has not been investigated. Secondly, the relevance of the anisotropic term should reflect as difference in the longitudinal and the transverse power-law behavior of the correlation functions. Further, unlike other commonly found Ising-like systems with relevant spatial bias \cite{janssen1986, schmittmann1996, bassler1994b,leung1986,becker1992speeding}, this model follows non-conserving dynamics. These factors raise several interesting questions. What are the similarities and the differences in the critical behavior of the model from that of the conserved ones?
Does the model exhibit common critical features such as faster decay of longitudinal fluctuations \cite{schmittmann1995}? Is the critical power-law decay of the response and the correlation functions spatially biased? Are the $\eta$-like exponents in the real space different from those in the momentum space?

Motivated by these questions, we explore the critical scaling behavior of this class. For this, we look at a simpler model obtained by replacing the cyclic permutation symmetry in the model introduced in Ref. \citep{DuttaPark2011} by permutation symmetry.  In other words, we consider the non-conserved strongly anisotropic permutation symmetric (\textit{NSAPS}) three-vector model.  It is sufficient to study this model and determine the critical exponents as it has the same \textit{NE} fixed point. A nontrivial correction to two of the independent exponents, the strong anisotropy exponent $\Delta$ and the correlation length exponent $\nu$, can be obtained at the one-loop order. However, there are other exponents, where a non-trivial correction appears only at the two-loop order. Hence we renormalize the theory to this order.

We organize this paper as follows. In Sec. \ref{the model}, we introduce the  \textit{NSAPS} three-vector model.   In Sec.~\ref{renormalization}, we first discuss the renormalization of the theory and then briefly describe the computational methods employed in the two-loop calculation.
 In Sec.~\ref{results}, we obtain the critical exponents to two-loop order in an expansion around the upper critical dimension $d_c=4$ and then discuss the various critical features of the model.  In Appendix~\ref{appcomp}, the computational methods used in obtaining and evaluating the Feynman diagrams are detailed, and in Appendices \ref{appg110} to \ref{appg131122}, the relevant 1PI diagrams and their divergences are listed. 

\section{The Model}\label{the model}
The most general field theory for non-conserved $N$-vector models subject to anisotropic forces with all the marginal perturbations in $4+1$ dimensions, was constructed in Ref \citep{DuttaPark2011}. The theory is written in Martin-Siggia-Rose (MSR) formalism \cite{martin1973} as 
\begin{align}
\mathcal{S}(\phi,\tilde{\phi})=
\int_{x}\left[\widetilde{\phi}_{a}\left(\partial_{t}-\nabla^{2}+r\right) \phi_{a}-\frac{1}{2} \mathcal{E}_{a b c} \widetilde{\phi}_{a} \phi_{b} \partial_{\|} \phi_{c}\right.
\nonumber
\\
\left.+\frac{1}{3 !} G_{a b c d} \widetilde{\phi}_{a} \phi_{b} \phi_{c} \phi_{d}-T \widetilde{\phi}_{a} \widetilde{\phi}_{a}\right],
\end{align}
where $x$ denotes the time and the space coordinates $\{t,\boldsymbol{x}\}$, $\int_x \equiv \int dt d^d \boldsymbol{x}$, $\tilde{\phi}$ is the auxillary field and $T$ is the noise strength. The fields $\phi$  and $\tilde{\phi}$ are functions of $x$ and the repeated indices are summed over.

It was shown in Ref. \citep{DuttaPark2011} that only when the number of components $N=3$, there can be anisotropic perturbations consistent with a single length scale. In the case of cyclic permutation symmetry, there are five allowed independent couplings namely, $G_{1111}$,  $G_{1122}$, $G_{1133}$, $\mathcal{E}_{123}$ and $\mathcal{E}_{132}$. Below the upper critical dimension $d_c=4$ this model has an infrared stable \textit{NE} fixed point at which the anisotropic coupling $\mathcal{E}_{123}$ and the couplings $G_{1111}$ and $G_{1122}$ are relevant, while the couplings $G_{1133}$ and $\mathcal{E}_{132}$ are irrelevant \citep{DuttaPark2011}. Thus, a non-conserved Ising-like model with a relevant anisotropic perturbation was constructed, identifying a new genuine non-equilibrium universality class. 

If we now restrict to full permutation symmetry, the \textit{NSAPS} three-vector model is obtained, where the number of allowed independent couplings reduce to three namely, $G_{1111}$, $G_{1122}$ and $\mathcal{E}_{123}$. This model has the same infrared stable fixed point as the cyclic permutation symmetric one \citep{DuttaPark2011}. Therefore, it is sufficient to study the critical scaling behavior of this model. The MSR action for this simpler case can be written as
\begin{align}\label{myaction}
S=\sum_{a=1}^{3} \int_{x}\left[\widetilde{\phi}_{a}\left(\partial_{t}-D\left(\nabla_{\perp}^{2}+\rho \partial_{\|}^{2}-r\right)\right) \phi_{a}-T \widetilde{\phi}_{a}^{2}\right.
\nonumber
\\
\left.+ \frac{u_{0}}{3 !} \widetilde{\phi}_{a} \phi_{a}^{3}
+\frac{u_{1}}{2 !} \widetilde{\phi}_{a} \phi_{a} \left( \phi_{a+1}^{2} + \phi_{a+2}^2 \right)
+e_{p} \phi_{a+1} \phi_{a+2} \partial_{\|} \widetilde{\phi}_{a}
\right] ,
\end{align}
where $u_0 \equiv G_{1111}$, $u_1 \equiv G_{1122}$, $e_p \equiv \mathcal{E}_{123}$,  $\phi_{i+3} \equiv \phi_{i}$  and $\widetilde{\phi}_{i+3} \equiv \widetilde{\phi}_{i}$. We split the $\nabla^2$ term into the longitudinal and the transverse components by introducing the coefficient $\rho$ as the theory is spatially anisotropic. 

We proceed to perform a two-loop \textit{RG} analysis on the \textit{NSAPS} three-vector model and extract the critical exponents associated to the response and the correlation functions.    
%-----------------------------------------------------------------

\section{\textit{RG} Analysis}\label{renormalization}
In this section, we first discuss the standard renormalization procedure (see, for example, the excellent text book by Tauber~\cite{tauber2014critical}), and apply it to the \textit{NSAPS} three-vector model, where we define the renormalization constants and state the renormalization conditions.  Then we briefly describe the computational techniques employed in the calculation, which are suitable when the diagrams are numerous. The computational packages FeynArts \cite{hahn2001} and FeynCalc \cite{mertig1991, shtabovenko2016new} are used with Mathematica~\cite{Mathematica} to obtain the Feynman diagrams and the package SecDec~\cite{carter2011secdec} is used for numerical dimensional regularization. 
\subsection*{Definitions and notations}

The effective action is written as
\begin{equation}
\Gamma[\psi, \widetilde{\psi}]=-\ln\mathcal{Z}[J,\widetilde{J}] + \sum_a \int_x  J_a(x) \psi_a(x) + \widetilde{J}_a(x) \widetilde{\psi}_a(x),
\end{equation}
where
$
\psi(x)=\frac{\delta \ln \mathcal{Z}}{\delta J(x)}, \text{ }\textbf{   } \widetilde{\psi}(x)=\frac{\delta \ln \mathcal{Z}}{\delta \widetilde{J}(x)},
$
and the generating functional for correlation functions 
$
\mathcal{Z}[J,\widetilde{J}]=\left\langle \exp{\sum_a \int_x  \phi_a(x) J_a(x) + \widetilde{\phi}_a(x) \widetilde{J}_a(x)} \right\rangle.
$
The 1PI diagrams are obtained by taking the functional derivaties of $\Gamma$,

\begin{align}
\Gamma_{\widetilde{n},n}^{\widetilde{a}_1...\widetilde{a}_{\widetilde{n}}a_1 ... a_n}(\widetilde{x}_1,..& \widetilde{x}_{\widetilde{n}}; x_1,..x_n) =
\nonumber
\\
& \prod_{i=1}^{\widetilde{n}}\frac{\delta}{\delta \widetilde{\psi}_{\widetilde{a}_i} (\widetilde{x}_i)}
\left.\prod_{j=1}^{n} \frac{\delta}{\delta \psi_{a_j}(x_j)} \Gamma[\widetilde{\psi}, \psi]\right|_{\tilde{\psi}=\psi=0}.
\end{align} 

The ultraviolet divergences are absorbed into the renormalization constants $Z_\phi$, $Z_{\widetilde{\phi}}$, $Z_D$, $Z_\rho$, $Z_T$, $Z_0$, $Z_1$ and $Z_p$, and the bare fields and the bare parameters are written in terms of their renormalized counterparts as

\begin{align}\label{rgconst}
\phi_a = {Z_\phi}^{1/2} \phi_{aR}, \textbf{   } \widetilde{\phi}_a={Z_{\widetilde{\phi}}}^{1/2} {\widetilde{\phi}_{aR}}, \textbf{   } D=\frac{Z_D}{Z} D_R,
\nonumber
\\
\rho=\frac{Z_\rho}{Z_D} \rho_R, \textbf{   } T=\frac{Z_T}{Z_{\widetilde{\phi}}} T_R, r=\frac{Z_r}{Z_D}\mu^2 r_R,
\nonumber
\\
u_0=\frac{Z_0}{Z Z_{\phi}} u_{0R}, \textbf{   } u_1=\frac{Z_1}{Z Z_\phi} u_{1R}, \textbf{   } e_p=\frac{Z_p}{Z {Z_{\phi}}^{1/2}} e_{pR},
\end{align}  
where $Z=\sqrt{Z_\phi Z_{\widetilde{\phi}}}$, the subscript $R$ denontes the renormalized quantities and the factor $\mu$ is introduced to make $r_R$ dimensionless. The renormalization constants are fixed by the following renormalization condtions with the minimal substraction scheme.
\begin{align} \label{zrcond}
&{\gammar^{11}}_{1,1} (q_i=0)=D_R r_R \mu^2, 
\\ \label{zrhocond}
& \left.\frac{\partial}{\partial q_{\|}^2} {\gammar^{11}}_{1,1}(q;q)\right|_{q=0}=D_R \rho_R,
\\\label{zcond}
& \left. \frac{\partial}{\partial i q_0}{\gammar^{11}}_{1,1}(q;q)\right|_{q=0}=1, 
\\ \label{zdcond}
&\left.\frac{\partial}{\partial q_{\perp}^2} {\gammar^{11}}_{1,1}(q;q)\right|_{q=0}=D_R,
\\ \label{ztcond}
&{\gammar^{11}}_{2,0}(q_i=0)=-2T_R, 
\\ \label{zpcond}
&\left. \frac{\partial}{\partial i q_{\|}} {\gammar^{123}}_{1,2}(-q,\frac{q}{2},\frac{q}{2}) \right|_{q=0}=e_{pR},
\\ \label{zu0cond}
&{\gammar^{1111}}_{1,3}(q_i=0)=u_{0R},
\\ \label{zu1cond}
&{\gammar^{1122}}_{1,3}(q_i=0)=u_{1R}.
\end{align}

\subsection*{Diagramatics and perturbative computation}

The unperturbed action in Fourier space is
\begin{equation}
S=\sum_a \int_{q} \widetilde{\phi}_a(-q) \left(- i q_0 + M(\boldsymbol{q}) \right) \phi_a(q),
\end{equation}
where $M(\boldsymbol{q})= D ({\boldsymbol{q}_{\perp}}^2 + \rho q_{\|}^2 + r)$ and $\int_{q} \equiv {\frac{1}{(2 \pi)^{d+1}}} \int dq_{0} d \boldsymbol{q}$. The Fourier transform of a function $f(x)$ is defined by the relation $f(x)=\int_q f(q) e^{-i q.x}$, where $q.x = q_0 x_0 -\boldsymbol{q} .\boldsymbol{x}$. The subscripts $\perp$, $\|$ and $0$ denote the transverse, the longitudnal and the temporal directions, respectively. 

The two non-vanishing unperturbed two-point correlations are
\begin{align}
\langle \phi_a (q_1) \widetilde{\phi}_b (q_2) \rangle_0= \frac{\delta_{ab} \bar{\delta}(q_1+q_2) }{-i q_0 + M(\boldsymbol{q})}=\delta_{ab} \bar{\delta}(q_1+q_2) G_0(q_1),
\nonumber
\\
\langle \phi_a (q_1) \phi_b (q_2) \rangle_0= 2T\frac{\delta_{ab} \bar{\delta}(q_1+q_2) }{ q_0^2 + M(\boldsymbol{q})^2}=\delta_{ab} \bar{\delta}(q_1+q_2) C_0(q_1),
\end{align}
where $ \bar{\delta} (q) \equiv (2 \pi)^{d+1} {\delta}(q)$. 

\begin{figure}
\begin{center}
\includegraphics[scale=0.4]{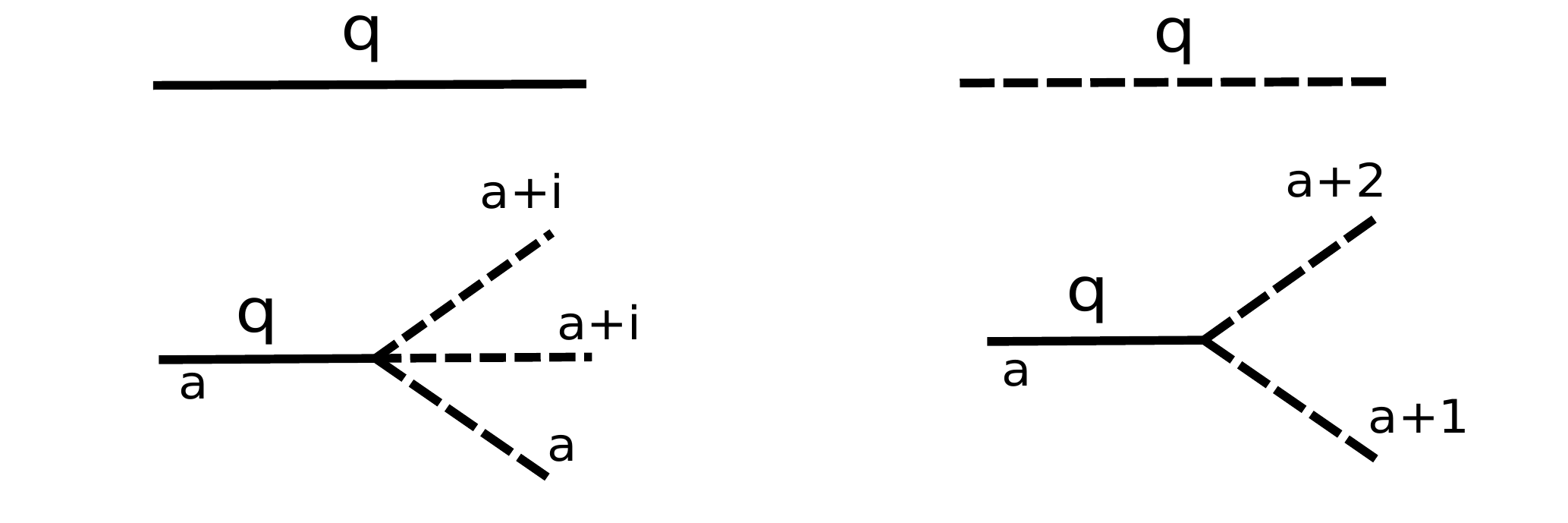}

\caption{The straight line represents $G_0(q)$ and the dotted line $C_0(q)$.  The four-point vertex takes the value $u_0/6$ if $i=0$ and $u_1/2$ if $i\neq 0$. The three-point vertex takes the value $i q_{\|} e_p$. The dotted branch becomes a straight line when it is hit with an auxiliary field and remains dotted otherwise. We choose the convention that the external $\phi$ fields hit from the left and the external $\widetilde{\phi}$ fields hit from the right. This makes the arrow which is usually attached to the propagator redundant and is hence not explicitly shown.} 

\label{basic diagrams}
\end{center}
\end{figure}

The diagrammatic representations of the two-point Gaussian correlation functions and the perturbations are illustrated in Fig.~\ref{basic diagrams}. With the help of these building blocks, we perform the perturbative expansions of the vertex functions to two loops and extract the divergences. This is implemented computationally in the following steps. 

\begin{enumerate}
\item For a given vertex function, we obtain all the contributing Feynman diagrams and the corresponding expressions to two-loops using the packages FeynArts~\cite{hahn2001} and FeynCalc~\cite{mertig1991,shtabovenko2016new}.  For this, we first draw all the topologically distinct irreducible diagrams (topologies) with $l$ external legs that can be constructed with three-point and four-point vertices to two loops, where $l$ is determined by the vertex function we evaluate.  Then, we draw all possible realizations of the topologies that can be obtained using the building blocks shown in Fig.~\ref{basic diagrams}. Each such realization (Feynman diagram) correspond to an expression of the form  
\begin{align}\label{from a diagram}
A_f \int_{ \{ q_i \} } \! \! \! \! \! \!p_{1_\|}p_{2_\|}... G_0(k_1)G_0(k_2)... C_0(k_{m})C_0(k_{m+1})...,
\nonumber
\\
\end{align}
where $\{ q_i \}$ is the set of internal momenta, $p_i$ and $k_i$ are in general linear combinations of the internal and the external momenta and $A_f$ is the overall factor associated with each diagram. 

\item  Derivatives and limits are now applied to isolate the divergences in the expressions obtained in step~1. Integrating out $q_{i0}$s subsequently leads to an expression of the form 
\begin{equation}
B_f \int_{\{ {\boldsymbol{q}_i} \}} \frac{ p_{1 \|}^{m_1} \; |\boldsymbol{p}_{2 \perp}|^{m_2} }{\mathcal{M}_1^{n_1} \mathcal{M}_2^{n_2} ...} + \text{similar terms},
\end{equation}
where $\mathcal{M}_i$ is either $M(\boldsymbol{k}_i)$ or sum of $M(\boldsymbol{k}_i)$s. The variables $\boldsymbol{p}_i$ and $\boldsymbol{k}_i$ now contain only the internal momenta. Once the parallel components of ${q_i}$s are scaled approprietly, $M(k_i)$ takes the form of the scalar propagator with a factor $D$.

\item The integrals obtained by the above procedure are regularized by the method of dimensional regularization and the UV-divergent parts are expanded as
\begin{equation}
\hat{\mathcal{D}} \left( \int_{\{ \boldsymbol{q_i} \}} \frac{ {p_1}_\|^{m_1} \; |\boldsymbol{p_2}_\perp|^{m_2} }{\mathcal{M}_1^{n_1} \mathcal{M}_2^{n_2} ...} \right)=\frac{\omega_1}{\epsilon^2}+\frac{\omega_2}{\epsilon},
\end{equation}
where the operator $\hat{D}$ is defined such that $\hat{D}$ acting on an intergral gives the poles of the corresponding dimensionally regularized integral, the parameter $\epsilon=4-d$, and $\omega_1$ and $\omega_2$ are real numbers. 
This step is implemented with the help of the package SecDec \cite{carter2011secdec, bruns2010normaliz, bruns2012power}.

The above steps are elucidated with the help of an example in Appendix \ref{appcomp}.
\item Once all the diagrams contributing to a given vertex function are evaluated and the divergences are obtained in negative powers of $\epsilon$, we apply the renormalization conditions with the minimal subtraction scheme to obtain the renormalization constants. 

The divergent 1PI Feynman diagrams contributing to the various relevant vertex functions, and the renormalization constants are given in Appendices \ref{appg110} to \ref{appg131122}. 
\end{enumerate}

There are $11$ distinct divergent Feynman diagrams contributing to the various relevant vertex functions at the one-loop order. The presence of both three-point and four-point vertices enhances the number of diagrams at the two-loop order enormously, where the number rises to $319$.  In the absence of the four-point vertices, the total number of two-loop diagrams reduces to $27$, while in the absence of the three-point vertices, the total number of two-loop diagrams reduces to $19$. For instance, there are $25$ two-loop diagrams contributing to $\Gamma_{1,1}^{11}(-q;q)$, as shown in Fig. \ref{everything}, of which only $6$ diagrams (diagrams $10$ to $15$) are constructed with three-point vertices alone, while only two diagrams (diagrams $20$ and $25$) are constructed with four-point vertices alone. 
Similarily, of the $83$ diagrams contributing to $ \left. \frac{\partial}{\partial i q_{\|}} {\Gamma^{123}}_{1,2}(-q,\frac{q}{2},\frac{q}{2}) \right|_{q=0}$, shown in Table \ref{g12diagrams}, only $21$ diagrams (diagrams $14$ to $25$ and $73$ to $81$) are constructed with three-point vertices alone, while none is constructed with four-point vertices alone. Of the $94$ diagrams contributing to $\Gamma_{1,3}^{1111}(q_i=0)$, shown in Table \ref{g13adiagrams}, none is constructed with three-point vertices alone, while only $8$ diagrams (diagrams $47$ to $53$ and  $94$) are constructed with four-point vertices alone. Of the $116$ diagrams contributing to $\Gamma_{1,3}^{1122}(q_i=0)$, shown in Table \ref{g13bdiagrams}, none is constructed with three-point vertices alone and only $8$ diagrams (diagrams $67$ to $73$ and $116$) are constructed with four-point vertices alone.  The only one diagram contributing to $\Gamma_{2,0}^{11}(0)$, shown in Table \ref{g20}, is constructed with four-point vertices alone.

\section{The critical exponents of the \textit{NSAPS} three-vector model}\label{results}
We proceed to write down and solve the \textit{RG} equation to obtain the scaling form of the vertex functions at the \textit{NE} fixed point as the temperature approaches the critical value. In particular, we analyze the scaling behavior of the dynamic structure factor and the dynamic susceptibility and extract the exponents associated with them. 

For notational simplicity, the subscript $R$ is suppressed in this section, and the following dimensionless couplings are employed    
\begin{align}\label{lambdas}
&\lambda_{0}=\frac{1}{8 \pi^2}\frac{T}{{D}^2 {\rho}^{1/2}}u_{0} \mu^{-\epsilon},
\nonumber
\\
&\lambda_{1}=\frac{1}{8 \pi^2}\frac{T}{{D}^2 {\rho}^{1/2}}{u_{1}}\mu^{-\epsilon} \textbf{   },
\nonumber
\\
&\lambda_{2}=\frac{1}{8 \pi^2}\frac{T}{{D}^3 {\rho}^{3/2}}{e_{p}}^2\mu^{-\epsilon}.
\end{align}
The beta functions are 
\begin{align}\label{betadef}
\beta_{i}= \mu \frac{d\lambda_{i}}{d \mu},
\end{align}
for $i=0,1$ and $2$, and Wilson's flow functions are
\begin{align}\label{gammadef}
&{\gamma}_{\phi} = -\mu \frac{\partial}{\partial \mu} \ln{Z_{\phi}}, \; {\gamma}_{\tilde{\phi}} = -\mu \frac{\partial}{\partial \mu} \ln{Z_{\tilde{\phi}}}, \;
{\gamma}_D = \mu \frac{\partial}{\partial \mu} \ln{D},
\nonumber
\\
&{\gamma}_{\rho} = \mu \frac{\partial}{\partial \mu} \ln{\rho}, \;
{\gamma}_T = \mu \frac{\partial}{\partial \mu} \ln{T}, \;
{\gamma}_r = \mu \frac{\partial}{\partial \mu} \ln{r},
\end{align}
where the derivatives are to be taken keeping the bare parameters and couplings constant. Since all the UV divergences can be absorbed into the eight renormalization constants $Z_\phi$, $Z_D$, $Z_\rho$, $Z_T$, $Z_r$, $Z_0$, $Z_1$ and $Z_p$,  the auxiliary field renormalization constant $Z_{\widetilde{\phi}}$ is set to unity which implies that $\gamma_{\widetilde{\phi}}=0$. 

\begin{widetext}
We now write down the \textit{RG} equation, which follows from the fact that the bare vertex functions are independent of $\mu$,
\begin{align} \label{rgeq}
\left[ \mu  \frac{\partial }{\partial \mu} +  \gamma_\phi \frac{n}{2}  +   \sum_i \gamma_{s_i} s_i \frac{\partial}{\partial s_i} 
%+    \gamma_{\rho} \rho \frac{\partial}{\partial \rho}
%+    \gamma_{T} T\frac{\partial}{\partial T}
%+    \gamma_{r} r\frac{\partial}{\partial r}
+ \sum_i \beta_i \frac{\partial}{\partial \lambda_{i}} \right]{\Gamma}_{\widetilde{n},n}(q_i, s_i, \lambda_i, \mu)=0,
\end{align} 
where $q_i$ denotes the external momenta and $s_i$ denotes the elements of the set of parameters $\{ D, \rho, T, r \}$. 

The beta functions are obtained by using the renormalization constants given in Appendices~\ref{appg110} to \ref{appg131122}, in  equations~(\ref{zr}), (\ref{zrho}), (\ref{z}), (\ref{zd}), (\ref{zt}), (\ref{zp}), (\ref{z0}) and (\ref{z1}), together with Eq.~(\ref{betadef}), and are explicitly written as
\begin{align}
\beta_{0}=&- \epsilon  \lambda_{0}+1.5 \lambda_{0}^2 +3 \lambda_{1}^2+0.375 \lambda_{0} \lambda_{2}-1.41667 \lambda_{0}^3-6 \lambda_{1}^3+0.336482  \lambda_{2} \lambda_{0}^2-2.5 \lambda_{1}^2 \lambda_{0} +0.156108 \lambda_{2}^2 \lambda_{0} 
\nonumber
\\
&-0.341841 \lambda_{0} \lambda_{1} \lambda_{2}-0.0163937 \lambda_{1} \lambda_{2}^2+0.991439 \lambda_{1}^2 \lambda_{2}, 
\nonumber
\\
\beta_{1}=&- \epsilon  \lambda_{1}+2.5 \lambda_{1}^2+ \lambda_{0} \lambda_{1}+0.375
   \lambda_{1} \lambda_{2}-4.5 \lambda_{1}^3-0.0359603 \lambda_{0}^2 \lambda_{2}-3 \lambda_{0} \lambda_{1}^2+0.700121 \lambda_{2} \lambda_{1}^2-0.416667 \lambda_{0}^2 \lambda_{1} 
\nonumber
\\
&+0.158841 \lambda_{2}^2 \lambda_{1}+0.135281 \lambda_{0} \lambda_{2} \lambda_{1} -0.00149424 \lambda_{0} \lambda_{2}^2, 
\nonumber
\\
\beta_{2}=&- \epsilon  \lambda_{2}+1.125 \lambda_{2}^2+3.5 \lambda_{1} \lambda_{2}+0.475028 \lambda_{2}^3-0.181794 \lambda_{0} \lambda_{2}^2+1.38638 \lambda_{1} \lambda_{2}^2+0.125 \lambda_{0}^2 \lambda_{2}-4.05652 \lambda_{1}^2 \lambda_{2}
\nonumber
\\
& -2.90455 \lambda_{0} \lambda_{1} \lambda_{2}.
\end{align}
Similarly, we obtain Wilson's flow functions by using the renormalization constants given in the Appendices~\ref{appg110} to \ref{appg131122} together with Eq.~(\ref{gammadef}) and are explicitly written as
\begin{align}\label{gammas}
\gamma_\phi & =-0.143841 \lambda _0^2-0.863046 \lambda _1^2,
\nonumber
\\
\gamma_D &=-0.0416667 \lambda _0^2-0.25 \lambda _1^2-0.025463 \lambda _1 \lambda _2,
\nonumber
\\
\gamma_\rho & =-0.75 \lambda_2-0.0416667 \lambda _0^2-0.25 \lambda _1^2-0.312217 \lambda _2^2-0.308408 \lambda _1 \lambda _2,
\nonumber
\\
\gamma_T & =-0.0719205 \lambda _0^2-0.431523 \lambda _1^2,
\nonumber
\\
\gamma_r & =-2+0.5 \lambda _0+\lambda _1-0.25 \lambda _0^2-1.5 \lambda _1^2+0.112161 \lambda _2 \lambda _0+0.224321 \lambda _1 \lambda _2.
\end{align}
\end{widetext}
The set of equations
\begin{align}\label{beta=0}
 \beta_i=0,
\end{align}
leads to the critical points. 
For $\epsilon<0$, the equilibrium Gaussian fixed point is stable. For $\epsilon>0$, the following \textit{NE} fixed point is stable,
\begin{align}\label{thefp}
&{\lambda_{0}}^*=0.461538 \epsilon+0.173639 \epsilon ^2,
\nonumber
\\
&{\lambda_{1}}^*=0.153847 \epsilon+0.0837608 \epsilon ^2, 
\\
\nonumber
&{\lambda_{2}}^*=0.410255 \epsilon-0.133947 \epsilon ^2,
\end{align}
where the superscript $^*$ denotes the fixed point values of the couplings $\lambda_{i}$. The above result agrees with the one-loop calculations in Ref.~\cite{DuttaPark2011} to that order. By substituting Eq.~(\ref{thefp}) in Eq.~(\ref{gammas}), we further obtain Wilson's flow functions at this fixed point as
\begin{align}
&{\gamma_D}^*=-0.0164001 \epsilon^2,
\nonumber
\\
&{\gamma_\rho}^*=-0.307691 \epsilon+0.0136525 \epsilon ^2,
\nonumber
\\
&{\gamma_r}^*=-2+0.384616 \epsilon+0.117218 \epsilon ^2, 
\nonumber
\\
&{\gamma_T}^*=-0.025534 \epsilon ^2, 
\nonumber
\\
&{\gamma_\phi}^*=-0.051068 \epsilon ^2.
\end{align}
We now solve the \textit{RG} equation~(\ref{rgeq}) using the method of characteristics (see, for instance, Ref.~\cite{tauber2014critical}). To this end, we define $\mu'(\sigma)=\mu \sigma$, where $\sigma$ is a dimensionless real parameter, and introduce the running parameters $s'_i(\sigma)$ and the couplings $\lambda'_i(\sigma)$ which respect following relations, 
\begin{align}\label{runningpars}
\sigma \frac{d s'_i(\sigma)}{d \sigma} =s'_i(\sigma) \gamma_{s_i} (\sigma), \; s'(1)=s_i,
\\
\sigma \frac{d \lambda'_i(\sigma)}{d \sigma} =\lambda'(\sigma) \beta_i (\sigma), \; \lambda'(1)=\lambda_i.
\end{align} 
The \textit{RG} equation (\ref{rgeq}) together with the above relations yields, 
\begin{align}\label{rgsoln}
& \Gamma_{\widetilde{n},n}(q_i, s_i, \lambda_i, \mu)  =  
\nonumber
\\ 
& \exp{ \left( \int_1^\sigma \frac{d\sigma'}{\sigma'} \frac{n}{2} \gamma_\phi(\sigma) \right)} \,
\Gamma_{\widetilde{n},n}(q_i, s'_i(\sigma), \lambda'_i (\sigma), \mu \sigma).
\end{align}
At the fixed points, the solution to Eq.~(\ref{runningpars}) gives simple power-law behavior, and at the \textit{NE} fixed point we obtain,
\begin{align}\label{sistar}
s_i'(\sigma) \approx s_i \, \sigma^{\gamma_{s_i}^*}.
\end{align}
Using the above result in Eq.~(\ref{rgsoln}), we obtain the critical scaling form of the vertex functions at the \textit{NE} fixed point,  
\begin{align}\label{vertexscale}
{\Gamma}_{\widetilde{n},n}(q_i,s_i, \mu) & = \sigma^{\frac{n}{2} \gamma_\phi^*} 
\Gamma_{\widetilde{n},n}(q_i, s_i \, \sigma^{\gamma_{s_i}^*}, \mu \sigma),
\end{align}
where we have not shown the arguments of $\Gamma$ which are not affected by rescaling. In the limit $r \rightarrow 0$, the parameter $\sigma$ scales as $\sigma \propto r^{-1/\gamma_r^*}$, as can be seen from Eq.~(\ref{sistar}).
 
From Eq.~(\ref{vertexscale}) the scaling forms of the dynamic structure factor $S(\boldsymbol{q},t)=\int_{q_0} e^{-i q_0 t} \Gamma_{2,0}(q)/|\Gamma_{1,1}(q)|^2$ and the dynamic susceptibility $\chi (q)=1/\Gamma_{1,1}(q)$ follow as
 
\begin{align}\label{structurefactor0}
 S(\boldsymbol{q_\perp},q_\|,t,r)  =&  \sigma^{-2 + \gamma_T^*- \gamma_\phi^*- \gamma_D^*} \nonumber
\\
& S( \frac{\boldsymbol{q}_\perp}{ \sigma}, \frac{ q_\|}{ \sigma^{1-\gamma_\rho^*/2}},  t\sigma^{2+\gamma_D^*}, \frac{r}{\sigma^{-\gamma_r^*}}),
\\
 \chi(\boldsymbol{q_\perp},q_\|,t,r)  =&\sigma^{-2 -\gamma_\phi^*/2 -\gamma_D^*} 
 \nonumber
 \\
 &\chi( \frac{\boldsymbol{q}_\perp}{ \sigma}, \frac{ q_\|}{ \sigma^{1-\gamma_\rho^*/2}},  t\sigma^{2+\gamma_D^*}, \frac{r}{\sigma^{-\gamma_r^*}}). \label{susceptibility0}
\end{align}
Comparing equations~(\ref{structurefactor0}) and (\ref{susceptibility0}) with the standard scaling forms~\citep{schmittmann1995},
\begin{align}\label{structurefactor}
& S(\boldsymbol{q_\perp},q_\|,t,r)=  \sigma^{-2 + \eta} \; S( \frac{\boldsymbol{q}_\perp}{ \sigma}, \frac{ q_\|}{ \sigma^{1+\Delta}},  t\sigma^{z}, \frac{r}{\sigma^{1/\nu}}),
\\
& \chi(\boldsymbol{q_\perp},q_\|,t,r)=\sigma^{-z+\widetilde{\eta}/2 + \eta/2} \; \chi( \frac{ \boldsymbol{q}_\perp}{ \sigma}, \frac{ q_\|}{ \sigma^{1+\Delta}}, \frac{q_0}{\sigma^{z}}, \frac{r}{\sigma^{1/\nu}}), \label{susceptibility}
\end{align}
we obtain the exponents   
\begin{align}\label{indexp}
\eta&=\gamma_T^* - \gamma_D^* -\gamma_\phi^*=0.0419341 \epsilon^2,
\nonumber
\\
\Delta&=-\gamma_\rho^*/2=0.1538455 \epsilon - 0.00682625 \epsilon^2,
\nonumber
\\
z&=2+\gamma_D^*=2-0.0164001 \epsilon^2,
\nonumber
\\
\nu&=-1/\gamma_r^*=0.5+0.192308 \epsilon + 0.0955914 \epsilon^2,
\nonumber
\\
\widetilde{\eta}&=\gamma_D^*-\gamma_T^*=0.0091339 \epsilon^2,
\end{align}
which are correct to second order in $\epsilon$.

The other standard anisotropy exponents~\cite{schmittmann1995} can be written in terms of the above five exponents. The transverse dynamic exponent $ z_\perp=z$ and the longitudinal dynamic exponent $z_\|=z/(1+\Delta)$. Now, from Eq.~(\ref{indexp}), it follows that $z_\|<z_\perp$ which implies that the longitudinal fluctuations decay faster than the transverse fluctuations. The transverse correlation length exponent $\nu_\perp=\nu$, while the longitudinal correlation length exponent  $\nu_\|=\nu(1+\Delta)$.  As the strong anisotropy exponent $\Delta >0$, $\nu_\|>\nu_\perp$. 
This implies that the longitudinal correlation length diverges faster than the transverse correlation length as the temperature approaches the critical value.     

There are four $\eta$-like exponents, two in the momentum space, and two in the real space. The two momentum space $\eta$-like exponents, $\eta_\perp^{MS}=\eta$ and $\eta_\|^{MS}=(\eta +2\Delta)/(1+\Delta)$, determine the anisotropic power-law behavior of the dynamic structure factor in momentum space and the two real space $\eta$-like exponents, $\eta_\perp^{RS}=\eta+\Delta$ and $\eta_\|^{RS}=(\eta-\Delta)/(1+\Delta)$, determine the anisotropic power-law behavior of the dynamic structure factor in real space. The relations obtained above are the same as the standard scaling relations observed in models that exhibit strong anisotropy \cite{schmittmann1995}.

The transverse and the longitudnal susceptibilities scale as $\chi_\perp \sim r^{-\gamma_\perp}$ and $\chi_\| \sim r^{-\gamma_\|}$, where $\chi_\perp \equiv \chi(\boldsymbol{q}_\perp \rightarrow 0, q_\| = 0)$ and $\chi_\| \equiv \chi(\boldsymbol{q}_\perp = 0, q_\| \rightarrow 0)$.
From Eq. (\ref{susceptibility}) we obtain
\begin{align}
 \gamma_\perp=\gamma_\|=\nu(z-\widetilde{\eta}/2 -\eta/2).
 \end{align}
As opposed to the strongly anisotropic models that follow conserving dynamics, the susceptibility exponents $\gamma_\perp$ and $\gamma_\|$ are equal \cite{schmittmann1995}.  

To summarize, we studied the critical scaling behavior of the  \textit{NSAPS} three-vector model, which belongs to a new genuine non-equilibrium universality class. We obtained the critical exponents, which characterize the anisotropic power-law behavior of the dynamic structure factor and the dynamic susceptibility, to two-loop order. Among them is the important strong anisotropy exponent $\Delta$ that captures the effects of the spatially biased drive. We briefly mentioned the similarities and the dissimilarity in the critical behavior of the model to that of strongly anisotropic models that follow conserving dynamics.  

\section*{Acknowledgment}

I extend my sincere gratitude to Sreedhar B. Dutta, who suggested this problem and has spent hours engaging in fruitful discussions during the progress of the work and the preparation of the manuscript.

\appendix

\begin{widetext}
\section{Generating and evaluating Feynman diagrams: an example}\label{appcomp}
To obtain the Feynman diagrams that contribute to a vertex function, we first draw all the relevant topologies. For instance, to obtain the two-loop Feynman diagrams contributing to $\Gamma_{1,1}^{11}$, we draw all the distinct irreducible two-loop topologies with two external legs that can be constructed with three-point and four-point vertices, as shown in Fig. \ref{tops}. Any two-loop contribution to $\Gamma_{1,1}^{11}$ must be topologically similar to one of these diagrams. Now, we use the building blocks shown in Fig. \ref{basic diagrams} to construct all the possible realizations of these topologies. This leads to the Feynman diagrams shown in Fig. \ref{everything}. 

\begin{figure}
\begin{center}
\includegraphics[scale=0.57]{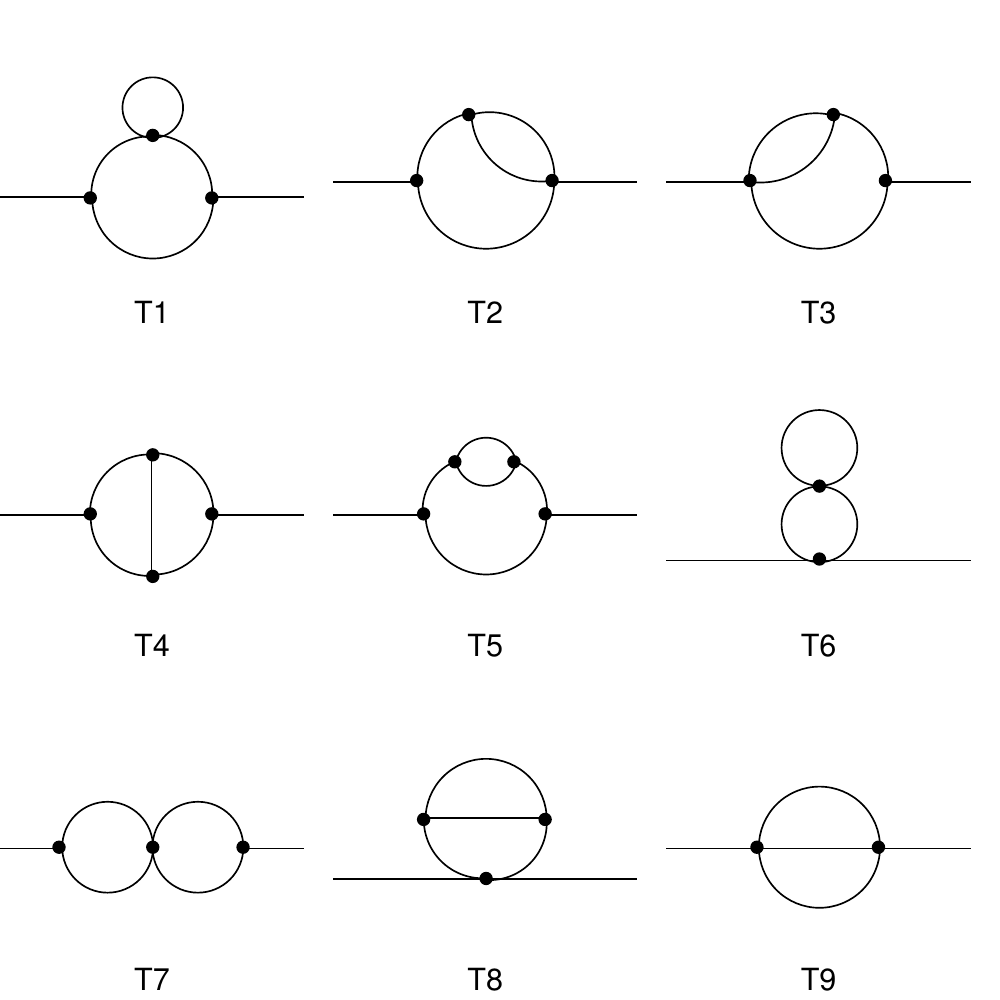}
\caption{All possible irreducible two-loop topologies with two external lines constructible with three-point and four-point vertices.} 
\label{tops}
\end{center}
\end{figure}

Each of the diagrams in Fig. \ref{everything} corresponds to an expression that is of the form given in Eq. (\ref{from a diagram}). For instance, diagram 22 corresponds to 
\begin{align*}
\mathcal{I}_{22}=12  T^2 u_0 e_p^2  \int_{q_1,q_2} ({q_1}_{\|}+{q_2}_{\|}){q_2}_{\|}
C_0\left(q_1\right) C_0\left(q_1+q_2\right) 
G_0\left(-q_1-q_2\right) G_0\left(-q_2\right). 
\end{align*}
All the above steps were implemented in Mathematica \cite{Mathematica} with the help of FeynCalc \cite{mertig1991,shtabovenko2016new} and FeynArts \cite{hahn2001}.
\begin{figure}
\begin{center}
\includegraphics[scale=1.5]{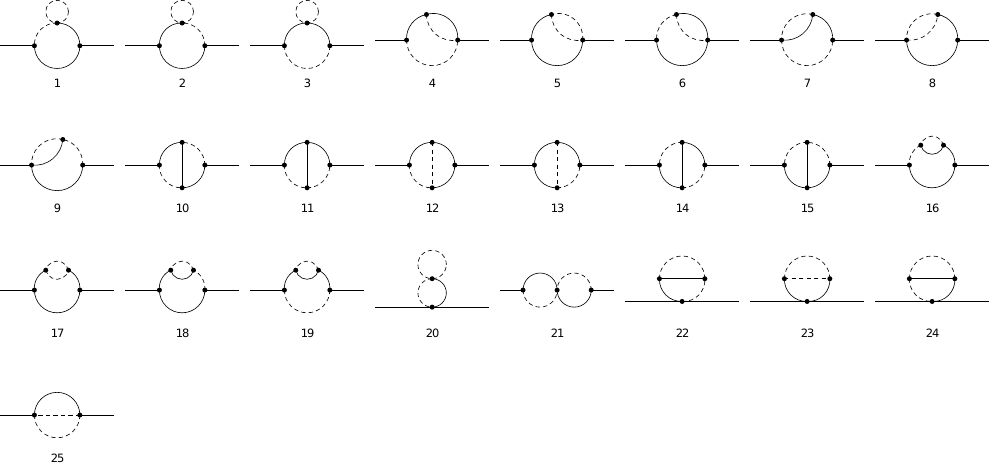}
\caption{Two-loop Feynman diagrams contributing to $\Gamma_{1,1}^{11}(q)$.} 
\label{everything}
\end{center}
\end{figure}

We can now proceed to extract the divergences. For instance, setting the external momenta to zero we obtain the quadratically divergent part which renormalizes the mass parameter $r$. The above integral does not depend on the external momenta and remains unchanged. Integrating out $q_{10}$ and $q_{20}$ from the above integral and making the transformation, $\{ \boldsymbol{q}_{i \perp} ,\sqrt{\rho} {q_i}_{\|} \} \rightarrow \{ \sqrt{r} \boldsymbol{q}_{i \perp}, \sqrt{r} {q_i}_\| \}$ we obtain, 
\begin{align}\label{intexample}
 \mathcal{I}_{22}=  \frac{3}{2}  \frac{T^2 u_0 e_p^2}{D^4 \rho^2} r^{d-3}  \int_{\boldsymbol{q}_1,\boldsymbol{q}_2}
\frac{({q_1}_{\|}+{q_2}_{\|}){q_2}_{\|}}{N(\boldsymbol{q}_1) N(\boldsymbol{q}_1+\boldsymbol{q}_2)^2 \left(N(\boldsymbol{q}_1)+N(\boldsymbol{q}_2)+N(\boldsymbol{q}_1+\boldsymbol{q}_2)\right)}
\end{align}
where, $N(\boldsymbol{q})= \boldsymbol{q}^2+1$. The UV-divergent parts of the above integral are expanded in powers of $\frac{1}{\epsilon}$ by employing dimensional regularization scheme with the help of the package SecDec\cite{carter2011secdec, bruns2010normaliz, bruns2012power}.
\begin{align}\label{div1}
\widehat{\mathcal{D}} & \left( \int_{\boldsymbol{q}_1,\boldsymbol{q}_2}  \frac{{q_1}_{\|}{q_2}_{\|} }{N(\boldsymbol{q}_1) N(\boldsymbol{q}_1+\boldsymbol{q}_2)^2 \left(N(\boldsymbol{q}_1)+N(\boldsymbol{q}_2)+N(\boldsymbol{q}_1+\boldsymbol{q}_2)\right)} \right)
 = \frac{1}{256 \pi^4}\left( \frac{0.125}{\epsilon ^2}-\frac{0.563592}{\epsilon } \right)
 \nonumber
\\
\text{and} \nonumber
\\
\widehat{\mathcal{D}} & \left( \int_{\boldsymbol{q}_1,\boldsymbol{q}_2}  \frac{{q_2}_{\|}^2 }{N(\boldsymbol{q}_1) N(\boldsymbol{q}_1+\boldsymbol{q}_2)^2 \left(N(\boldsymbol{q}_1)+N(\boldsymbol{q}_2)+N(\boldsymbol{q}_1+\boldsymbol{q}_2)\right)} \right)
 = \frac{1}{256 \pi^4}\left( -\frac{0.75}{\epsilon ^2}+\frac{0.290792}{\epsilon } \right).
\end{align}
Eq. (\ref{intexample}) together with Eq.~(\ref{div1}) gives the UV-divergent parts of diagram~22,
\begin{equation}
\widehat{\mathcal{D}} \left( \mathcal{I}_{22} \right) =-\frac{3}{512 \pi^4} \frac{T^2 u_0 e_p^2}{D^4 \rho^2} r^{d-3} \left( \frac{0.625}{\epsilon^2}+ \frac{0.2728}{\epsilon}\right).
\end{equation}

\section{$\Gamma_{1,1}^{11}(0)$} \label{appg110}

Table \ref{g1101ltable} shows the only one-loop diagram contributing to $\Gamma_{1,1}^{11}(0)$ and its divergent contribution.

\begin{longtable}{|ccL|}
\hline
Diagrams & & \text{Divergence in $\epsilon$-expansion} \\
\hline
\raisebox{-0.5\height}{\includegraphics[scale=0.2]{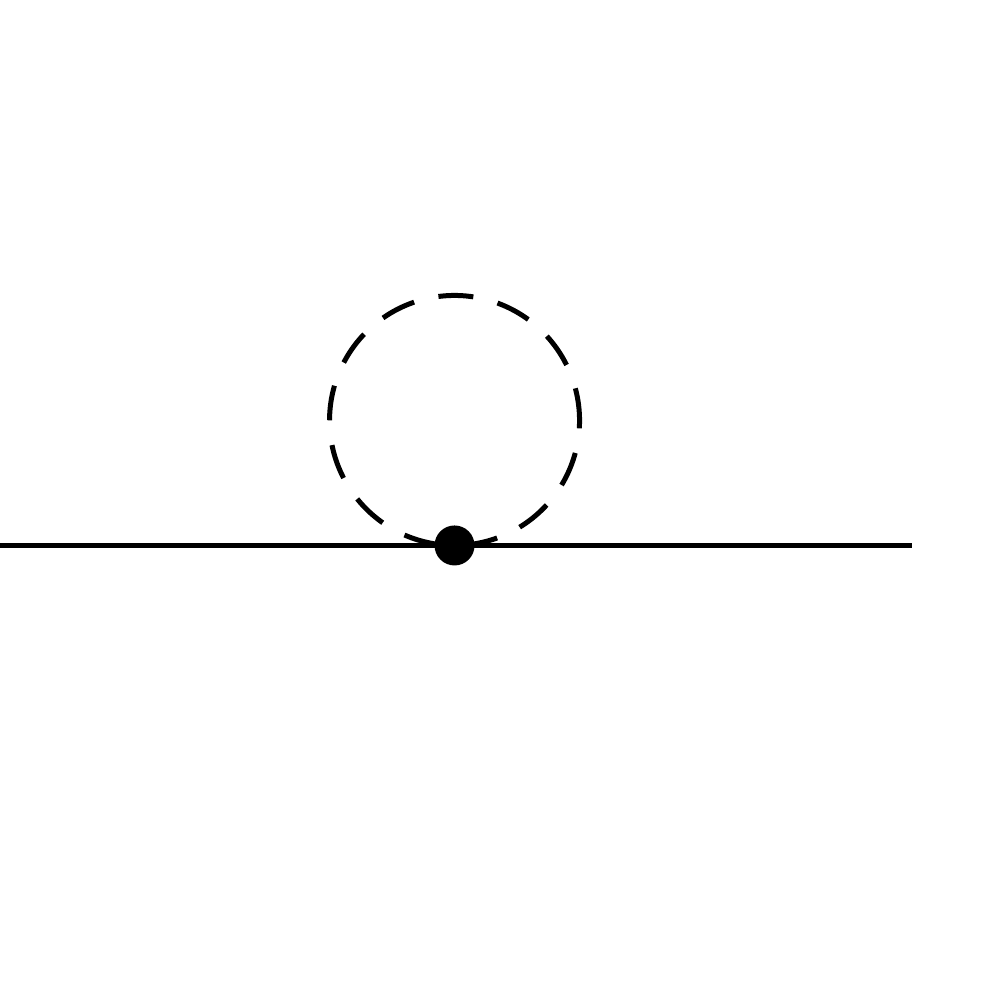}}& & D r^{1-\epsilon/2} \left( \frac{g_0}{2}+g_1 \right) \frac{1}{\epsilon} \\
\hline
\caption{\label{g1101ltable} One-loop contribution to $\Gamma_{1,1}^{11}(0)$}\\
\end{longtable}
The divergent parts of the two-loop diagrams contributing to $\Gamma_{1,1}^{11}(0)$ have the general form,
\begin{align}
\widehat{\mathcal{D}} \left( \mathcal{I}(q=0) \right)= r^{1-\epsilon} D \mathcal{A} \left( \frac{n}{\epsilon}+\frac{m}{\epsilon^2} \right)
\end{align}
where, $\mathcal{I}$ is the integral that a diagram represents and $n$ and $m$ are real numbers. The factor $\mathcal{A}$ is a function of the modified couplings  
\begin{align}
g_0 \equiv \frac{T}{8 \pi^2 D^2 \rho^{1/2}} u_0, \text{  } g_1 \equiv \frac{T}{8 \pi^2 D^2 \rho^{1/2}}u_1 \textbf{  } \text{and} \textbf{   }
g_2 \equiv \frac{T}{8 \pi^2 D^3 \rho^{3/2}}e_p^2.  
\end{align}
Table \ref{g110table} shows the two-loop diagrams and their respective contributions.
\begin{longtable}{|ccc|ccc|}
\hline
Diagram & & $\mathcal{A},n,m$ & Diagram & &  $\mathcal{A},n,m$ \\
\hline
%1
\raisebox{-0.5\height}{\includegraphics[scale=0.13]{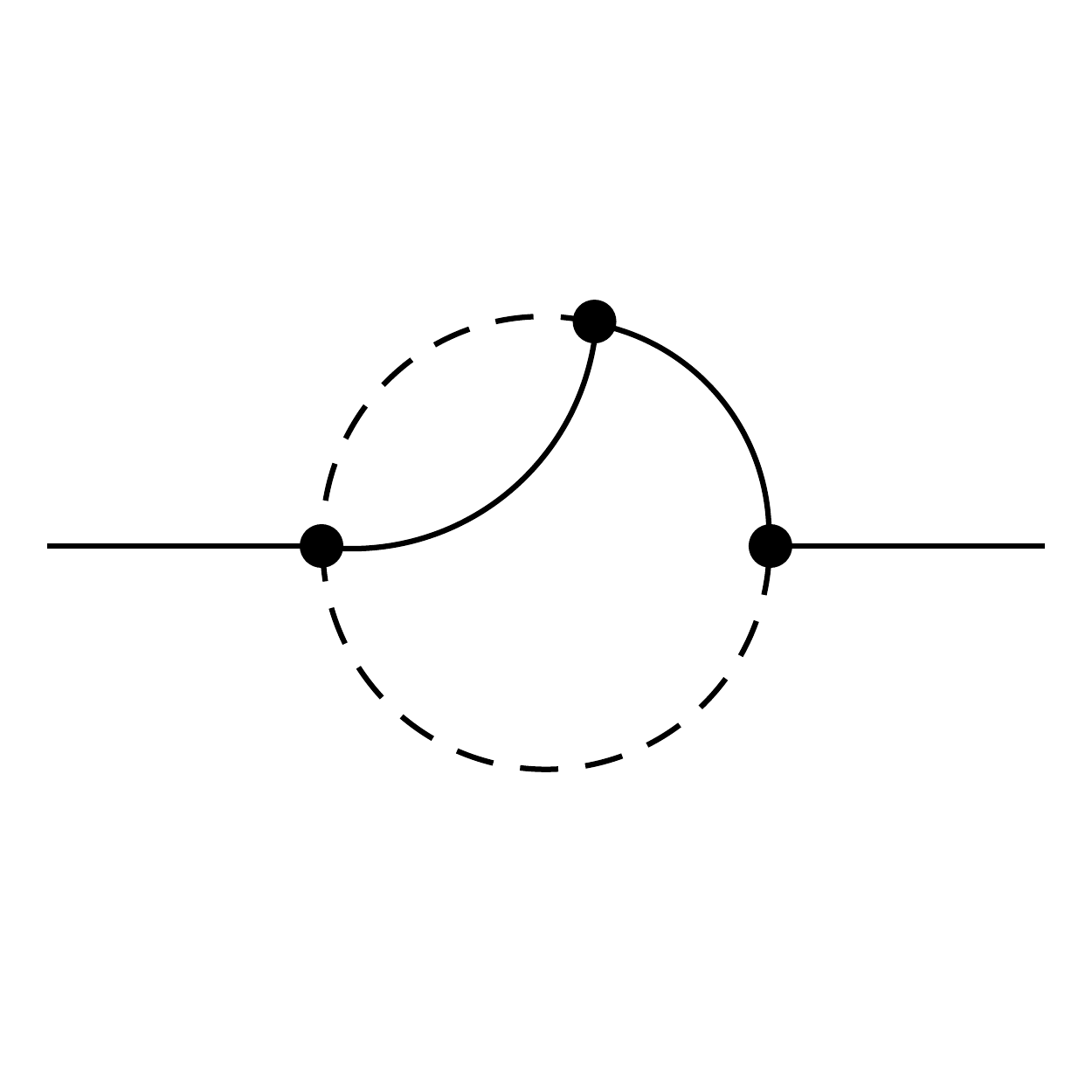}} & & $g_1 g_2,-0.1364,-0.3125$  & 
\raisebox{-0.5\height}{\includegraphics[scale=0.13]{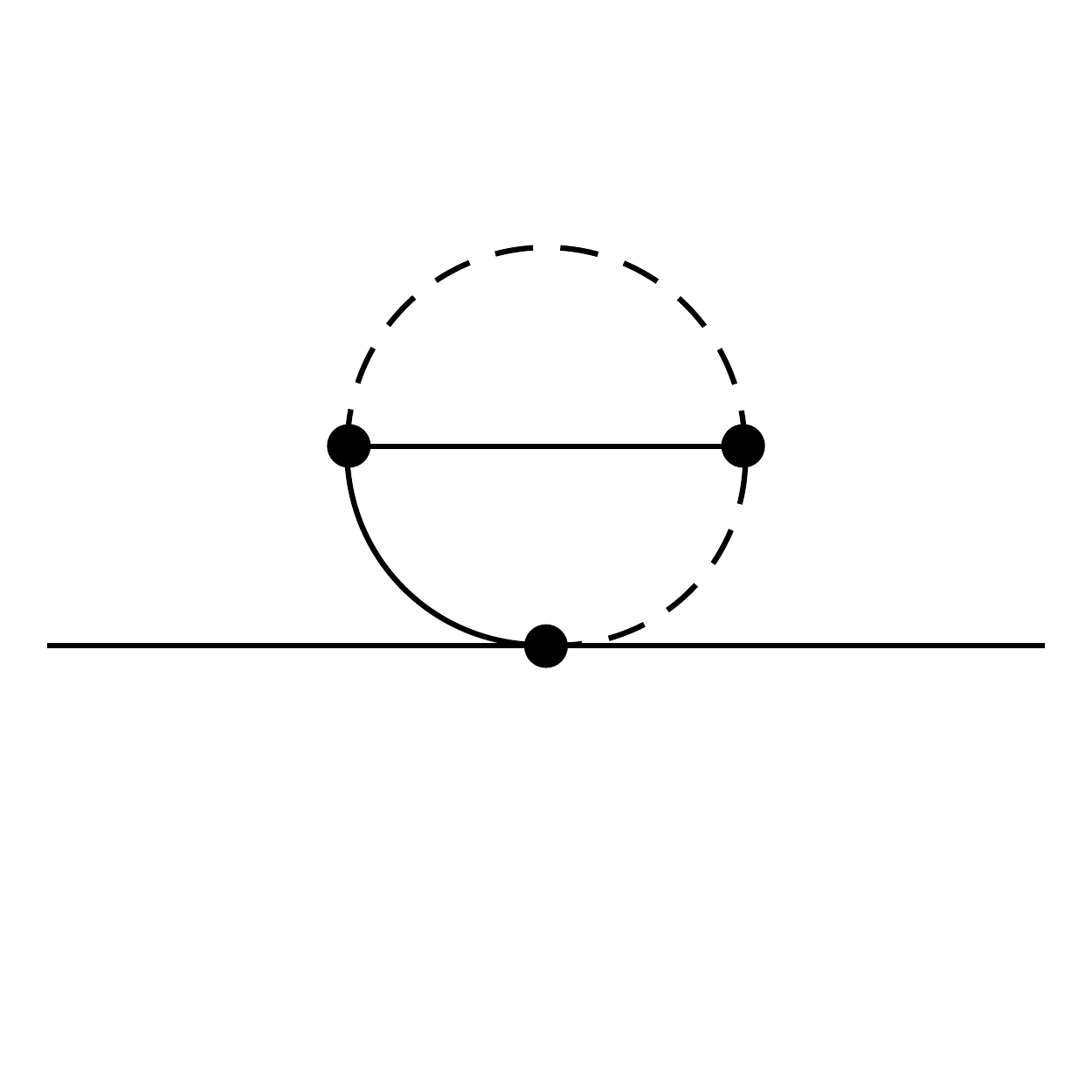}}  &   &$\left(g_0+2 g_1\right) g_2,-0.0341,-0.078125$\\
%2
\raisebox{-0.5\height}{\includegraphics[scale=0.13]{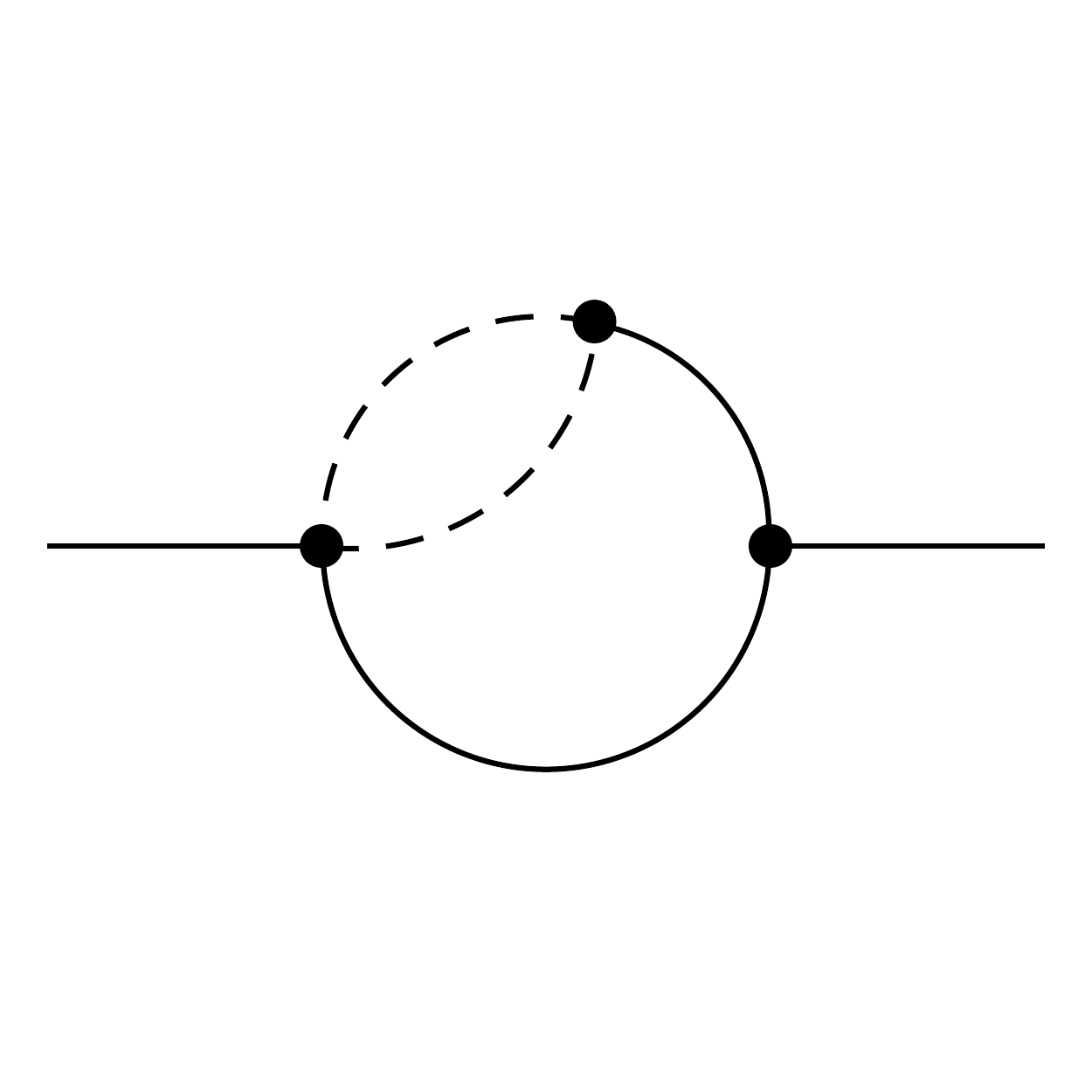}}  &  & $g_1 g_2,-0.133328,-0.125$ & 
\raisebox{-0.5\height}{\includegraphics[scale=0.13]{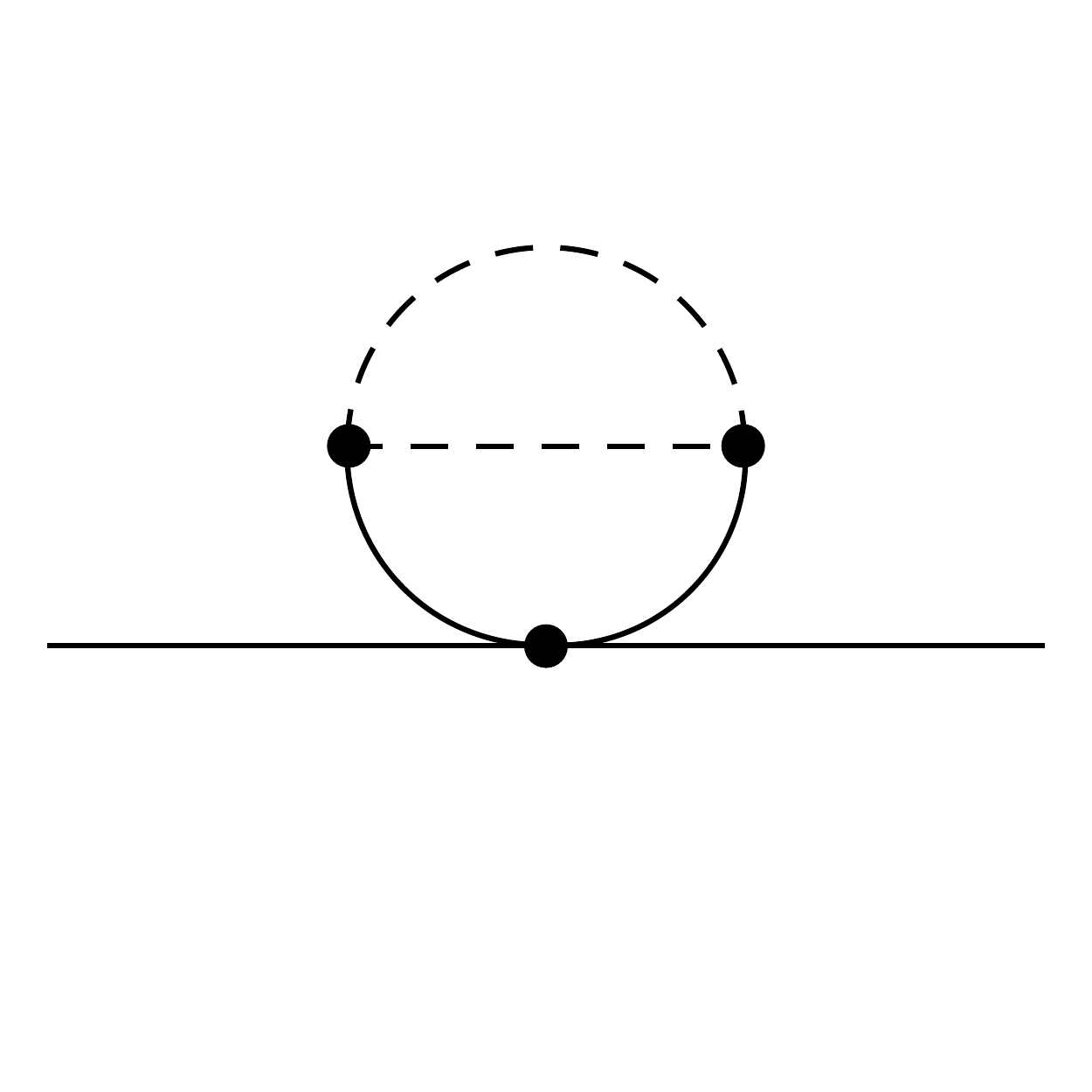}}  &  & $\left(g_0+2 g_1\right) g_2,0.0846443,0.0625$\\
%3
\raisebox{-0.5\height}{\includegraphics[scale=0.13]{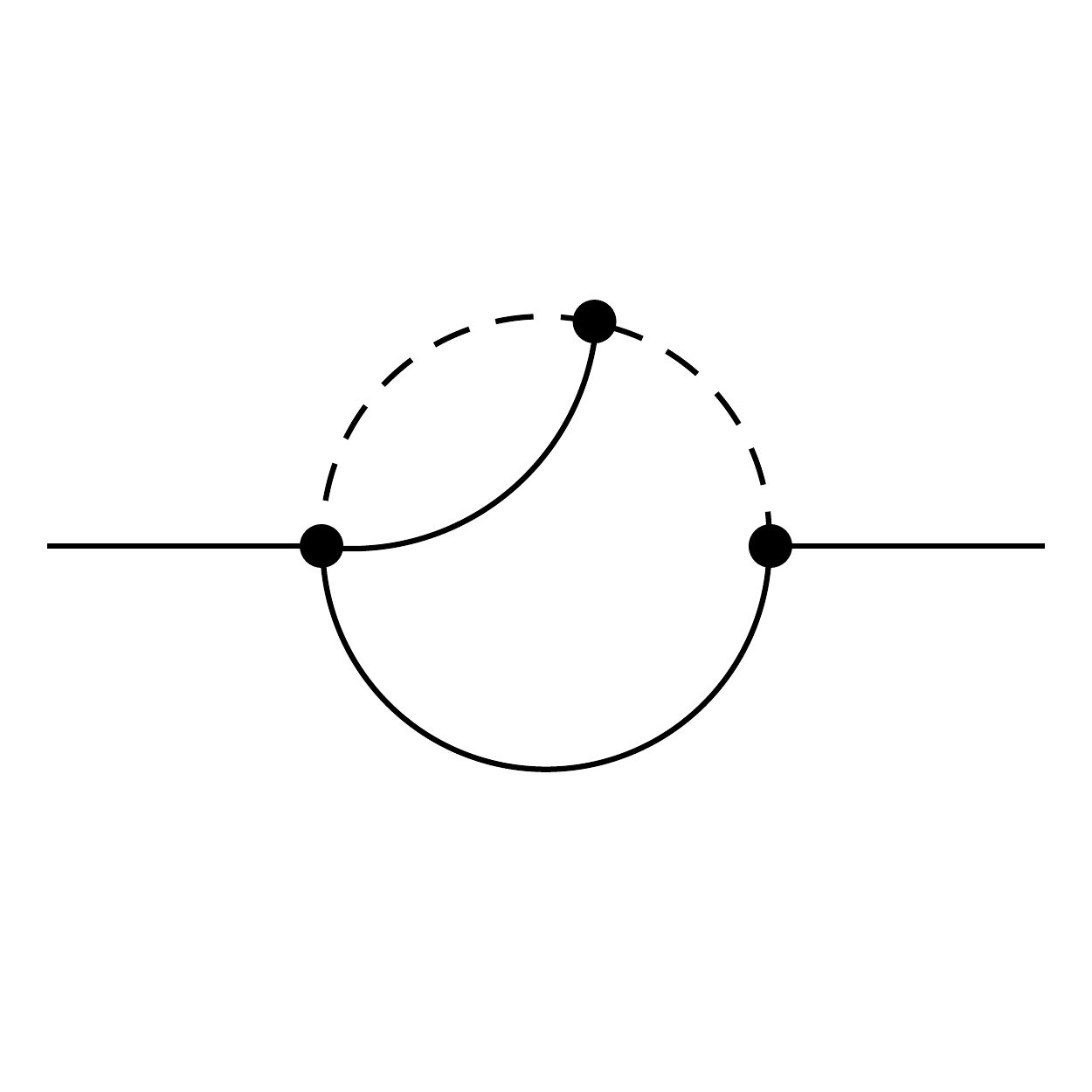}}  &   &
$g_1 g_2,0.269728,0.4375$ &
\raisebox{-0.5\height}{\includegraphics[scale=0.13]{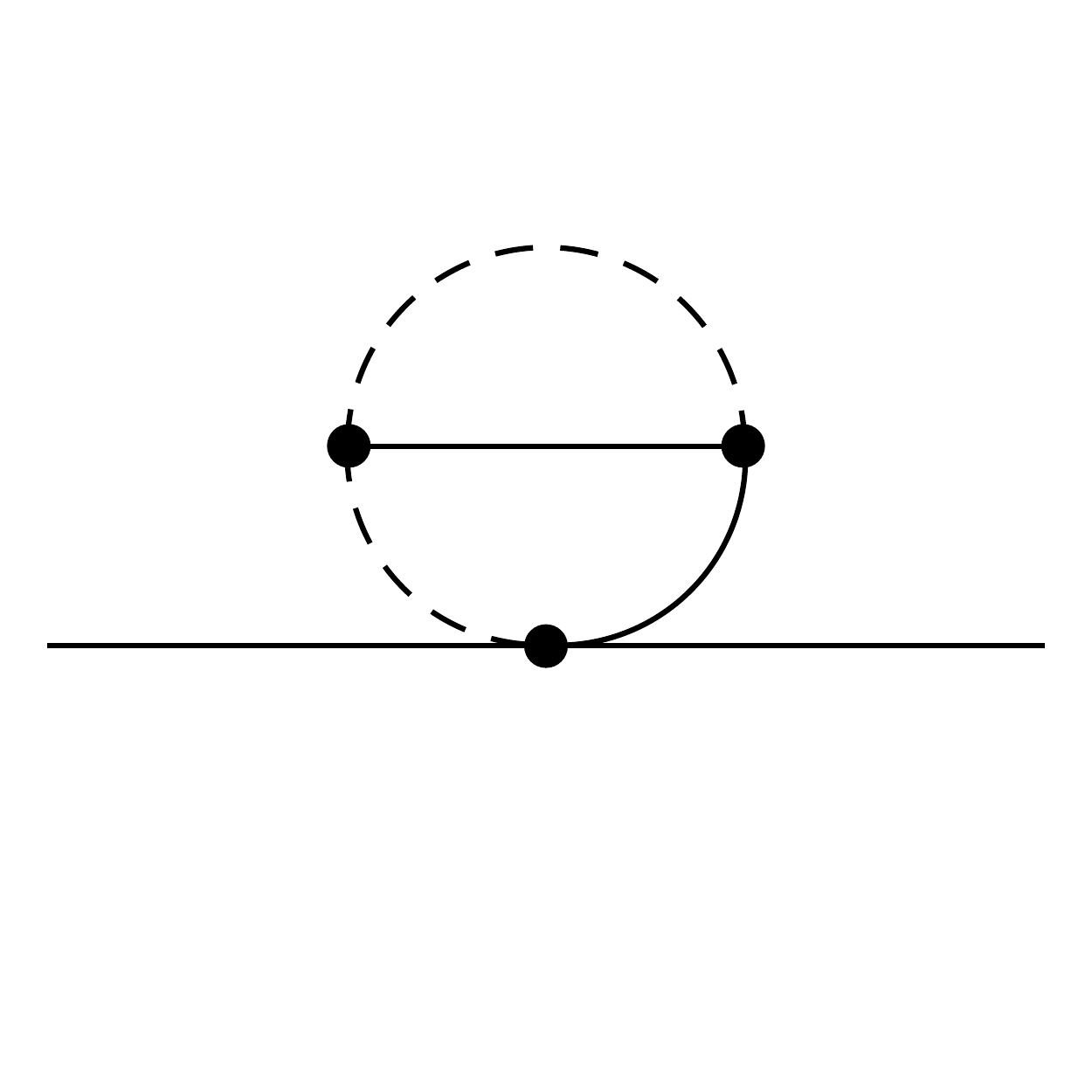}} &   & $\left(g_0+2 g_1\right) g_2,-0.0341,-0.078125$ \\
%4
\raisebox{-0.5\height}{\includegraphics[scale=0.13]{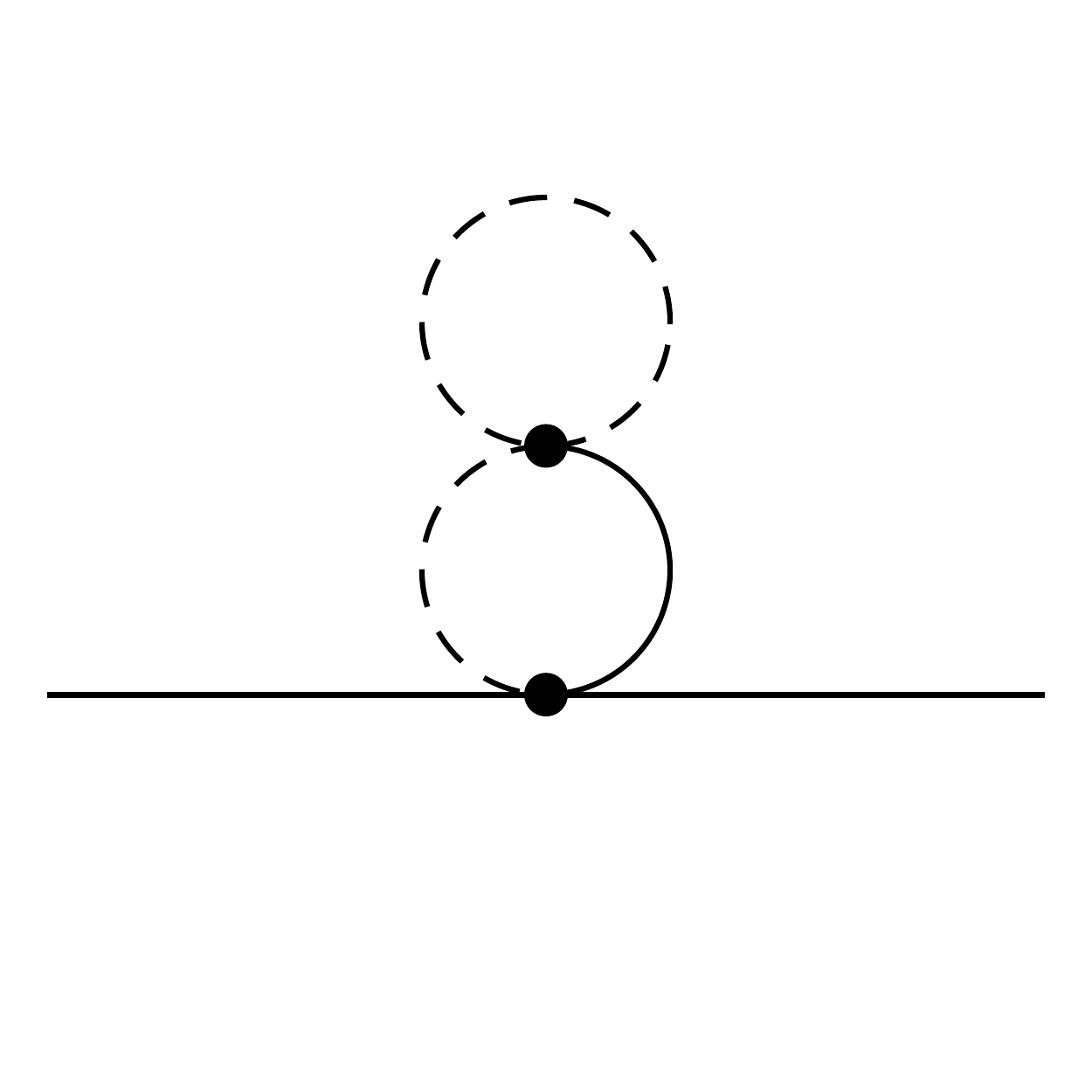}}  &   & $\left(g_0+2 g_1\right){}^2,0.0193039,-0.25$ &
\raisebox{-0.5\height}{\includegraphics[scale=0.13]{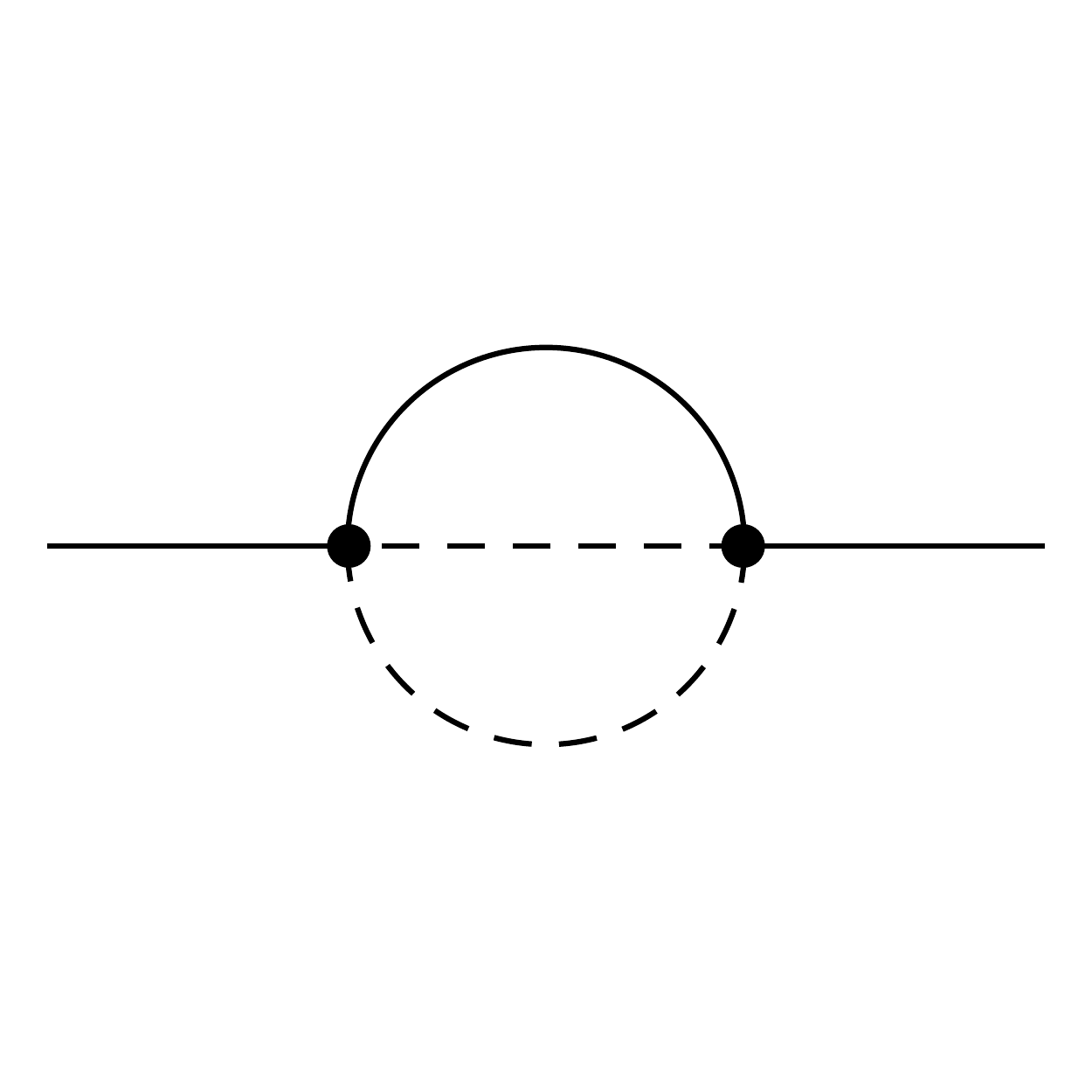}} &  & $g_0^2+6 g_1^2,-0.230696,-0.25$ \\
\hline
\caption{\label{g110table}Two-loop contributions to $\Gamma_{1,1}^{11}(0)$}\\
\end{longtable}

%Using Eq.~(\ref{rgconst}) and the renormalization condition, ${\gammar^{11}}_{1,1} (q_i=0)=D_R r_R \mu^2$, we can now write down the renormalization constant $Z_r$ to two-loop order as
Collecting the divergences from all the above diagrams and applying the renormalization condition~(\ref{zrcond}) we obtain the renormalization constant $Z_r$. In terms of the dimensionless renormalized couplings $\lambda_i$, which are defined in Eq.~(\ref{lambdas}), $Z_r$ can be written explicitly as 
\begin{align}\label{zr}
Z_r=& 1+ \frac{1}{\epsilon} \left(\lambda _1 +0.5 \lambda _0 -0.125 \lambda _0^2-0.75 \lambda _1^2+0.0560803 \lambda _0 \lambda _2+0.112161 \lambda _1 \lambda _2 \right)
+ \frac{1}{\epsilon^2} \left( 0.5 \lambda _0^2+2.5 \lambda _1^2+\lambda _0 \lambda _1 \right.
\nonumber
\\
 & \left. +0.09375 \lambda _0 \lambda _2 +0.1875 \lambda _1 \lambda _2
\right)
\end{align}

%\subsection{two-loop contributions to $ \left.\frac{d}{di q_{0}} _{q=0}\Gamma_{1,1}(q)\right|_{q=0}$}

\section{$\frac{\partial}{\partial q_\|^2}\Gamma_{1,1}^{11}  (q) \left. \right|_{q =0} $} 
Table \ref{g11pL1ltable} shows the only one-loop diagram contributing to  $\frac{\partial}{\partial q_\|^2}\Gamma_{1,1}^{11}  (q) \left. \right|_{q =0} $ and its divergent contribution.
\begin{longtable}{|ccL|}
\hline
Diagrams & & \text{Divergence in $\epsilon$-expansion} \\
\hline
\raisebox{-0.5\height}{\includegraphics[scale=0.25]{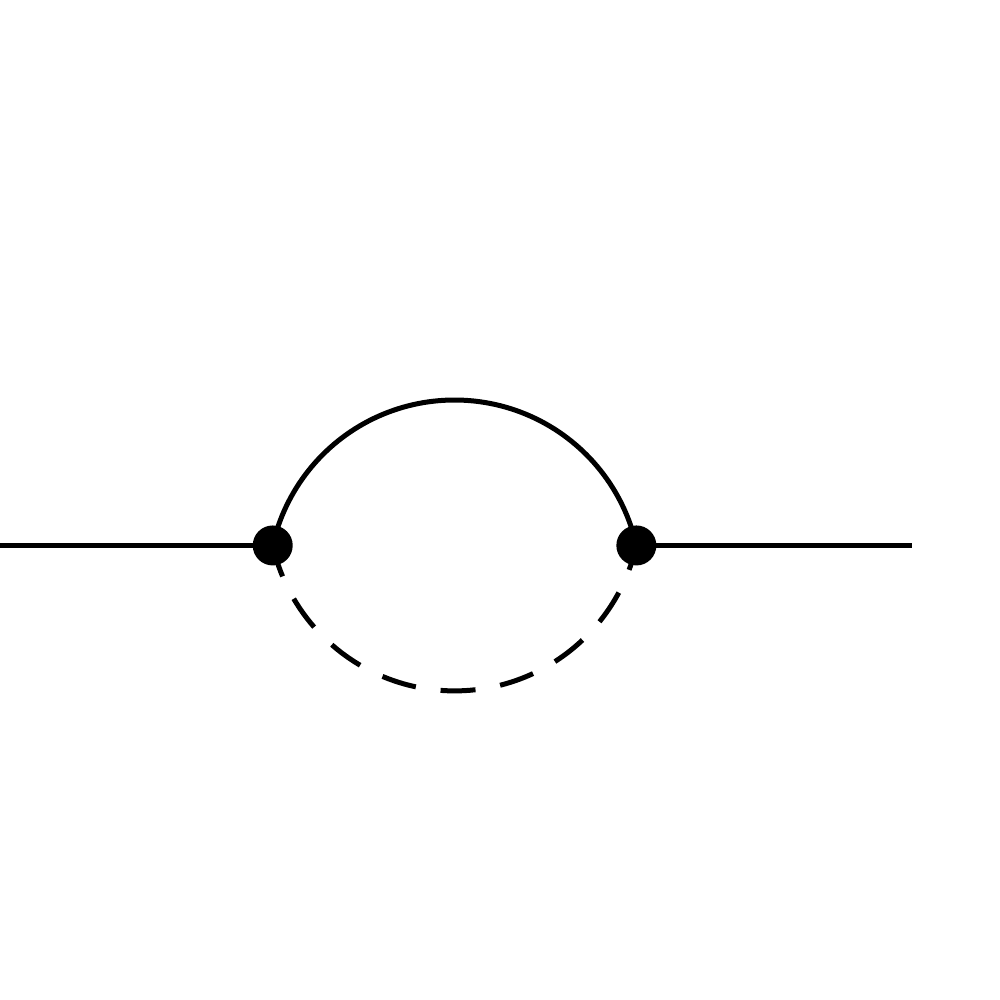}}& & -r^{-\epsilon/2} D \rho g_2 \left(\frac{0.75}{\epsilon}\right)  \\
\hline
\caption{\label{g11pL1ltable} One-loop contribution to $\frac{\partial}{\partial q_\|^2}\Gamma_{1,1}^{11}  (q) \left. \right|_{q =0} $}\\
\end{longtable}

The divergent parts of the two-loop diagrams contributing to $\frac{\partial}{\partial q_\|^2}\Gamma_{1,1}^{11}  (q) \left. \right|_{q =0} $ have the general form $\reps D \rho \mathcal{A} \left(\frac{n}{\epsilon} + \frac{m}{\epsilon^2} \right)$. Table \ref{g11pLtable} shows these diagrams and their respective divergences. 

\begin{center}
\begin{longtable}{|ccc|ccc|} 
\hline
\text{Diagram} &  & $\mathcal{A}, n, m$ & \text{Diagram } &  & $\mathcal{A}, n, m$ \\
\hline
%\endfirsthead
%1
\raisebox{-0.5\height}{\includegraphics[scale=0.15]{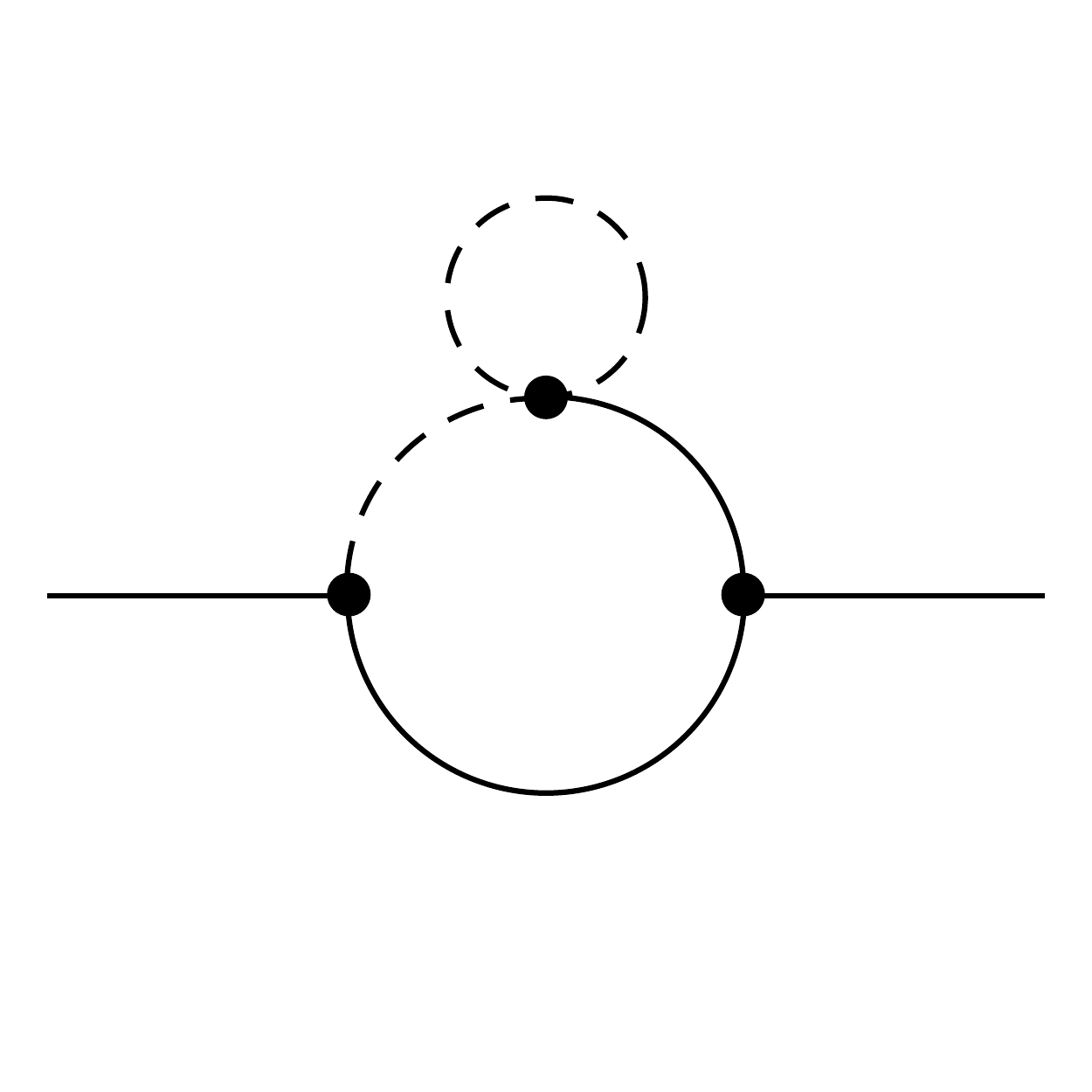}} &  & $g_0 g_2+2 g_1 g_2,-0.0520833,0$  & 
\raisebox{-0.5\height}{\includegraphics[scale=0.15]{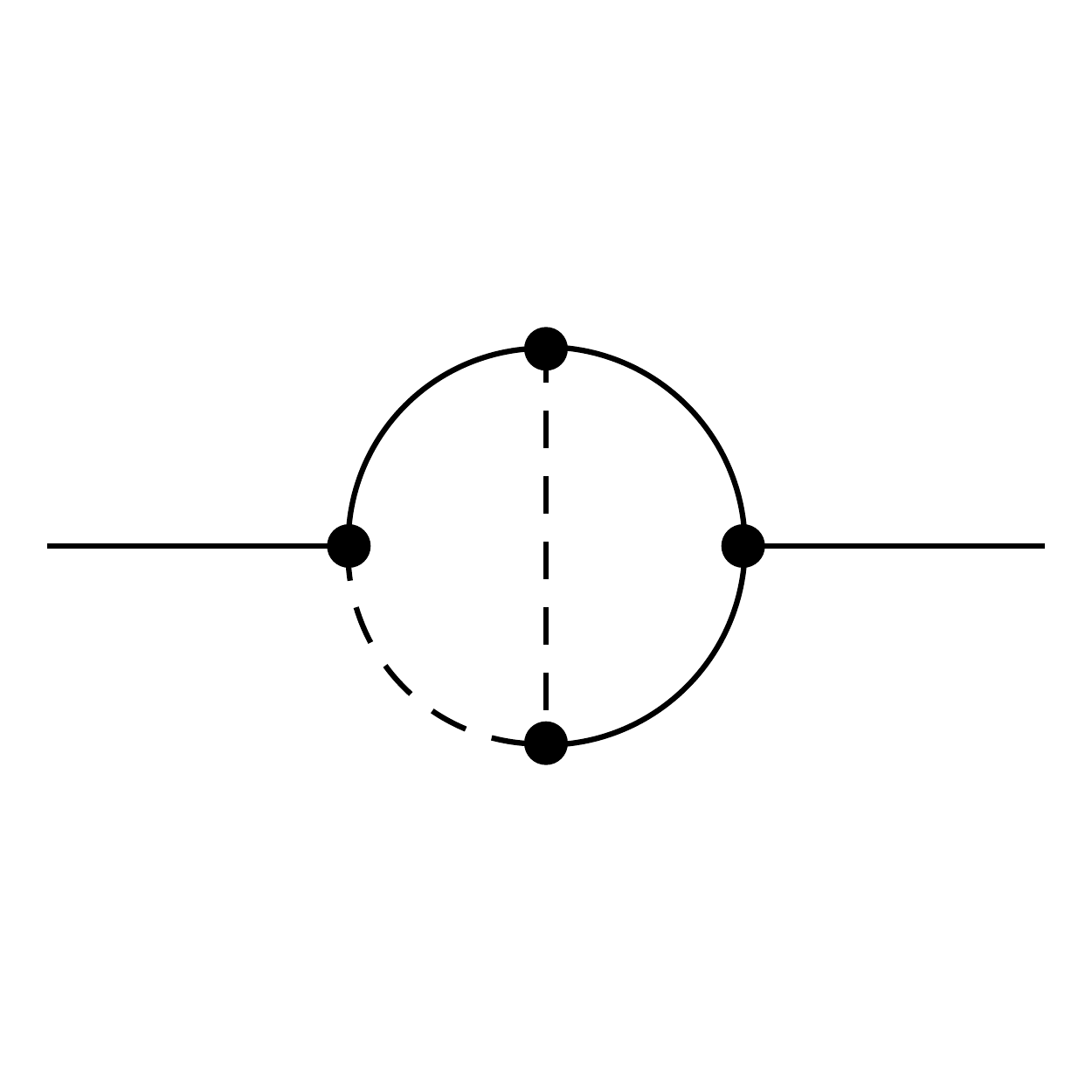}}  &   &$g_2^2,-0.00835706,0.0234375$\\
%2
\raisebox{-0.5\height}{\includegraphics[scale=0.15]{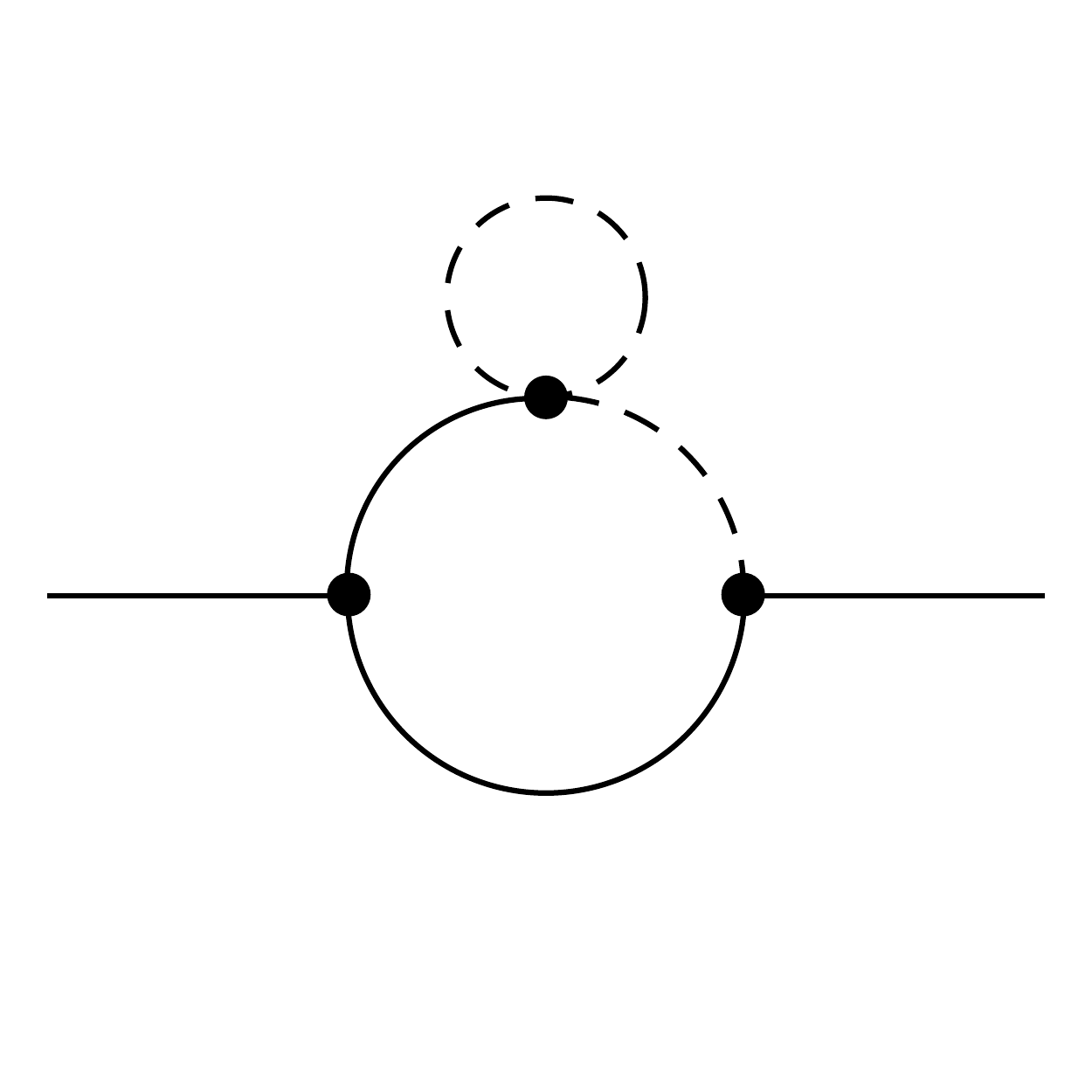}}  &   & $g_0 g_2+2 g_1 g_2,-0.09375,0$ & 
\raisebox{-0.5\height}{\includegraphics[scale=0.15]{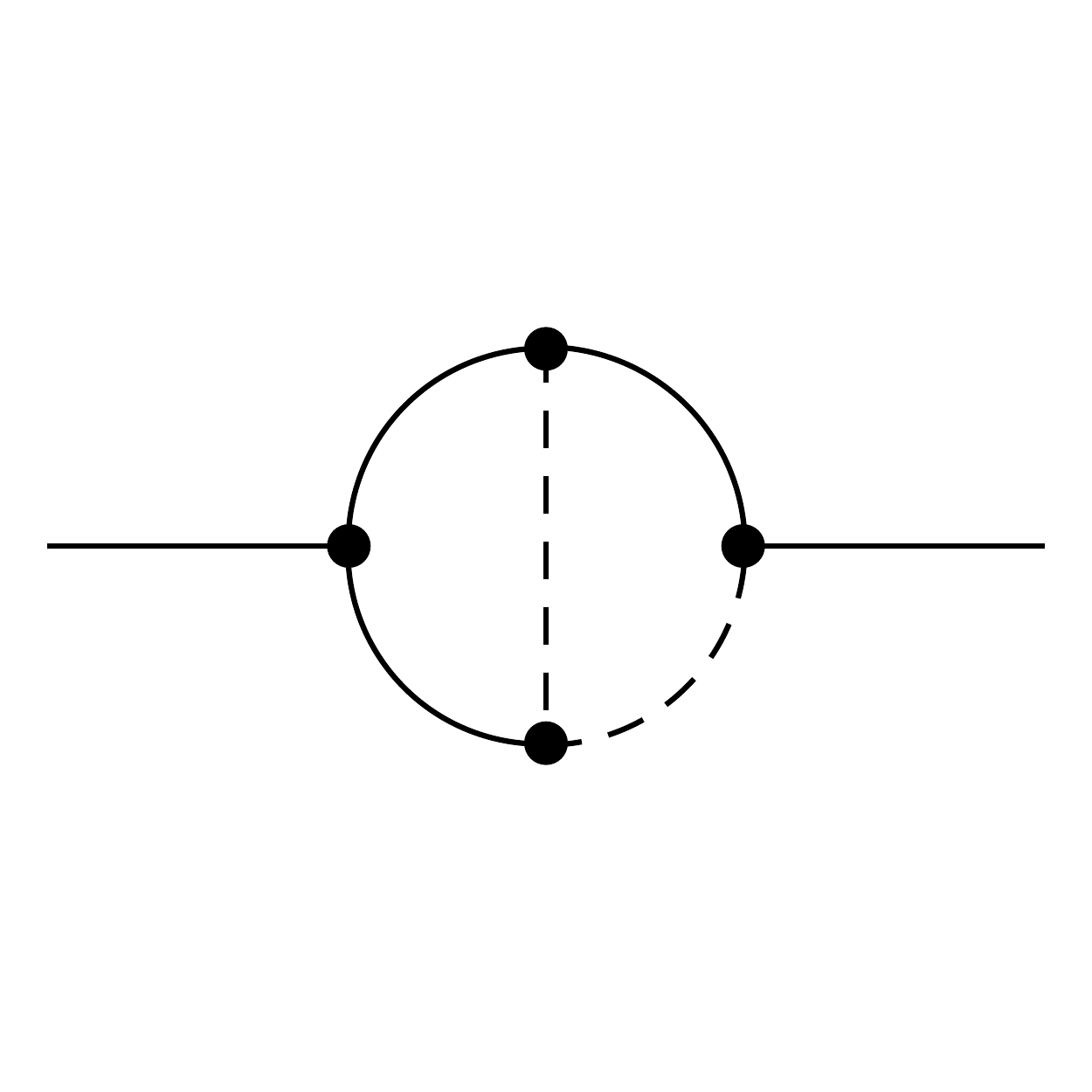}}  &  & $g_2^2,0.0020853,-0.046875 $\\
%3
\raisebox{-0.5\height}{\includegraphics[scale=0.15]{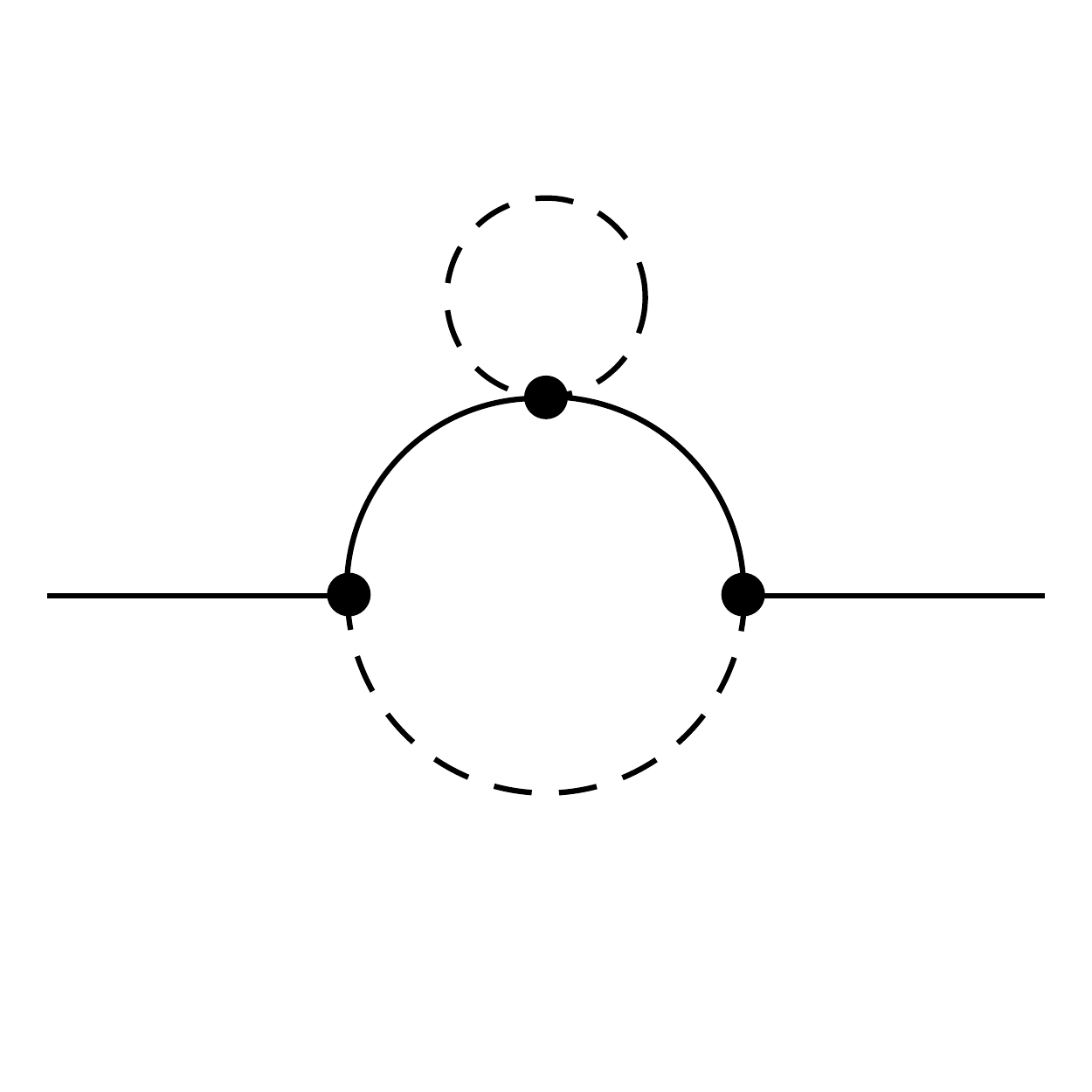}}  &   &
$g_0 g_2+2 g_1 g_2,-0.0416667,0 $ &
\raisebox{-0.5\height}{\includegraphics[scale=0.15]{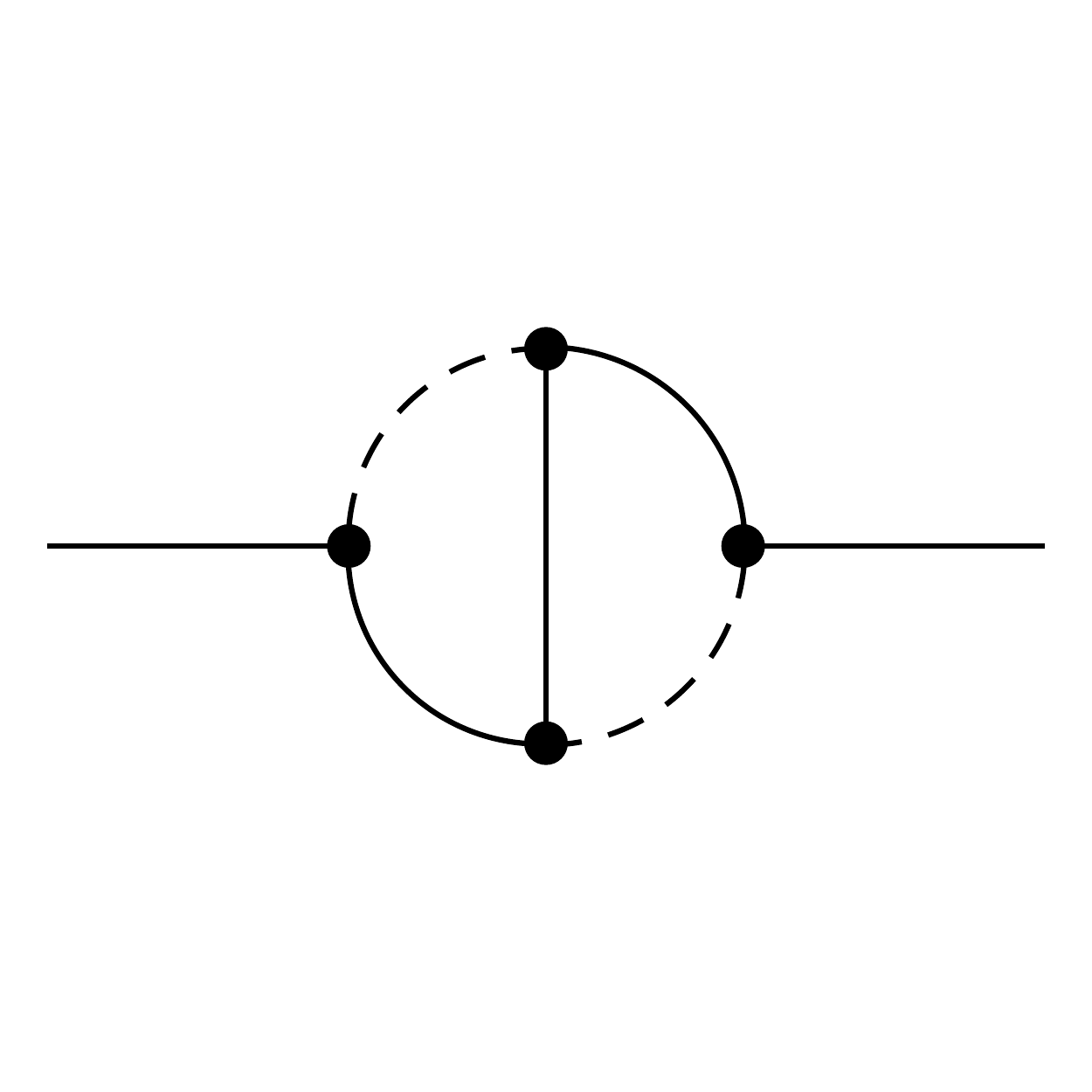}} &   & $g_2^2,-0.0180942,0.0234375 $ \\
%4
\raisebox{-0.5\height}{\includegraphics[scale=0.15]{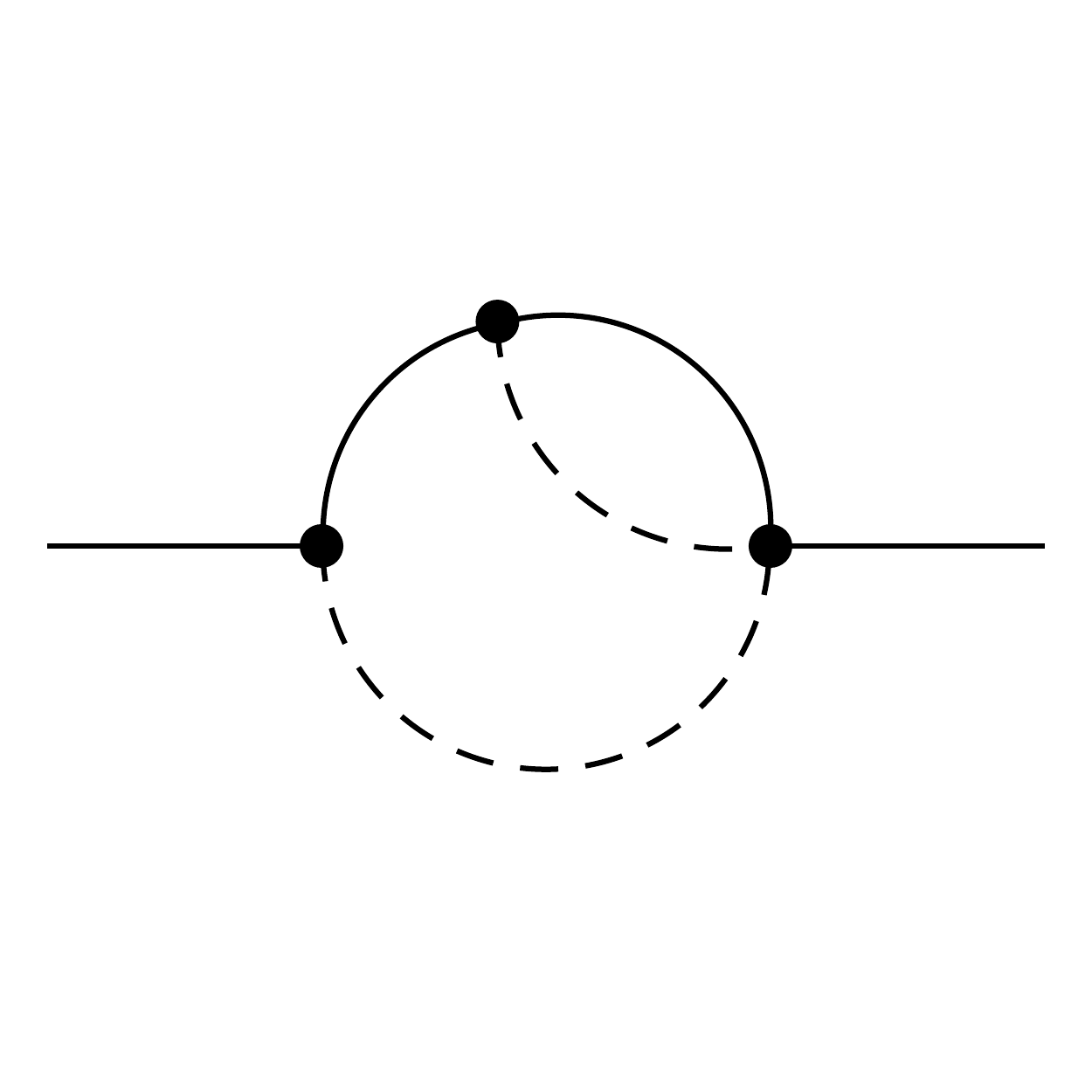}}  &   & $g_1 g_2,-0.306357,0.375$ &
\raisebox{-0.5\height}{\includegraphics[scale=0.15]{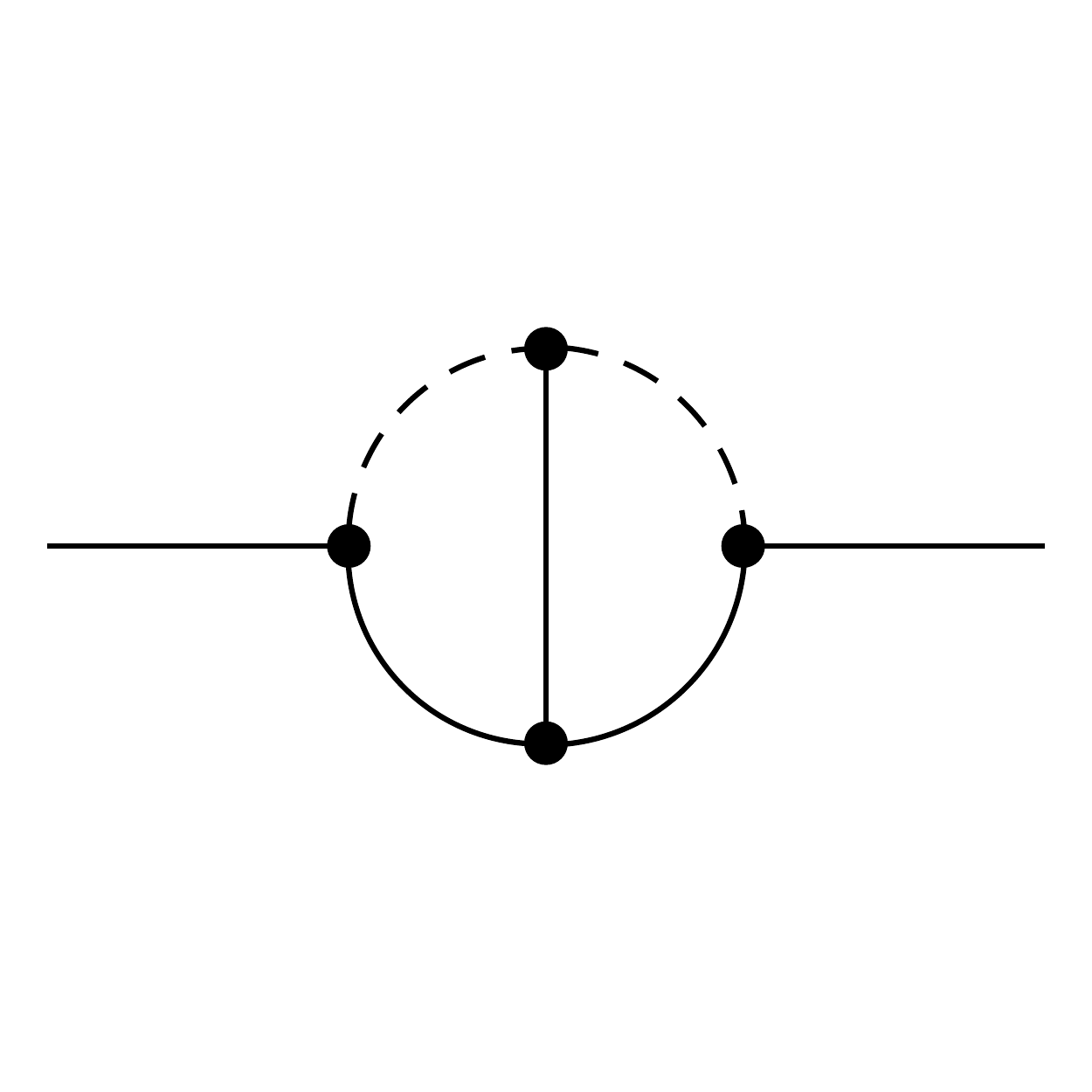}} &  & $g_2^2,0.018626,-0.0117188$ \\
%5
\raisebox{-0.5\height}{\includegraphics[scale=0.15]{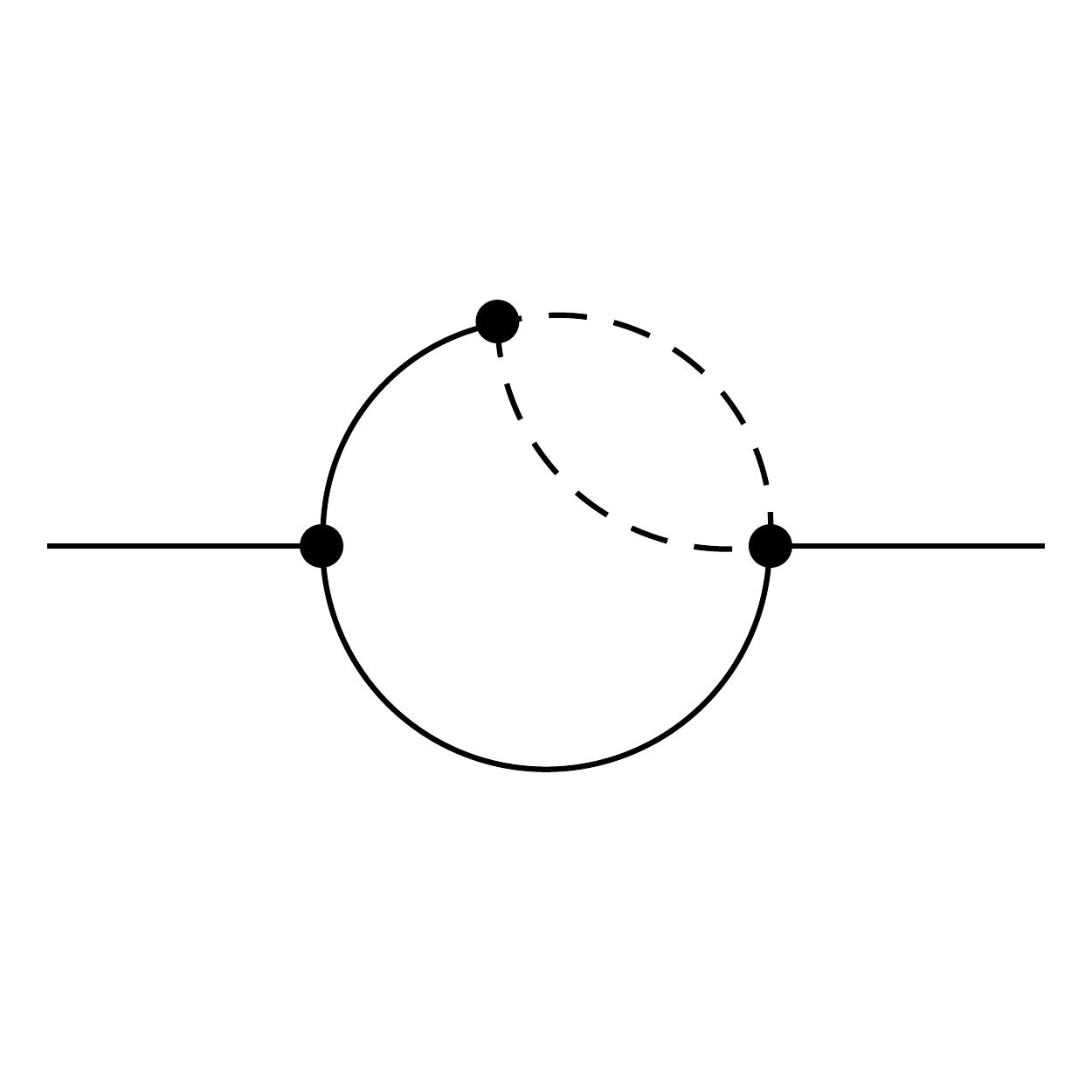}} &  & $g_1 g_2,0.143841,0$  & 
\raisebox{-0.5\height}{\includegraphics[scale=0.15]{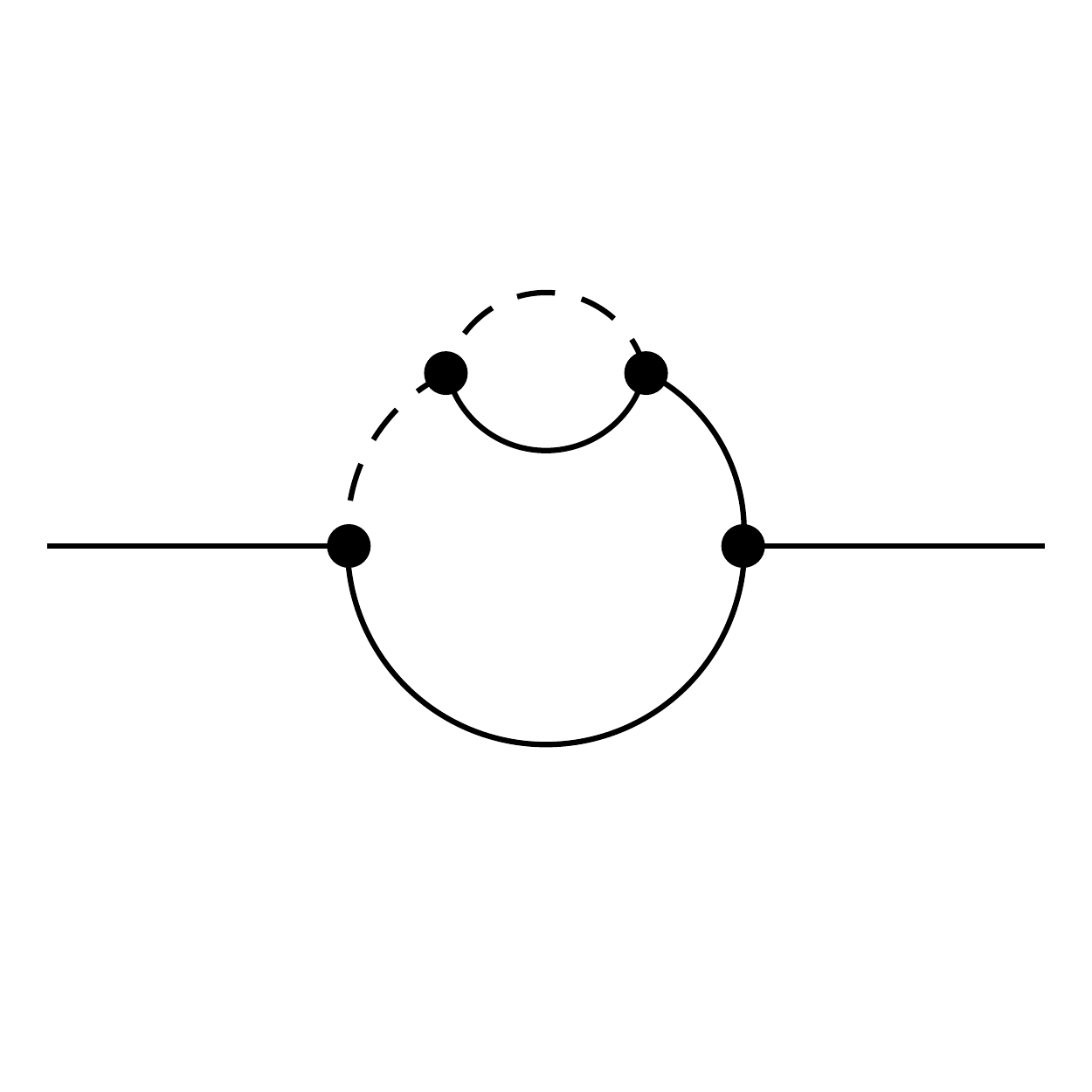}}  &   &$g_2^2,-0.0401051,0.0234375$\\
%6
\raisebox{-0.5\height}{\includegraphics[scale=0.15]{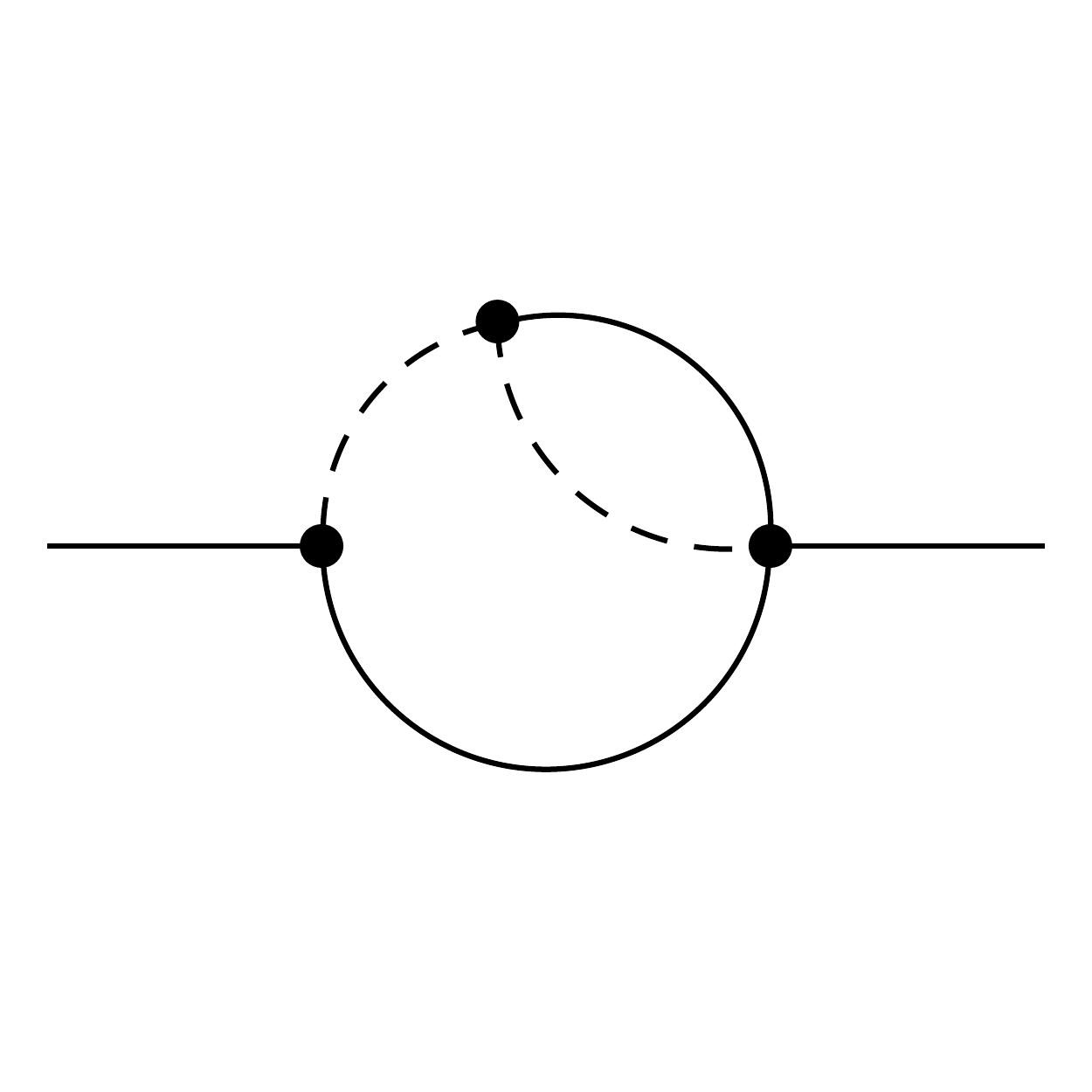}}  &   & $g_1 g_2,0.0875039,-0.09375$ & 
\raisebox{-0.5\height}{\includegraphics[scale=0.15]{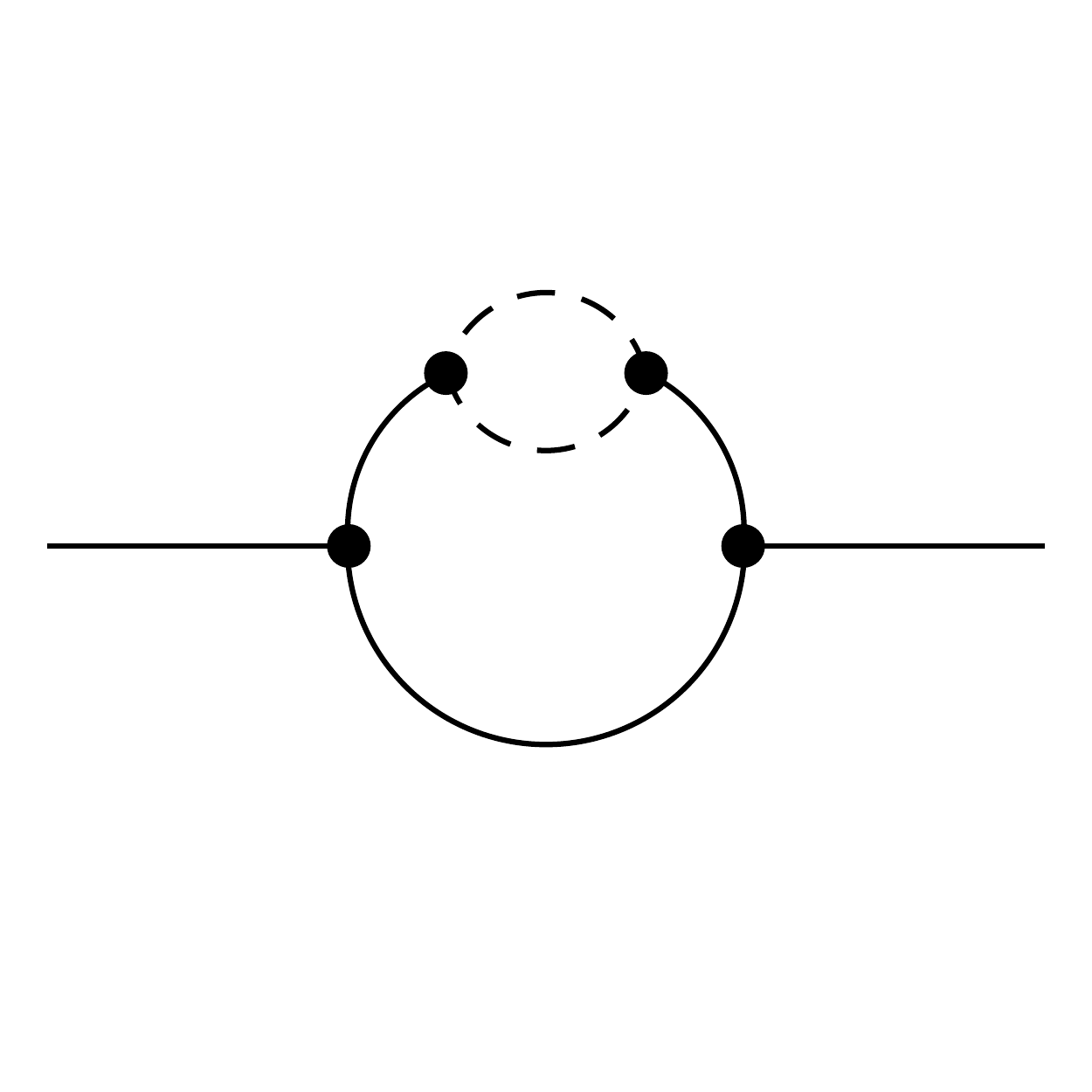}}  &  & $g_2^2,-0.0283968,0$\\
%7
\raisebox{-0.5\height}{\includegraphics[scale=0.15]{top3_69.pdf}}  &   &
$g_1 g_2,-0.216176,0.1875 $ &
\raisebox{-0.5\height}{\includegraphics[scale=0.15]{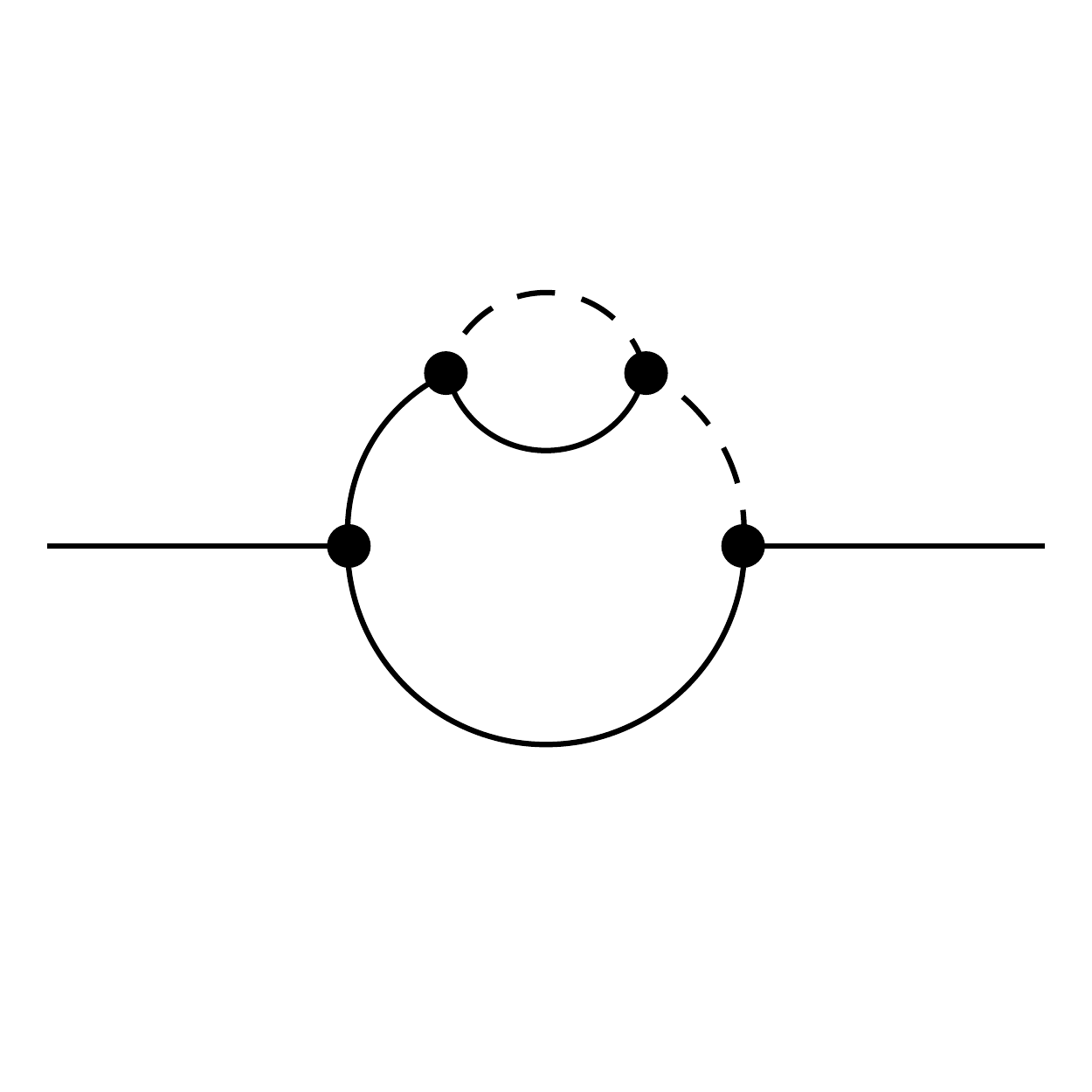}} &   & $g_2^2,-0.0580853,0.0234375 $ \\
%8
\raisebox{-0.5\height}{\includegraphics[scale=0.15]{top3_73.pdf}}  &   & $g_1 g_2,-0.0313306,0$ &
\raisebox{-0.5\height}{\includegraphics[scale=0.15]{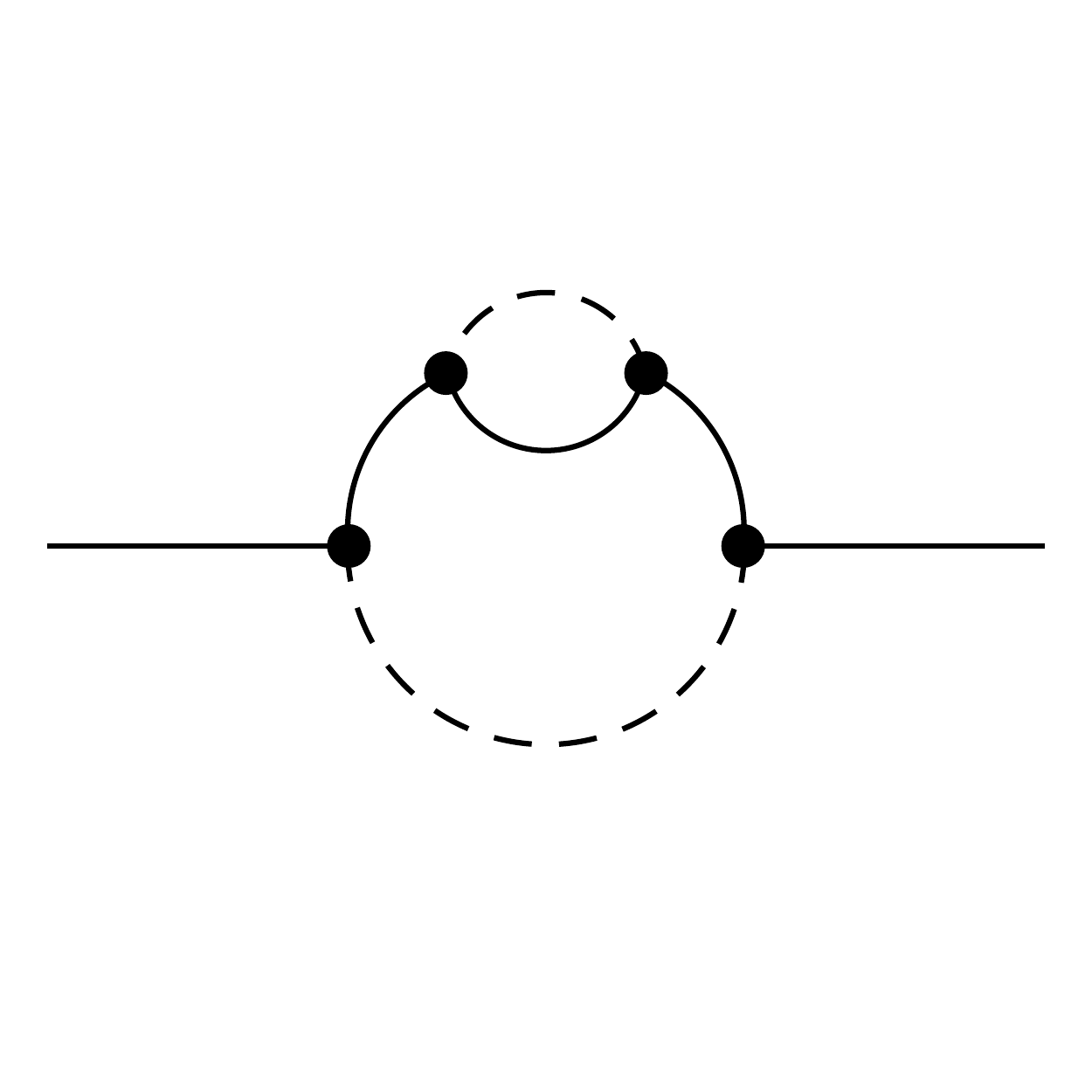}} &  & $g_2^2,-0.0794423,0.09375$ \\
%9
\raisebox{-0.5\height}{\includegraphics[scale=0.15]{top3_75.pdf}}  &   & $g_1 g_2,-0.000119615,0.09375$ &
\raisebox{-0.5\height}{\includegraphics[scale=0.15]{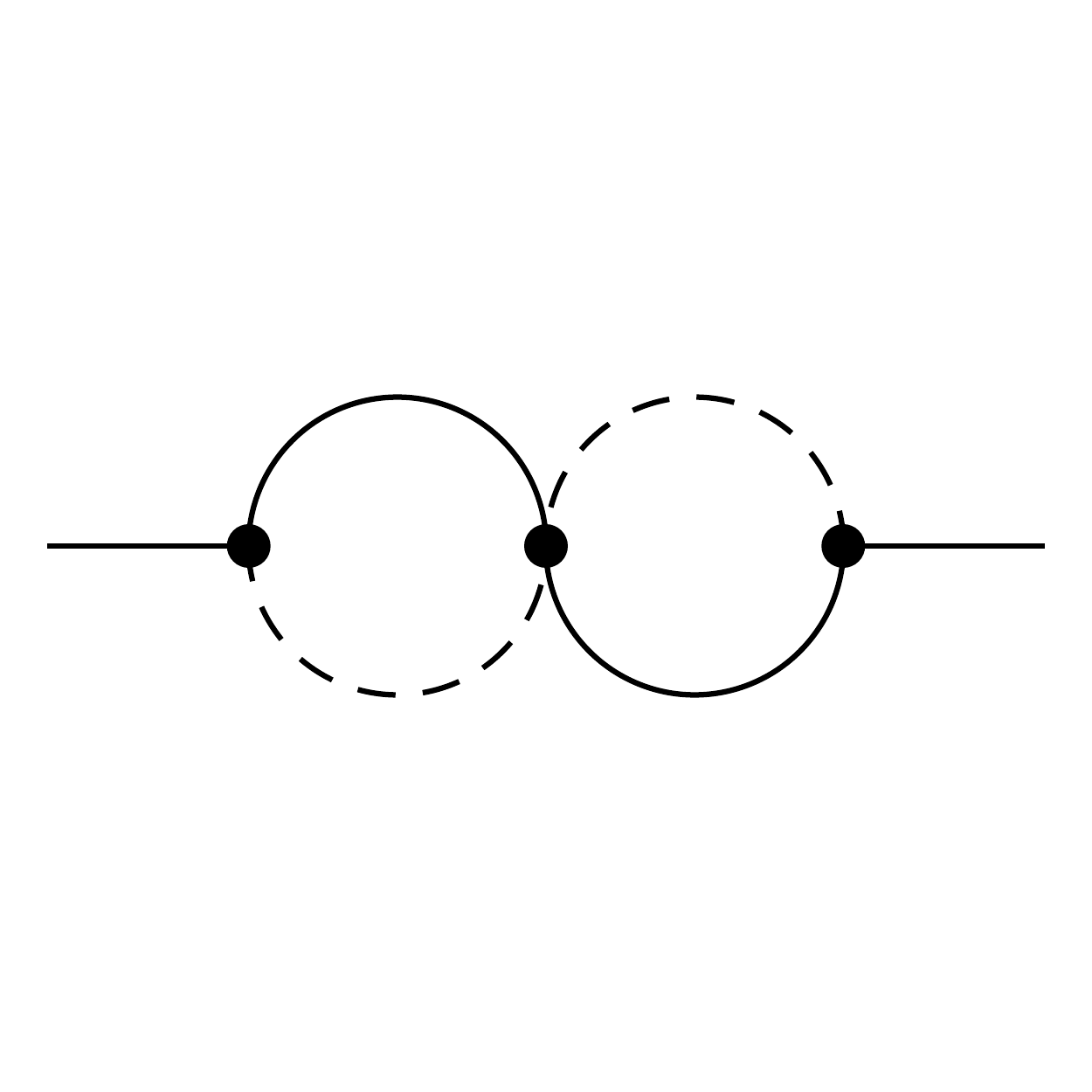}} &  & $g_1 g_2,-0.370412,0.75$ \\
%10
\raisebox{-0.5\height}{\includegraphics[scale=0.15]{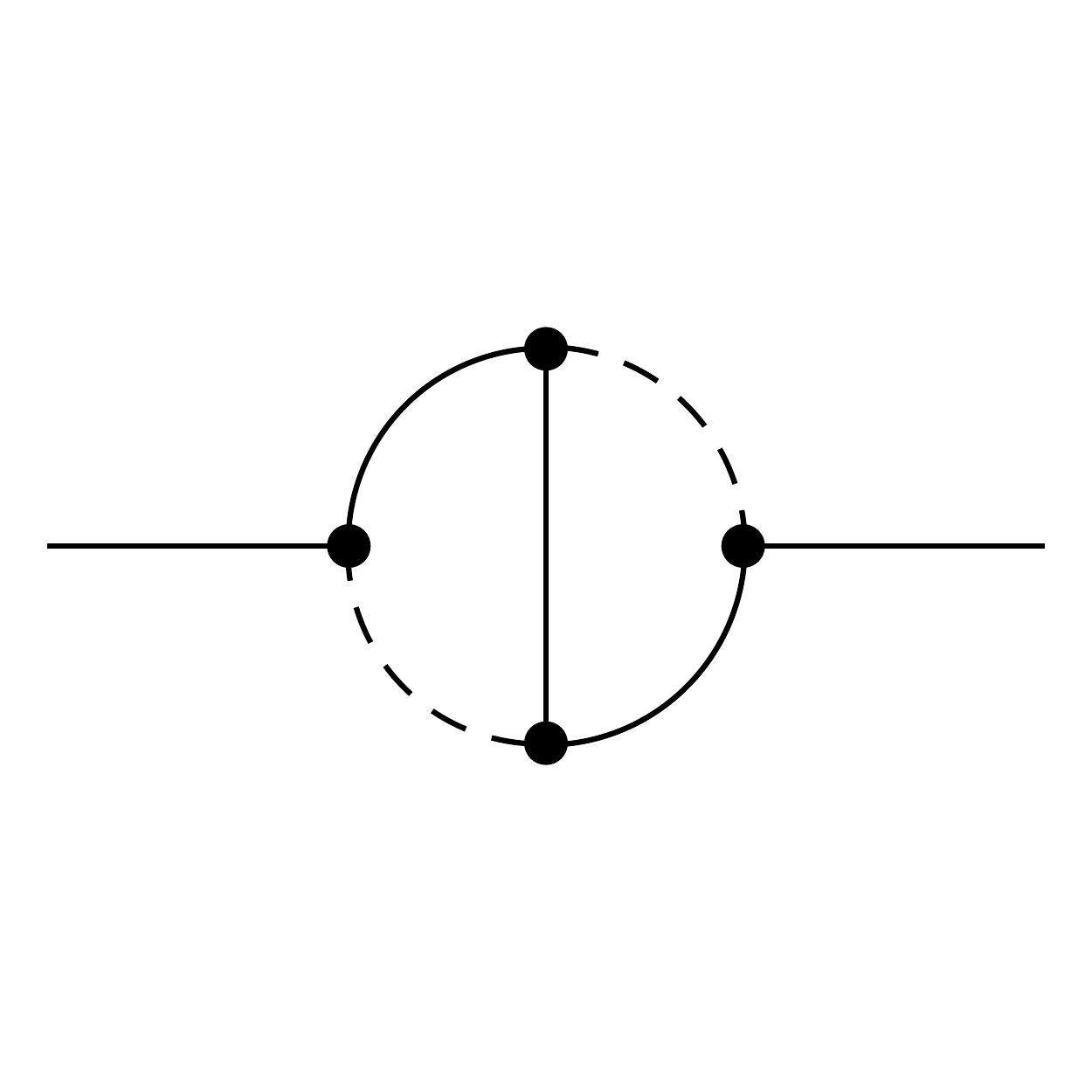}}  &   & $g_2^2,-0.0180942,0.0234375$ &
\raisebox{-0.5\height}{\includegraphics[scale=0.15]{top9_203.pdf}} &  & $g_0^2+6 g_1^2,-0.0208333,0 $ \\
%11
\raisebox{-0.5\height}{\includegraphics[scale=0.15]{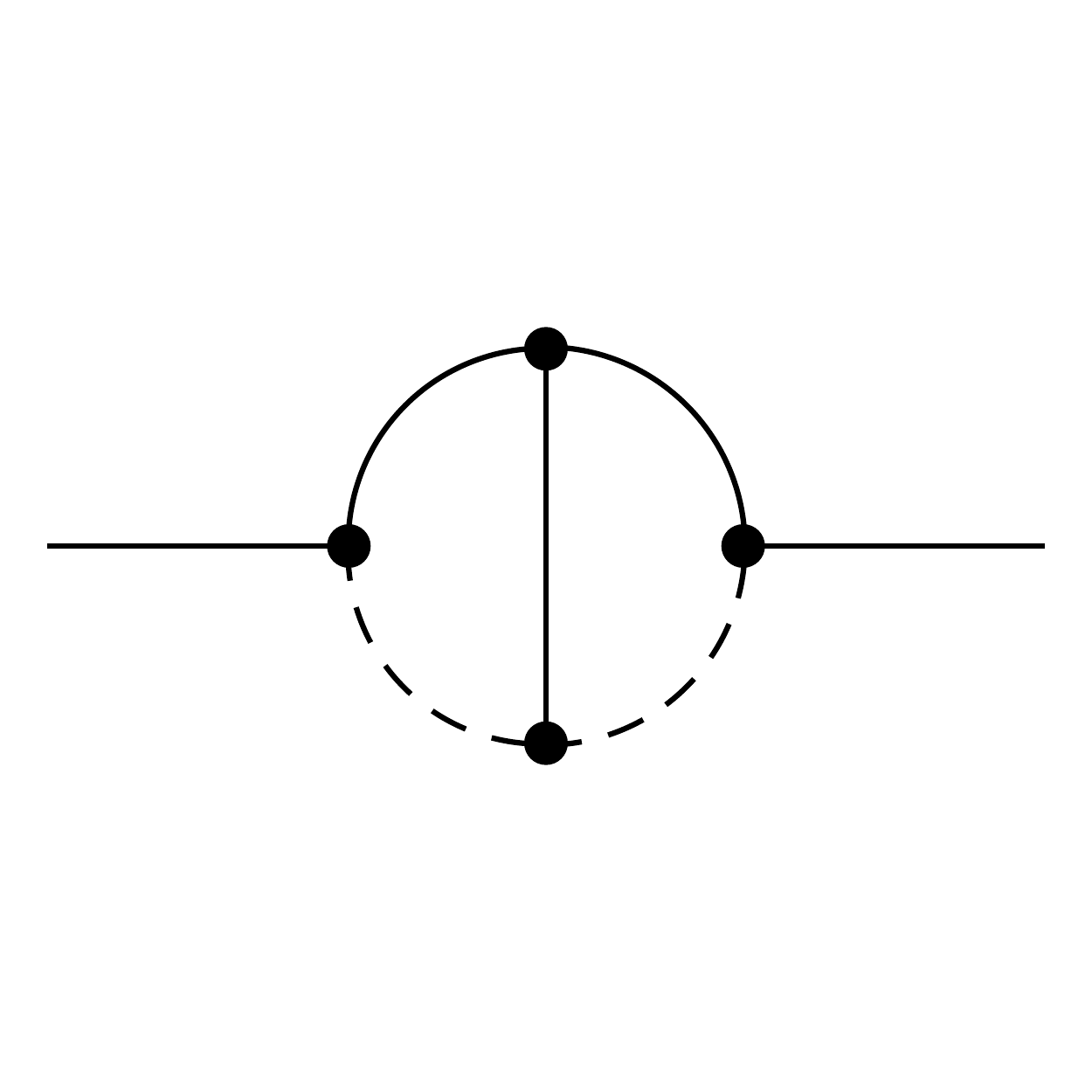}}  &   & $g_2^2,0.0160218,-0.0117188$ &
 & -& -\\
\hline
\caption{\label{g11pLtable} Two-loop contributions to $\frac{\partial}{\partial q_\|^2}\Gamma_{1,1}^{11}  (q) \left. \right|_{q =0} $}
\end{longtable}
\end{center}
%Using Eq.~(\ref{rgconst}) and the renormalization condition, $\left.\frac{\partial}{\partial q_{\|}^2} {\gammar^{11}}_{1,1}(q;q)\right|_{q=0}=D_R \rho_R$, we can now write down the renormalization constant $Z_\rho$ to two-loop order as

Collecting the divergences from all the above diagrams and applying the renormalization condition~(\ref{zrhocond}) yields, 
\begin{align}\label{zrho}
Z_{\rho}=1+ \frac{1}{\epsilon} \left(-0.75 \lambda _2 -0.0208333 \lambda _0^2-0.125 \lambda _1^2-0.156109 \lambda _2^2-0.154204 \lambda _1 \lambda _2 \right)
+ \frac{1}{\epsilon^2}\left( -0.140625 \lambda _2^2-1.3125 \lambda _1 \lambda _2
\right)
\end{align}

\section{$\frac{\partial}{\partial i q_0}\Gamma_{1,1}  (q) \left. \right|_{q =0} $}

There is no one-loop contribution to $\frac{\partial}{\partial i q_0}\Gamma_{1,1}  (q) \left. \right|_{q =0} $. The divergent parts of the two-loop diagrams have the general form $\reps \mathcal{A} \left( \frac{n}{\epsilon} + \frac{m}{\epsilon^2} \right) $. Table \ref{g11d0} shows these diagrams and their respective divergences. 

\begin{center}
\begin{longtable}{|ccc|ccc|}
\hline
Diagram & & $\mathcal{A},n,m$ & Diagram & &$\mathcal{A},n,m$ \\
\hline
\raisebox{-0.5\height}{\includegraphics[scale=0.15]{top3_69.pdf}} &  & $ g_1 g_2, 0.0172122,-0.0468752$  & 
\raisebox{-0.5\height}{\includegraphics[scale=0.15]{top3_75.pdf}}  &   &$g_1 g_2,- 0.00208572, 0.0468752  $\\
\raisebox{-0.5\height}{\includegraphics[scale=0.15]{top3_73.pdf}}  &   & $g_1g_2, -0.01512692,0 $& 
\raisebox{-0.5\height}{\includegraphics[scale=0.15]{top9_203.pdf}}  &  & $g_0^2 + 6 g_1^2, -0.03596025,0  $\\
\hline
\caption{\label{g11d0} Two-loop contributions to  $\frac{\partial}{\partial i q_0}\Gamma_{1,1}  (q) \left. \right|_{q = 0} $.}\\
\end{longtable}
\end{center}
%Using Eq.~(\ref{rgconst}) and the renormalization condition, $\left.\frac{\partial}{\partial i q_0}{\gammar^{11}}_{1,1}(q;q)\right|_{q=0}=1$, we can now write down the renormalization constant $Z$ to two-loop order as

Collecting the divergences from all the above diagrams and applying the renormalization condition~(\ref{zcond}) yields,
\begin{align}\label{z}
Z=1+ \frac{1}{\epsilon} \left( -0.0359603 \lambda _0^2-0.215762 \lambda _1^2 \right).
\end{align}

\section{$\frac{\partial}{\partial \boldsymbol{q}_\perp^2}\Gamma_{1,1}^{11}  (q) \left. \right|_{q = 0}$}

There is no one-loop contribution to $\frac{\partial}{\partial \boldsymbol{q}_\perp^2}\Gamma_{1,1}^{11}  (q) \left. \right|_{q = 0}$. The divergent parts of the two-loop diagrams have the general form $\reps D \mathcal{A} \left(\frac{n}{\epsilon}+\frac{m}{\epsilon^2}\right)$. Table \ref{gamma11pr} shows these diagrams and their respective divergences. 
\newpage
\begin{center}
\begin{longtable}{|ccc|ccc|}
\hline
Diagram & & $\mathcal{A},n,m$ & Diagram & & $\mathcal{A},n,m$  \\
\hline
\raisebox{-0.5\height}{\includegraphics[scale=0.15]{top3_69.pdf}} &  & $ g_1 g_2,-0.00224807,-0.03125$  & 
\raisebox{-0.5\height}{\includegraphics[scale=0.15]{top3_75.pdf}}  &   &$g_1 g_2,-0.0000398716,0.03125 $\\
\raisebox{-0.5\height}{\includegraphics[scale=0.15]{top3_73.pdf}}  &   & $g_1 g_2, -0.01044354,0 $ & 
\raisebox{-0.5\height}{\includegraphics[scale=0.15]{top9_203.pdf}}  &  & $ g_0^2 + 6 g_1^2, -0.0208333,0 $\\
\hline
\caption{\label{gamma11pr} Two-loop contributions to  $\frac{\partial}{\partial \boldsymbol{q}_\perp^2}\Gamma_{1,1}^{11}  (q) \left. \right|_{q = 0} $.}\\
\end{longtable}
\end{center}

Collecting the divergences from all the above diagrams and applying the renormalization condition~(\ref{zdcond}) yields, 
\begin{align}\label{zd}
Z_D=1+ \frac{1}{\epsilon} \left( -0.0208333 \lambda _0^2-0.125 \lambda _1^2-0.0127315 \lambda _1 \lambda _2 \right)
\end{align}

\section{$\Gamma_{2,0}^{11}  (q) \left. \right|_{q = 0} $}
There is no one-loop contribution to $\Gamma_{2,0}^{11}  (q) \left. \right|_{q = 0} $. The only two-loop contribution is given in Table \ref{g20}. 
\begin{table}[h!]
\centering
\begin{tabular}{|ccc|}
\hline
Diagram & & Divergence  \\
\hline
\raisebox{-0.5\height}{\includegraphics[scale=0.15]{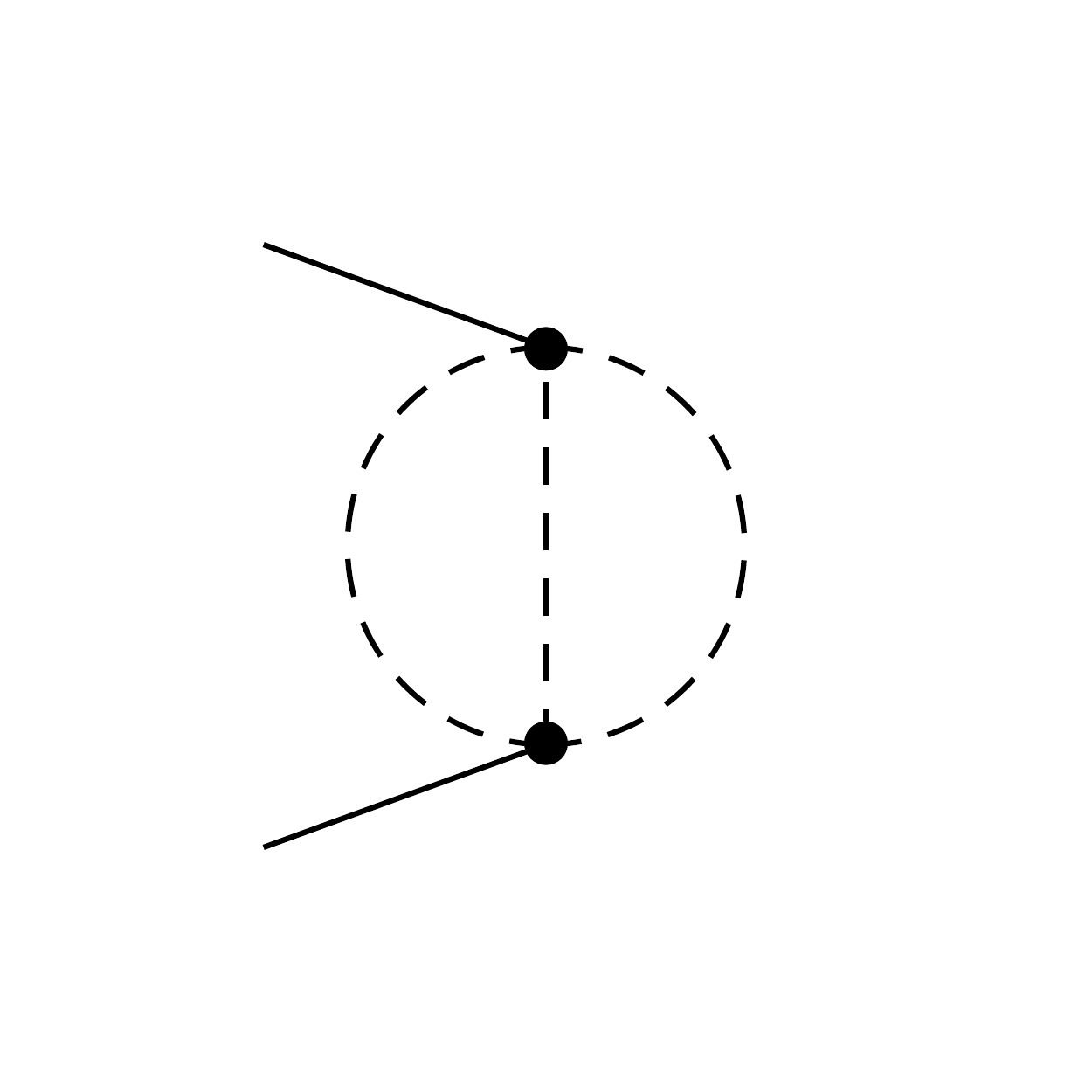}} &  & $\reps T  \left(\frac{g_0^2}{6}+ g_1^2\right) \left( \frac{0.4315235}{\epsilon } \right)   $ \\
\hline
\end{tabular}
\caption{\label{g20} Two-loop contribution to  $\Gamma_{2,0}^{11}  (q) \left. \right|_{q = 0} $.}
\end{table}

Collecting the divergence from the above diagram and applying the renormalization conditon~(\ref{ztcond}) yields, 
\begin{align}\label{zt}
Z_T=1+ \frac{1}{\epsilon} \left( -0.0359603 \lambda _0^2-0.215762 \lambda _1^2 \right)
\end{align}
\section{$\left. \frac{\partial}{\partial i q_{\|}} {\Gamma^{123}}_{1,2}(-q,\frac{q}{2},\frac{q}{2}) \right|_{q=0}$}
Table \ref{gamma121l} shows the one-loop diagrams contributing to $\left. \frac{\partial}{\partial i q_{\|}} {\Gamma^{123}}_{1,2}(-q,\frac{q}{2},\frac{q}{2}) \right|_{q=0}$ and their respective divergent contributions. The divergent part of the one-loop diagrams have the general form $ r^{-\epsilon/2} e_p \mathcal{A} \left( \frac{n}{\epsilon}\right)$.

\begin{center}
\begin{longtable}{|ccc|ccc|ccc|ccc|}
\hline
Diagram & & \textbf{   }\textbf{   }\textbf{   }\textbf{   }$\mathcal{A},n$ \textbf{   }\textbf{   }\textbf{   } & Diagram & & \textbf{   }\textbf{   }\textbf{   }$\mathcal{A},n$ \textbf{   }\textbf{   }\textbf{   }& Diagram & & \textbf{   }\textbf{   }\textbf{   }\textbf{   }$\mathcal{A},n$ \textbf{   }\textbf{   }\textbf{   } & Diagram & & \textbf{   }\textbf{   }\textbf{   }$\mathcal{A},n$ \textbf{   }\textbf{   }\textbf{   } \\
\hline
\raisebox{-0.5\height}{\includegraphics[scale=0.19]{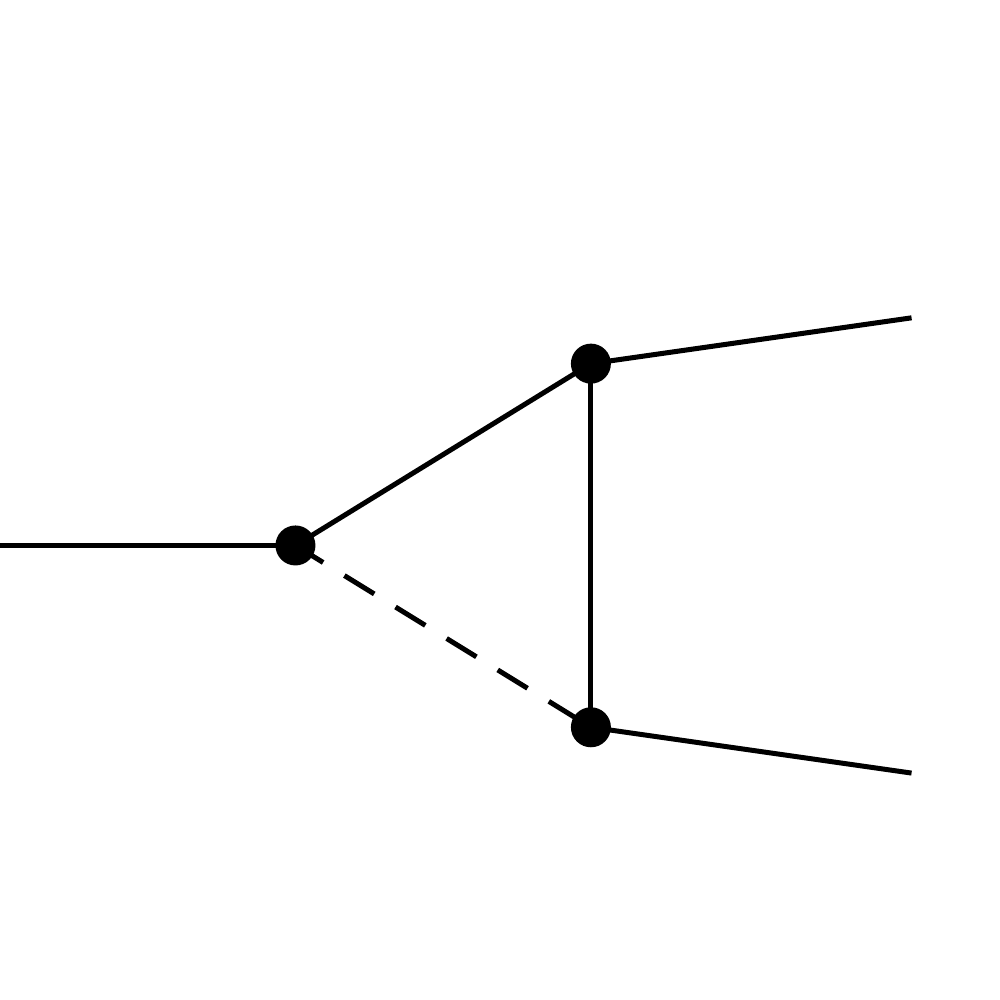}} &  & $ g_2, 0.125$  
& \raisebox{-0.5\height}{\includegraphics[scale=0.19]{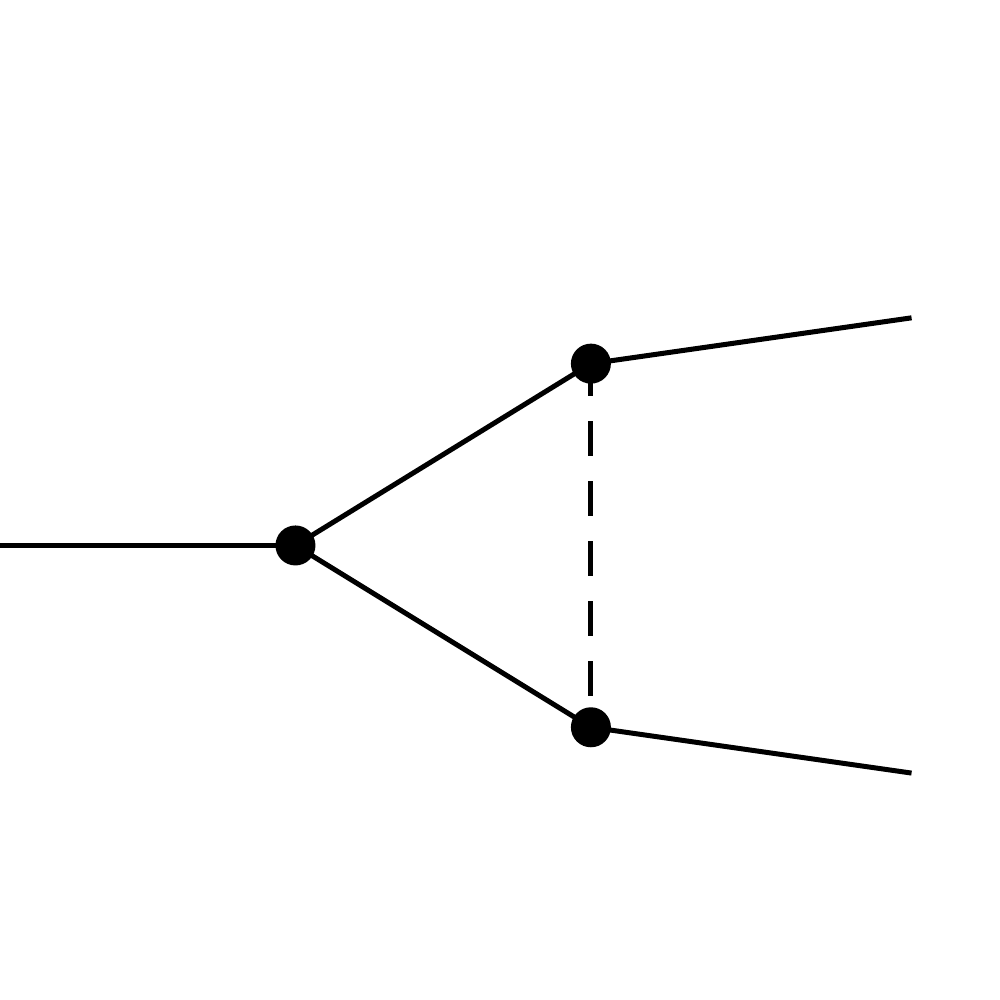}}  &   & $ g_2,-0.125 $  
& \raisebox{-0.5\height}{\includegraphics[scale=0.19]{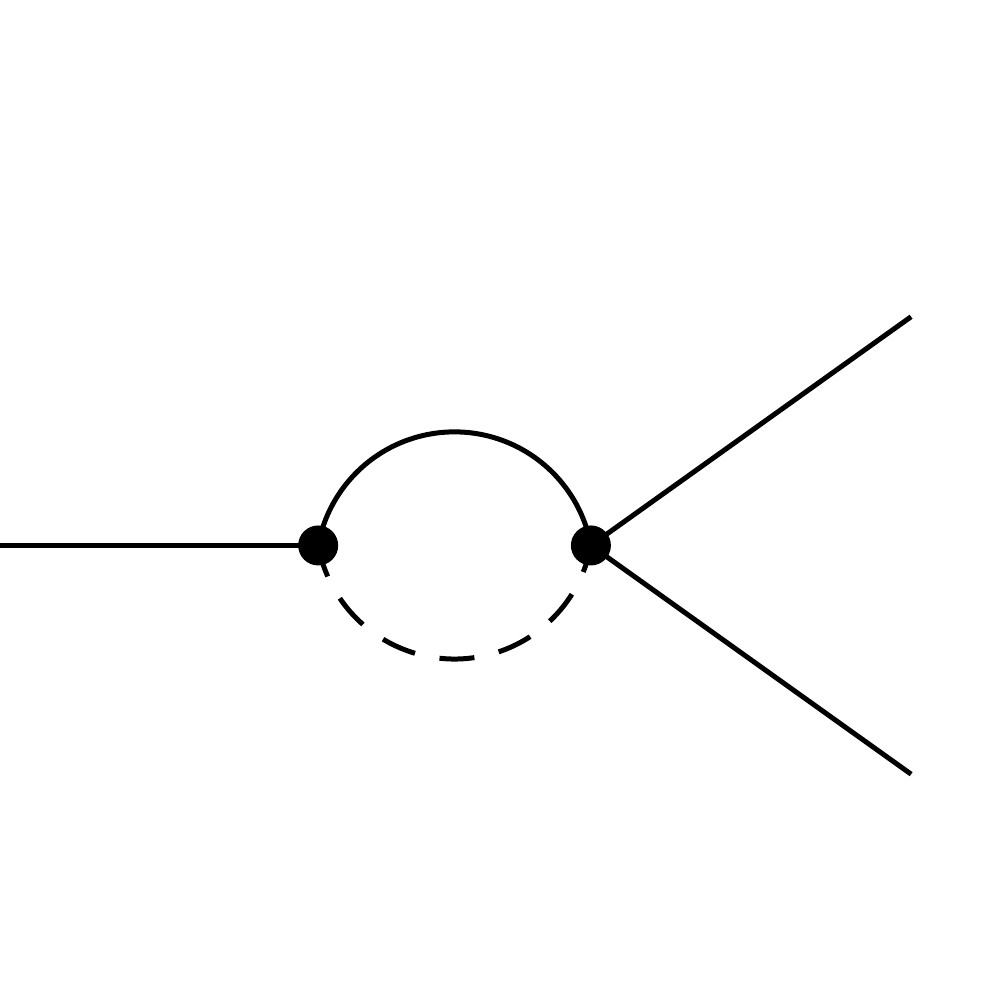}}  &   &$g_1,1 $
& \raisebox{-0.5\height}{\includegraphics[scale=0.19]{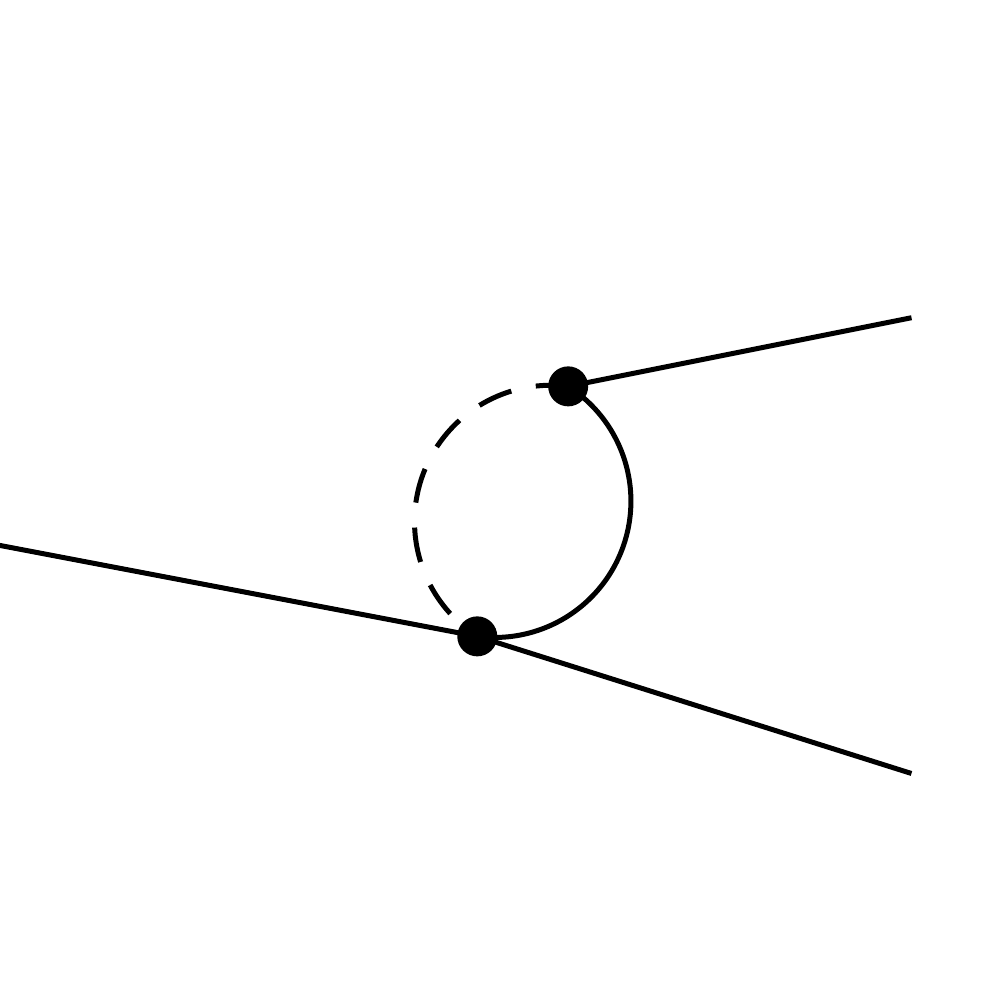}}  &  & $ g_1, 0.75 $\\
\hline
\caption{\label{gamma121l} One-loop contributions to  $\left. \frac{\partial}{\partial i q_{\|}} {\Gamma^{123}}_{1,2}(-q,\frac{q}{2},\frac{q}{2}) \right|_{q=0}$.}\\
\end{longtable}
\end{center}
As the first and the second diagrams in Table \ref{gamma121l} cancel each other there is no $g_2$ contribution to the renormalization constant $Z_p$ at the one-loop order. 

Table \ref{g12diagrams} shows the two-loop diagrams contributing to $\left. \frac{\partial}{\partial i q_{\|}} {\Gamma^{123}}_{1,2}(-q,\frac{q}{2},\frac{q}{2}) \right|_{q=0}$.

\begin{center}
\begin{longtable}{|c|}

\endfirsthead
\hline
\includegraphics[scale=1.75]{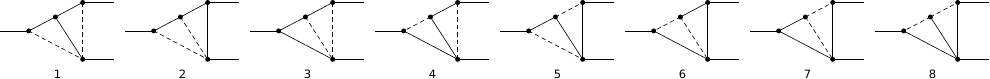}\\\\
\includegraphics[scale=1.75]{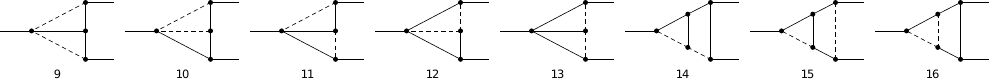}\\\\
\includegraphics[scale=1.75]{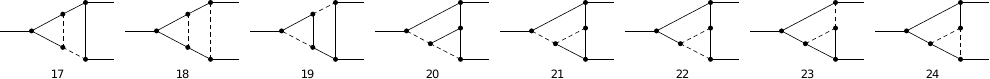}\\\\
\includegraphics[scale=1.75]{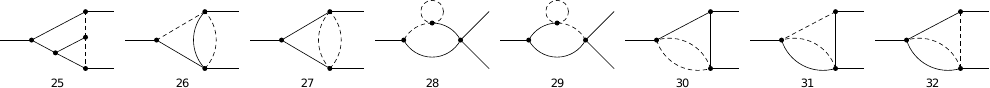}\\\\
\includegraphics[scale=1.75]{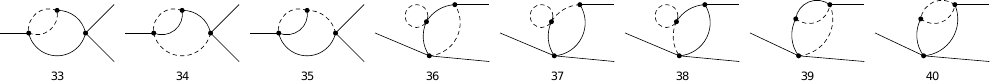}\\\\
\includegraphics[scale=1.75]{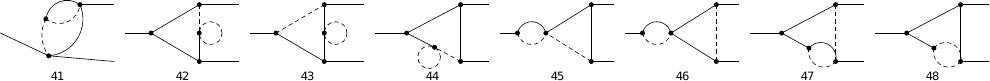}\\\\
\includegraphics[scale=1.75]{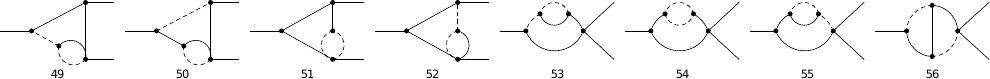}\\\\
\includegraphics[scale=1.75]{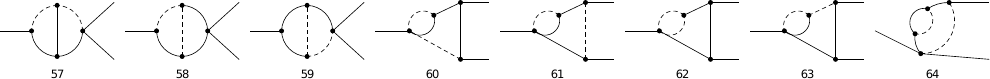}\\\\
\includegraphics[scale=1.75]{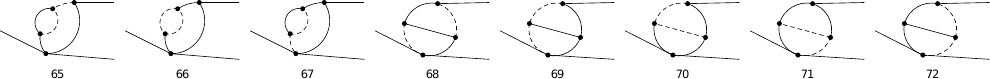}\\\\
\includegraphics[scale=1.75]{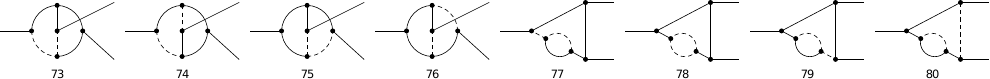}\\\\
\includegraphics[scale=1.75]{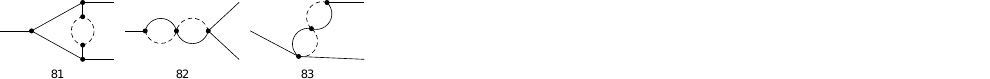}\\\\
\hline
\caption{\label{g12diagrams}Two-loop contributions to $\left. \frac{\partial}{\partial i q_{\|}} {\Gamma^{123}}_{1,2}(-q,\frac{q}{2},\frac{q}{2}) \right|_{q=0}$}
\end{longtable}
\end{center}

The divergent contributions of the above diagrams to $\left. \frac{\partial}{\partial i q_{\|}} {\Gamma^{123}}_{1,2}(-q,\frac{q}{2},\frac{q}{2}) \right|_{q=0}$ have the general form,
$ r^{-\epsilon} e_p \mathcal{A} \left( \frac{n}{\epsilon}+\frac{m}{\epsilon^2} \right)$. 
The parameters $\mathcal{A}$, $n$ and $m$ for each of the above diagrams are listed against their respective diagram numbers in the table below.
\end{widetext}

\begin{longtable}{|L|L|L|L|}
\hline
No. & \mathcal{A} & n & m \\
\hline
 1 & \left(g_0+2 g_1\right) g_2 & -0.00899006 & 0 \\
 2 & \left(g_0+2 g_1\right) g_2 & -0.0132699 & 0 \\
 3 & \left(g_0+2 g_1\right) g_2 & -0.0359603 & 0 \\
 4 & \left(g_0+2 g_1\right) g_2 & 0.0201779 & -0.03125 \\
 5 & \left(g_0+2 g_1\right) g_2 & 0.00427981 & 0 \\
 6 & \left(g_0+2 g_1\right) g_2 & 0.0206083 & -0.03125 \\
 7 & \left(g_0+2 g_1\right) g_2 & 0.0359603 & 0 \\
 8 & \left(g_0+2 g_1\right) g_2 & -0.0540561 & 0.0625 \\
 9 & \left(g_0+2 g_1\right) g_2 & 0.0125967 & -0.0117188 \\
 10 & \left(g_0+2 g_1\right) g_2 & 0.000396799 & -0.0117188 \\
 11 & \left(g_0+2 g_1\right) g_2 & -0.0297559 & 0.0117188 \\
 12 & \left(g_0+2 g_1\right) g_2 & 0.00759028 & -0.0117188 \\
 13 & \left(g_0+2 g_1\right) g_2 & -0.00502626 & 0.0234375 \\
 14 & g_2^2 & 0.00328934 & -0.00390625 \\
 15 & g_2^2 & -0.0160106 & 0.015625 \\
 16 & g_2^2 & -0.00344199 & 0 \\
 17 & g_2^2 & -0.000192796 & 0.0078125 \\
 18 & g_2^2 & 0.00208366 & -0.0078125 \\
 19 & g_2^2 & 0.0137743 & -0.0117188 \\
 20 & g_2^2 & -0.0216832 & 0.015625 \\
 21 & g_2^2 & 0.00586539 & -0.0078125 \\
 22 & g_2^2 & 0.0123075 & 0 \\
 23 & g_2^2 & -0.000657054 & 0.0078125 \\
 24 & g_2^2 & -0.0160893 & 0 \\
 25 & g_2^2 & 0.019079 & -0.015625 \\
 26 & g_1 \left(2 g_0+3 g_1\right) & 0.127185 & -0.25 \\
 27 & g_1 \left(2 g_0+3 g_1\right) & -0.107881 & 0 \\
 28 & g_1 \left(g_0+2 g_1\right) & -0.00449503 & 0 \\
 29 & g_1 \left(g_0+2 g_1\right) & -0.00449503 & 0 \\
 30 & g_1 \left(2 g_0+3 g_1\right) & -0.0492301 & 0 \\
 31 & g_1 \left(2 g_0+3 g_1\right) & 0.089859 & -0.09375 \\
 32 & g_1 \left(2 g_0+3 g_1\right) & -0.0398513 & -0.09375 \\
 33 & g_1^2 & 0.0302539 & 0 \\
 34 & g_1^2 & 0.209432 & -0.28125 \\
 35 & g_1^2 & 0.00417054 & -0.09375 \\
 36 & g_1 \left(g_0+2 g_1\right) & -0.0230808 & 0 \\
 37 & g_1 \left(g_0+2 g_1\right) & 0.09375 & 0 \\
 38 & g_1 \left(g_0+2 g_1\right) & 0.0520833 & 0 \\
 39 & g_1^2 & 0.306357 & -0.375 \\
 40 & g_1^2 & -0.143841 & 0 \\
 41 & g_1^2 & -0.0875039 & 0.09375 \\
 42 & \left(g_0+2 g_1\right) g_2 & -0.03125 & 0 \\
 43 & \left(g_0+2 g_1\right) g_2 & 0.015625 & 0 \\
 44 & \left(g_0+2 g_1\right) g_2 & 0.015625 & 0 \\
 45 & g_1 g_2 & 0.103402 & -0.125 \\
 46 & g_1 g_2 & -0.103402 & 0.125 \\
 47 & g_1 g_2 & -0.0901321 & 0.125 \\
 48 & g_1 g_2 & 0.0719206 & 0 \\
 49 & g_1 g_2 & 0.0875039 & -0.09375 \\
 50 & g_1 g_2 & 0.126092 & -0.125 \\
 51 & g_1 g_2 & -0.0719206 & 0 \\
 52 & g_1 g_2 & -0.0609641 & 0.09375 \\
 53 & g_1 g_2 & 0.0875039 & -0.09375 \\
 54 & g_1 g_2 & 0.0719206 & 0 \\
 55 & g_1 g_2 & 0.0609641 & -0.09375 \\
 56 & g_1 g_2 & 0.0201779 & -0.03125 \\
 57 & g_1 g_2 & 0.00690802 & -0.03125 \\
 58 & g_1 g_2 & -0.0132699 & 0 \\
 59 & g_1 g_2 & -0.0138161 & 0.0625 \\
 60 & g_1 g_2 & 0.0854314 & -0.0820313 \\
 61 & g_1 g_2 & -0.0333353 & 0.046875 \\
 62 & g_1 g_2 & -0.0121528 & 0 \\
 63 & g_1 g_2 & -0.0115466 & 0.0351563 \\
 64 & g_1 g_2 & 0.105484 & -0.09375 \\
 65 & g_1 g_2 & 0.00600193 & -0.0234375 \\
 66 & g_1 g_2 & 0.0283968 & 0 \\
 67 & g_1 g_2 & 0.0140635 & -0.0234375 \\
 68 & g_1 g_2 & 0.0361885 & -0.046875 \\
 69 & g_1 g_2 & -0.0160218 & 0.0117188 \\
 70 & g_1 g_2 & -0.00208526 & 0.046875 \\
 71 & g_1 g_2 & 0.0135654 & -0.0234375 \\
 72 & g_1 g_2 & -0.018626 & 0.0117187 \\
 73 & g_2^2 & 0.00806154 & 0 \\
 74 & g_2^2 & -0.00471026 & 0 \\
 75 & g_2^2 & 0.00378175 &0\\
 76 & g_2^2 & -0.00378175 & 0 \\
 77 & g_2^2 & 0.0445327 & -0.0351563 \\
 78 & g_2^2 & 0.0217619 & 0 \\
 79 & g_2^2 & 0.0274811 & -0.0351563 \\
 80 & g_2^2 & -0.0757955 & 0.0703125 \\
 81 & g_2^2 & -0.0179801 & 0 \\
 82 & g_1^2 & 0.577216 & -1 \\
 83 & g_1^2 & 0.370412 & -0.75 \\
\hline
\end{longtable}
\begin{widetext}
%Using Eq.~(\ref{rgconst}) and the renormalization condition, $ \left. \frac{\partial}{\partial i q_{\|}} {\gammar^{123}}_{1,2}(-q,\frac{q}{2},\frac{q}{2}) \right|_{q=0}=e_{pR}$, we can now write down the renormalization constant $Z_p$ to two-loop order as

Collecting the divergences from all the above diagrams and applying the renormalization condition~(\ref{zpcond}) yields,
\begin{align}\label{zp}
Z_p= & 1+ \frac{1}{\epsilon} \left( 1.75 \lambda _1-1.20163 \lambda _1^2+0.00167566 \lambda _2^2-0.726138 \lambda _0 \lambda _1+0.221394 \lambda _1 \lambda _2-0.0454484 \lambda _0 \lambda _2 \right)
+ \frac{1}{\epsilon^2} \left( 3.71875 \lambda _1^2 \right.
\nonumber
\\
& \left. +0.875 \lambda _0 \lambda _1+0.328125 \lambda _1 \lambda _2 \right)
\end{align}

\section{$\Gamma_{1,3}^{1111}(0)$}

Table \ref{gamma1311111l} shows the one-loop diagrams contributing to $\Gamma_{1,3}^{1111}(0)$ and their respective divergent contributions. The divergent part of the one-loop diagrams have the general form $ r^{-\epsilon/2} u_0 \mathcal{A} \left( \frac{n}{\epsilon}\right)$.

\begin{center}
\begin{longtable}{|ccc|ccc|ccc|}
\hline
Diagram & & \textbf{   }\textbf{   }\textbf{   }\textbf{   }$\mathcal{A},n$ \textbf{   }\textbf{   }\textbf{   } & Diagram & & \textbf{   }\textbf{   }\textbf{   }$\mathcal{A},n$ \textbf{   }\textbf{   }\textbf{   } & Diagram & & \textbf{   }\textbf{   }\textbf{   }$\mathcal{A},n$ \textbf{   }\textbf{   }\textbf{   } \\
\hline
\raisebox{-0.5\height}{\includegraphics[scale=0.2]{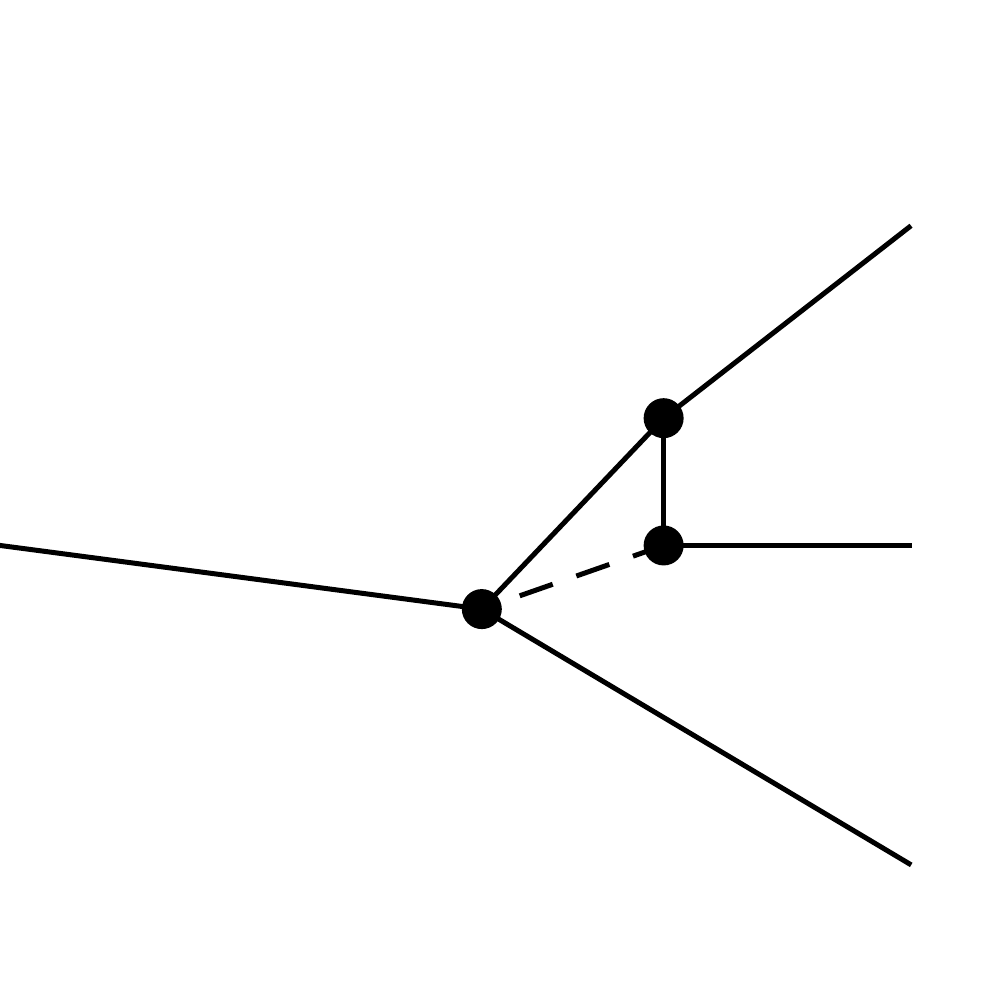}} &  & $\frac{g_1 g_2}{g_0}, 0.75 $ &
\raisebox{-0.5\height}{\includegraphics[scale=0.2]{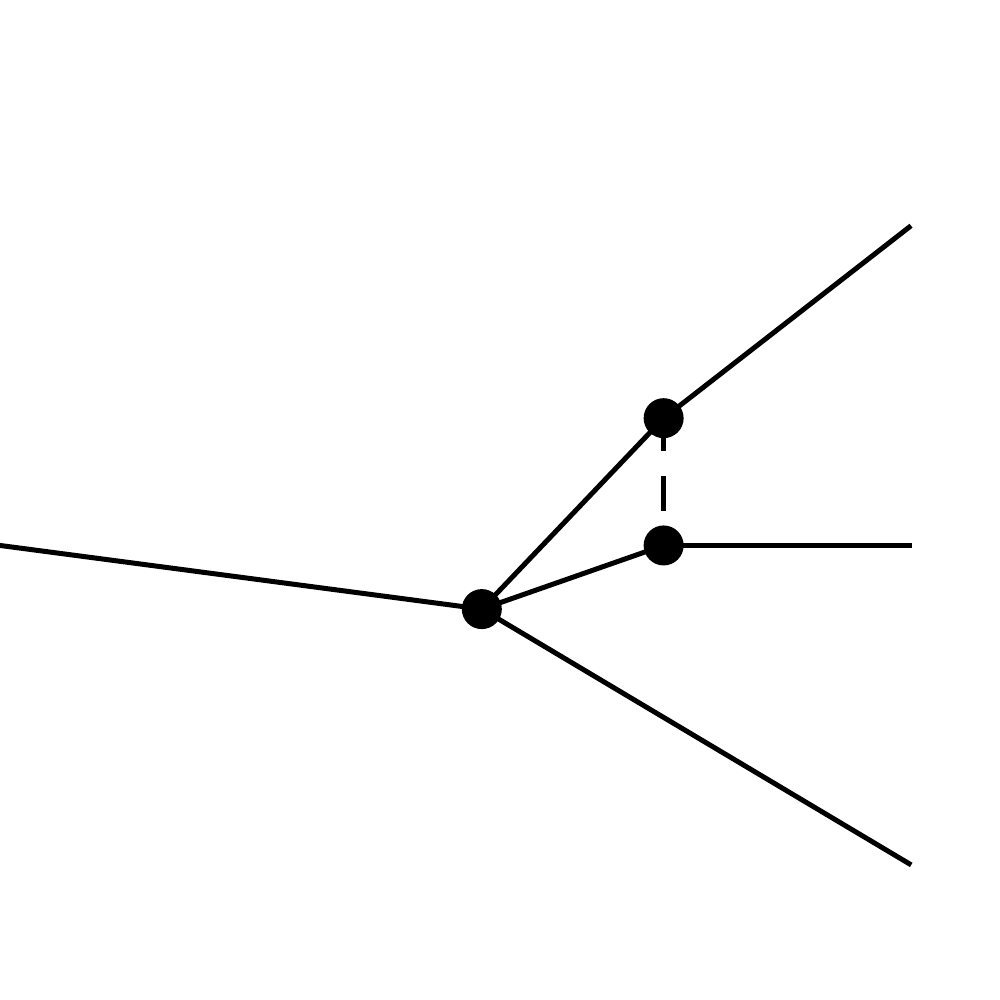}}  &   & $\frac{g_1 g_2}{g_0},-0.75 $ & 
\raisebox{-0.5\height}{\includegraphics[scale=0.2]{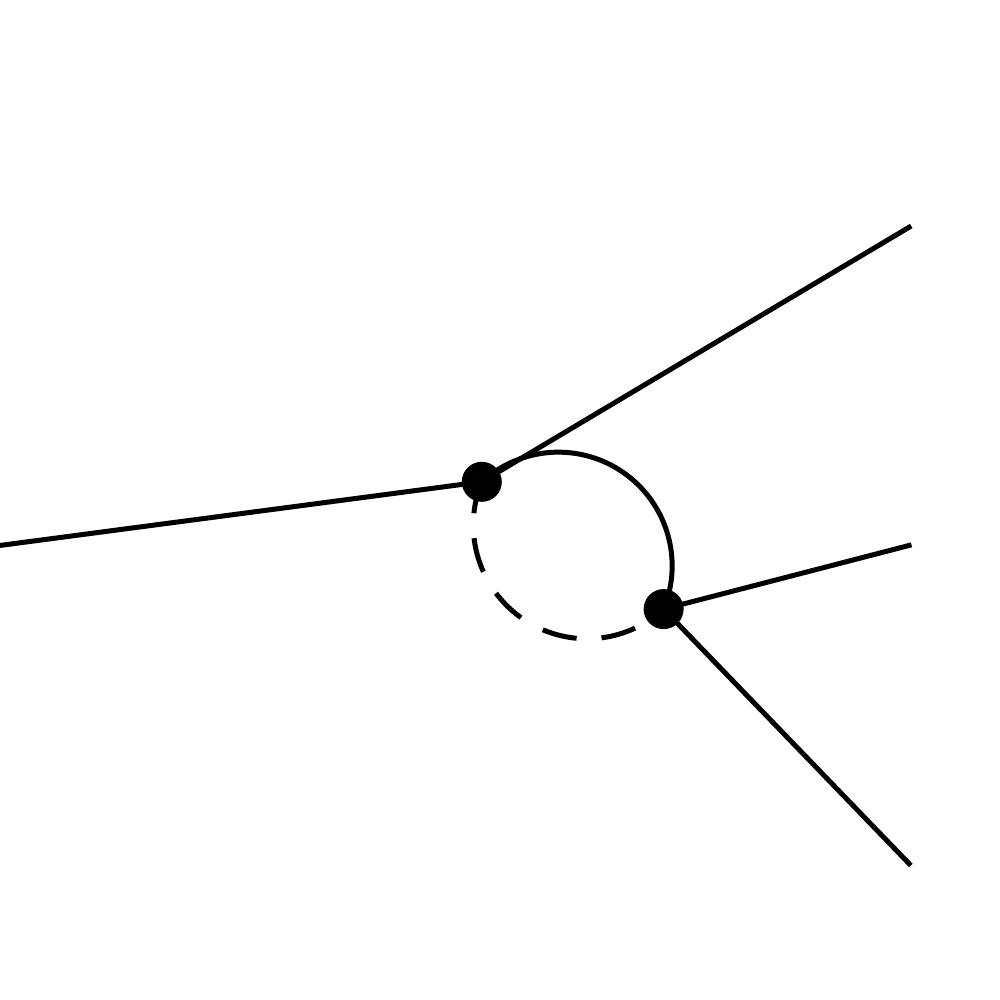}}  &   &$g_0+ \frac{2 g_1^2}{g_0}, 1.5$\\
\hline
\caption{\label{gamma1311111l} One-loop contributions to  $\Gamma_{1,3}^{1111}(0)$.}\\
\end{longtable}
\end{center}
As the first two diagrams in Table \ref{gamma1311111l} cancel each other there is no $g_2$ contribution to the renormalization constant $Z_0$ at the one-loop order. Table \ref{g13adiagrams} shows the two-loop diagrams contributing to $\Gamma_{1,3}^{1111}(0)$.

\begin{center}
\begin{longtable}{|c|}
\hline
\includegraphics[scale=1.75]{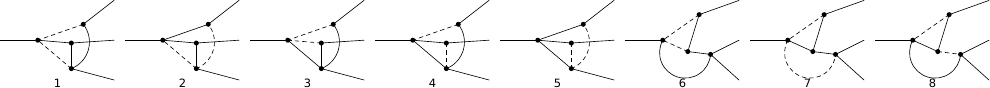}\\\\
\includegraphics[scale=1.75]{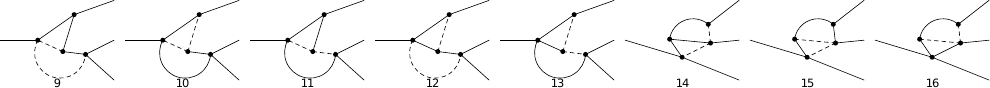}\\\\
\includegraphics[scale=1.75]{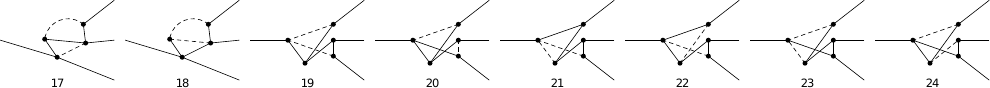}\\\\
\includegraphics[scale=1.75]{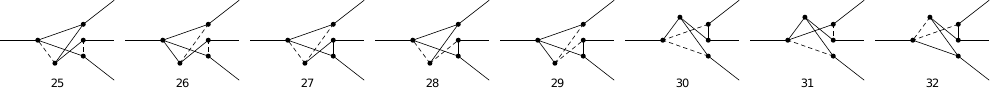}\\\\
\includegraphics[scale=1.75]{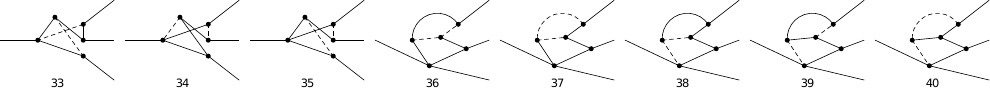}\\\\
\includegraphics[scale=1.75]{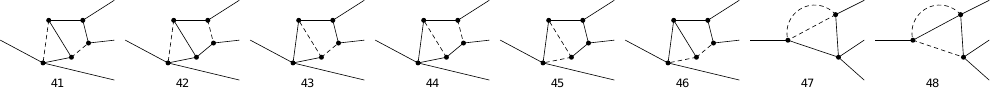}\\\\
\includegraphics[scale=1.75]{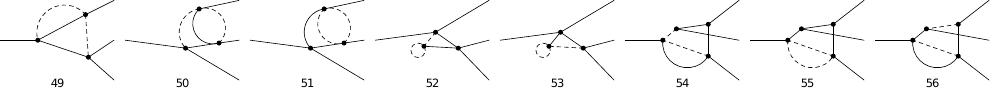}\\\\
\includegraphics[scale=1.75]{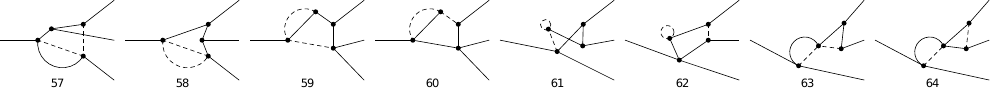}\\\\
\includegraphics[scale=1.75]{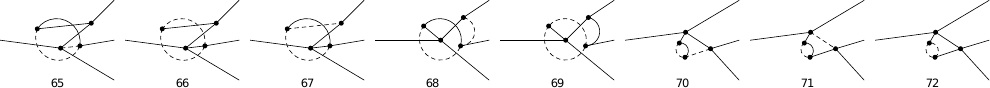}\\\\
\includegraphics[scale=1.75]{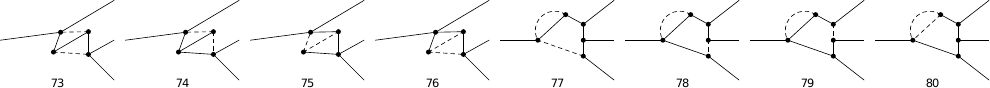}\\\\
\includegraphics[scale=1.75]{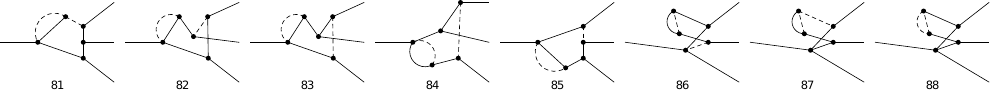}\\\\
\includegraphics[scale=1.75]{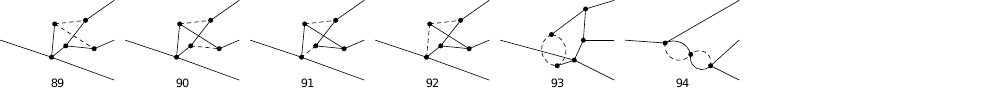}\\\\
\hline
\caption{\label{g13adiagrams}Two-loop diagrams contributing to $\Gamma_{1,3}^{1111}(0)$}\\
\end{longtable}
\end{center}

The divergent contributions of the above diagrams to $\Gamma_{1,3}^{1111}(0)$ have the general form
$ r^{-\epsilon} u_0 \mathcal{A} \left( \frac{n}{\epsilon}+\frac{m}{\epsilon^2} \right)$. 
The parameters $\mathcal{A}$, $n$ and $m$ for each of the above diagrams are listed against their respective diagram numbers in the table below. 
\end{widetext}
\begin{longtable}{|L|L|L|L|}
\hline
No. & \mathcal{A} & n & m \\
\hline 
 1 & \frac{g_1^2 g_2}{g_0} & -0.0796192 & 0 \\
 2 & \frac{g_1^2 g_2}{g_0} & 0.170381 & 0 \\
 3 & \frac{g_1^2 g_2}{g_0} & 0.0128394 & 0 \\
 4 & \frac{g_1^2 g_2}{g_0} & -0.0539404 & 0 \\
 5 & \frac{g_1^2 g_2}{g_0} & 0.0582202 & 0 \\
 6 & \frac{g_1 \left(g_0+g_1\right) g_2}{g_0} & -0.0398096 & 0 \\
 7 & \frac{g_1 \left(g_0+g_1\right) g_2}{g_0} & 0.181601 & -0.28125 \\
 8 & \frac{g_1 \left(g_0+g_1\right) g_2}{g_0} & 0.0207241 & -0.09375 \\
 9 & \frac{g_1 \left(g_0+g_1\right) g_2}{g_0} & 0.185475 & -0.28125 \\
 10 & \frac{g_1 \left(g_0+g_1\right) g_2}{g_0} & 0.0851904 & 0 \\
 11 & \frac{g_1 \left(g_0+g_1\right) g_2}{g_0} & -0.0289365 & -0.09375 \\
 12 & \frac{g_1 \left(g_0+g_1\right) g_2}{g_0} & -0.486505 & 0.5625 \\
 13 & \frac{g_1 \left(g_0+g_1\right) g_2}{g_0} & -0.0769779 & 0.1875 \\
 14 & \frac{g_1 \left(g_0+g_1\right) g_2}{g_0} & -0.0539404 & 0 \\
 15 & \frac{g_1 \left(g_0+g_1\right) g_2}{g_0} & -0.0796192 & 0 \\
 16 & \frac{g_1 \left(g_0+g_1\right) g_2}{g_0} & -0.215762 & 0 \\
 17 & \frac{g_1 \left(g_0+g_1\right) g_2}{g_0} & 0.0256788 & 0 \\
 18 & \frac{g_1 \left(g_0+g_1\right) g_2}{g_0} & 0.215762 & 0 \\
 19 & \frac{g_1 g_2^2}{g_0} & 0.0459143 & -0.0390625 \\
 20 & \frac{g_1 g_2^2}{g_0} & -0.0800529 & 0.078125 \\
 21 & \frac{g_1 g_2^2}{g_0} & 0.0293269 & -0.0390625 \\
 22 & \frac{g_1 g_2^2}{g_0} & -0.108416 & 0.078125 \\
 23 & \frac{g_1 g_2^2}{g_0} & -0.0206519 & 0 \\
 24 & \frac{g_1 g_2^2}{g_0} & 0.0197361 & -0.0234375 \\
 25 & \frac{g_1 g_2^2}{g_0} & -0.00657054 & 0.078125 \\
 26 & \frac{g_1 g_2^2}{g_0} & 0.19079 & -0.15625 \\
 27 & \frac{g_1 g_2^2}{g_0} & 0.0522647 & 0 \\
 28 & \frac{g_1 g_2^2}{g_0} & -0.0415526 & -0.0234375 \\
 29 & \frac{g_1 g_2^2}{g_0} & -0.0468204 & 0.046875 \\
 30 & \frac{g_1 g_2^2}{g_0} & 0.00918285 & -0.0078125 \\
 31 & \frac{g_1 g_2^2}{g_0} & -0.0160106 & 0.015625 \\
 32 & \frac{g_1 g_2^2}{g_0} & 0.00586539 & -0.0078125 \\
 33 & \frac{g_1 g_2^2}{g_0} & -0.0216832 & 0.015625 \\
 34 & \frac{g_1 g_2^2}{g_0} & -0.00131411 & 0.015625 \\
 35 & \frac{g_1 g_2^2}{g_0} & 0.038158 & -0.03125 \\
 36 & \frac{g_1 g_2^2}{g_0} & -0.0965356 & 0 \\
 37 & \frac{g_1 g_2^2}{g_0} & 0.0738452 & 0 \\
 38 & \frac{g_1 g_2^2}{g_0} & 0.0175962 & -0.0234375 \\
 39 & \frac{g_1 g_2^2}{g_0} & 0.0275486 & -0.0234375 \\
 40 & \frac{g_1 g_2^2}{g_0} & -0.0650495 & 0.046875 \\
 41 & \frac{\left(g_0+g_1\right) g_2^2}{g_0} & 0.0275486 & -0.0234375 \\
 42 & \frac{\left(g_0+g_1\right) g_2^2}{g_0} & -0.0480317 & 0.046875 \\
 43 & \frac{\left(g_0+g_1\right) g_2^2}{g_0} & -0.00115712 & 0.046875 \\
 44 & \frac{\left(g_0+g_1\right) g_2^2}{g_0} & 0.0125021 & -0.046875 \\
 45 & \frac{\left(g_0+g_1\right) g_2^2}{g_0} & -0.0206519 & 0 \\
 46 & \frac{\left(g_0+g_1\right) g_2^2}{g_0} & 0.0197361 & -0.0234375 \\
 47 & \frac{g_0^3+2 g_1^2 g_0+4 g_1^3}{g_0} & -0.107881 & 0 \\
 48 & \frac{g_0^3+2 g_1^2 g_0+4 g_1^3}{g_0} & 0.381554 & -0.75 \\
 49 & \frac{g_0^3+2 g_1^2 g_0+4 g_1^3}{g_0} & -0.0249845 & -0.375 \\
 50 & \frac{g_0^3+2 g_1^2 g_0+4 g_1^3}{g_0} & 0.190777 & -0.375 \\
 51 & \frac{g_0^3+2 g_1^2 g_0+4 g_1^3}{g_0} & -0.323643 & 0 \\
 52 & \frac{g_0^3+2 g_1 g_0^2+2 g_1^2 g_0+4 g_1^3}{g_0} & 0.1875 & 0 \\
 53 & \frac{g_0^3+2 g_1 g_0^2+2 g_1^2 g_0+4 g_1^3}{g_0} & 0.1875 & 0 \\
 54 & \frac{g_1^2 g_2}{g_0} & 0.378277 & -0.375 \\
 55 & \frac{g_1^2 g_2}{g_0} & -0.136142 & 0 \\
 56 & \frac{g_1^2 g_2}{g_0} & -0.540793 & 0.75 \\
 57 & \frac{g_1^2 g_2}{g_0} & 0.0375155 & -0.375 \\
 58 & \frac{g_1^2 g_2}{g_0} & -0.0453808 & 0 \\
 59 & \frac{g_1^2 g_2}{g_0} & 0.787535 & -0.84375 \\
 60 & \frac{g_1^2 g_2}{g_0} & -0.0125116 & 0.28125 \\
 61 & \frac{g_1 \left(g_0+2 g_1\right) g_2}{g_0} & 0.09375 & 0 \\
 62 & \frac{g_1 \left(g_0+2 g_1\right) g_2}{g_0} & -0.09375 & 0 \\
 63 & \frac{g_1 \left(2 g_0+g_1\right) g_2}{g_0} & 0.310206 & -0.375 \\
 64 & \frac{g_1 \left(2 g_0+g_1\right) g_2}{g_0} & -0.310206 & 0.375 \\
 65 & \frac{g_1^2 g_2}{g_0} & 0.756554 & -0.75 \\
 66 & \frac{g_1^2 g_2}{g_0} & -0.431524 & 0 \\
 67 & \frac{g_1^2 g_2}{g_0} & -0.365785 & 0.5625 \\
 68 & \frac{g_1^2 g_2}{g_0} & -0.540793 & 0.75 \\
 69 & \frac{g_1^2 g_2}{g_0} & 0.431524 & 0 \\
 70 & \frac{\left(g_0^2+2 g_1^2\right) g_2}{g_0} & 0.0914462 & -0.140625 \\
 71 & \frac{\left(g_0^2+2 g_1^2\right) g_2}{g_0} & 0.131256 & -0.140625 \\
 72 & \frac{\left(g_0^2+2 g_1^2\right) g_2}{g_0} & 0.107881 & 0 \\
 73 & \frac{g_1 \left(2 g_0+g_1\right) g_2}{g_0} & 0.0605337 & -0.09375 \\
 74 & \frac{g_1 \left(2 g_0+g_1\right) g_2}{g_0} & 0.0207241 & -0.09375 \\
 75 & \frac{g_1 \left(2 g_0+g_1\right) g_2}{g_0} & -0.0398096 & 0 \\
 76 & \frac{g_1 \left(2 g_0+g_1\right) g_2}{g_0} & -0.0414484 & 0.1875 \\
 77 & \frac{g_1 g_2^2}{g_0} & 0.267196 & -0.210938 \\
 78 & \frac{g_1 g_2^2}{g_0} & -0.227387 & 0.210938 \\
 79 & \frac{g_1 g_2^2}{g_0} & 0.0549622 & -0.0703125 \\
 80 & \frac{g_1 g_2^2}{g_0} & -0.0210752 & 0 \\
 81 & \frac{g_1 g_2^2}{g_0} & -0.00920429 & 0.0703125 \\
 82 & \frac{g_1 g_2^2}{g_0} & 0.24733 & -0.316406 \\
 83 & \frac{g_1 g_2^2}{g_0} & -0.151591 & 0.140625 \\
 84 & \frac{g_1 g_2^2}{g_0} & 0.0274811 & -0.0351563 \\
 85 & \frac{g_1 g_2^2}{g_0} & -0.0757955 & 0.0703125 \\
 86 & \frac{g_1 g_2^2}{g_0} & 0.0890654 & -0.0703125 \\
 87 & \frac{g_1 g_2^2}{g_0} & -0.227387 & 0.210938 \\
 88 & \frac{g_1 g_2^2}{g_0} & -0.107881 & 0 \\
 89 & g_2^2 & 0.0113452 & 0 \\
 90 & g_2^2 & -0.0113452 & 0 \\
 91 & g_2^2 & -0.0141308 & 0\\
 92 & g_2^2 & 0.0241846 & 0\\
 93 & \frac{g_1 g_2^2}{g_0} & 0.130571 & 0 \\
 94 & \frac{g_0^3+6 g_1^2 g_0+2 g_1^3}{g_0} & 0.432912 & -0.75 \\
 \hline
\end{longtable}

\begin{widetext}
%Using Eq.~(\ref{rgconst}) and the renormalization condition, $ {\gammar^{1111}}_{1,3}(q_i=0)=u_{0R}$, we can now write down the renormalization constant $Z_0$ to two-loop order as

Collecting the divergences from all the above diagrams and applying the renormalization condition~(\ref{zu0cond}) yields,
\begin{align}\label{z0}
Z_0= & 1+ \frac{1}{\epsilon} \left(1.5 \lambda _0-0.75 \lambda _0^2 -1.5 \lambda _1^2-0.26712 \lambda _1
   \lambda _2+0.168241 \lambda _0 \lambda _2 -\frac{3 \lambda _1^3}{\lambda _0}+\frac{0.49572 \lambda _2 \lambda _1^2}{\lambda _0}+\frac{3 \lambda _1^2}{\lambda _0}-\frac{0.00819685 \lambda _2^2 \lambda _1}{\lambda _0}\right)
\nonumber
\\
& +\frac{1}{\epsilon^2} \left(2.25 \lambda _0^2+7.5 \lambda _1^2-+0.28125 \lambda _0 \lambda _2 + \frac{7.5 \lambda _1^3}{\lambda _0}+\frac{0.5625 \lambda _2 \lambda _1^2}{\lambda _0}\right)
\end{align}

\section{$\Gamma_{1,3}^{1122}(0)$}\label{appg131122}

Table \ref{gamma1311221l} shows the one-loop diagrams contributing to $\Gamma_{1,3}^{1122}(0)$ and their respective divergent contributions. The divergent part of the one-loop diagrams have the general form $ r^{-\epsilon/2} u_1 \mathcal{A} \left( \frac{n}{\epsilon}\right)$.

\begin{center}
\begin{longtable}{|ccc|ccc|ccc|}
\hline
Diagram & & \textbf{   }\textbf{   }\textbf{   }\textbf{   }$\mathcal{A},n$ \textbf{   }\textbf{   }\textbf{   } & Diagram & & \textbf{   }\textbf{   }\textbf{   }$\mathcal{A},n$ \textbf{   }\textbf{   }\textbf{   } & Diagram & & \textbf{   }\textbf{   }\textbf{   }$\mathcal{A},n$ \textbf{   }\textbf{   }\textbf{   } \\
\hline
\raisebox{-0.5\height}{\includegraphics[scale=0.2]{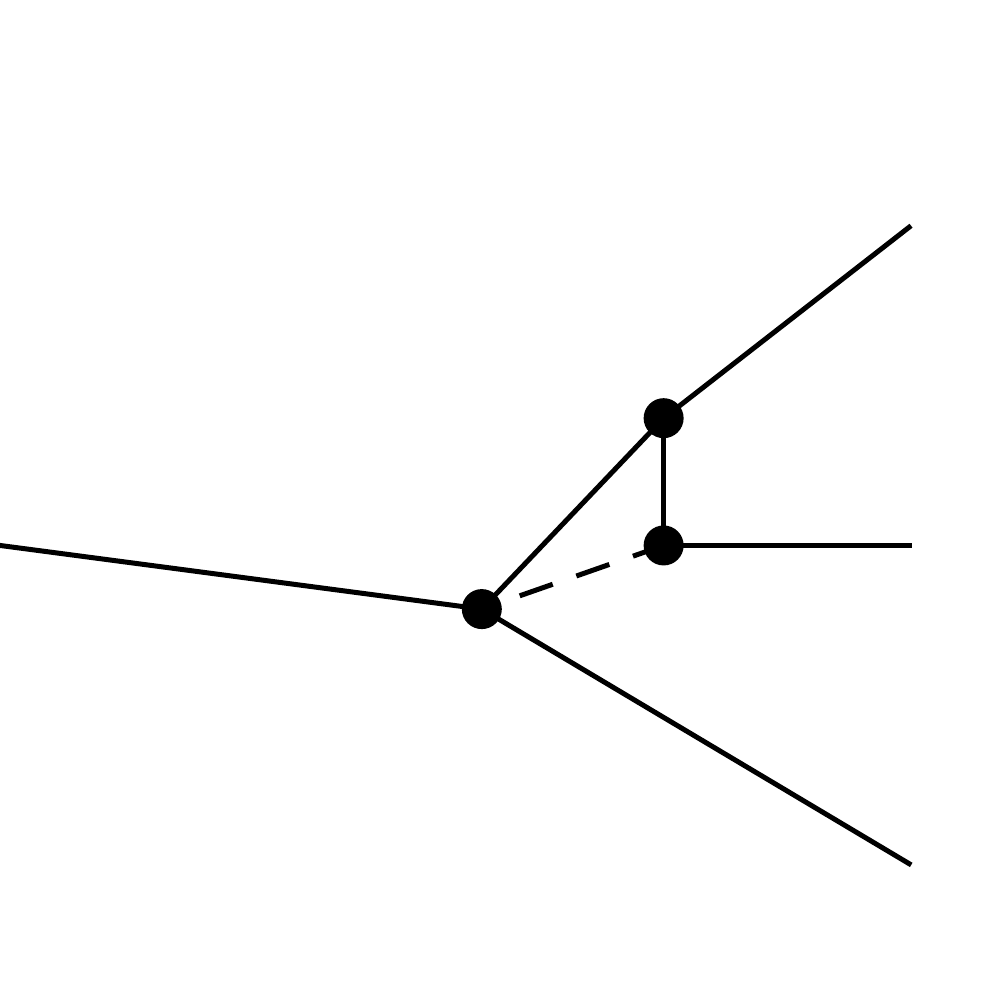}} &  & $\frac{g_0 g_2}{g_1}, 0.125 $  & 

\raisebox{-0.5\height}{\includegraphics[scale=0.2]{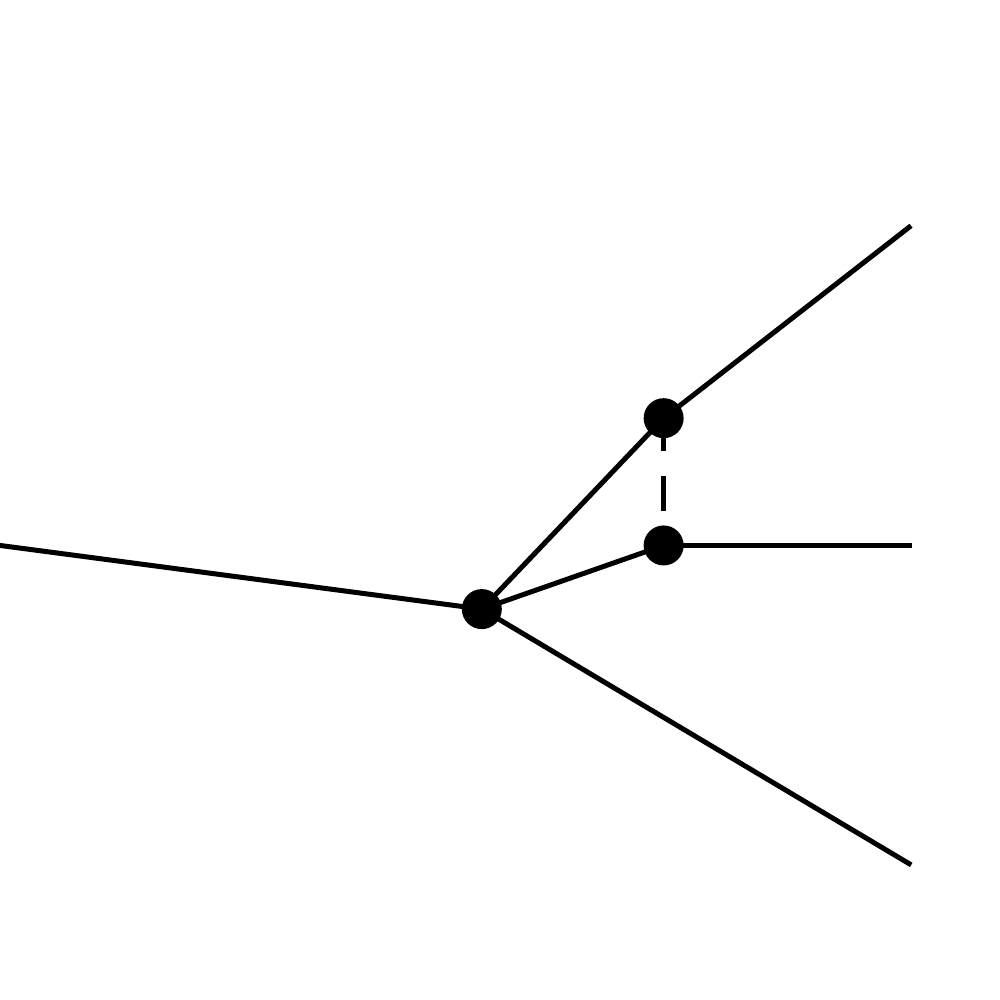}}  &   & $\frac{g_0 g_2}{g_1}, -0.125$ &
\raisebox{-0.5\height}{\includegraphics[scale=0.2]{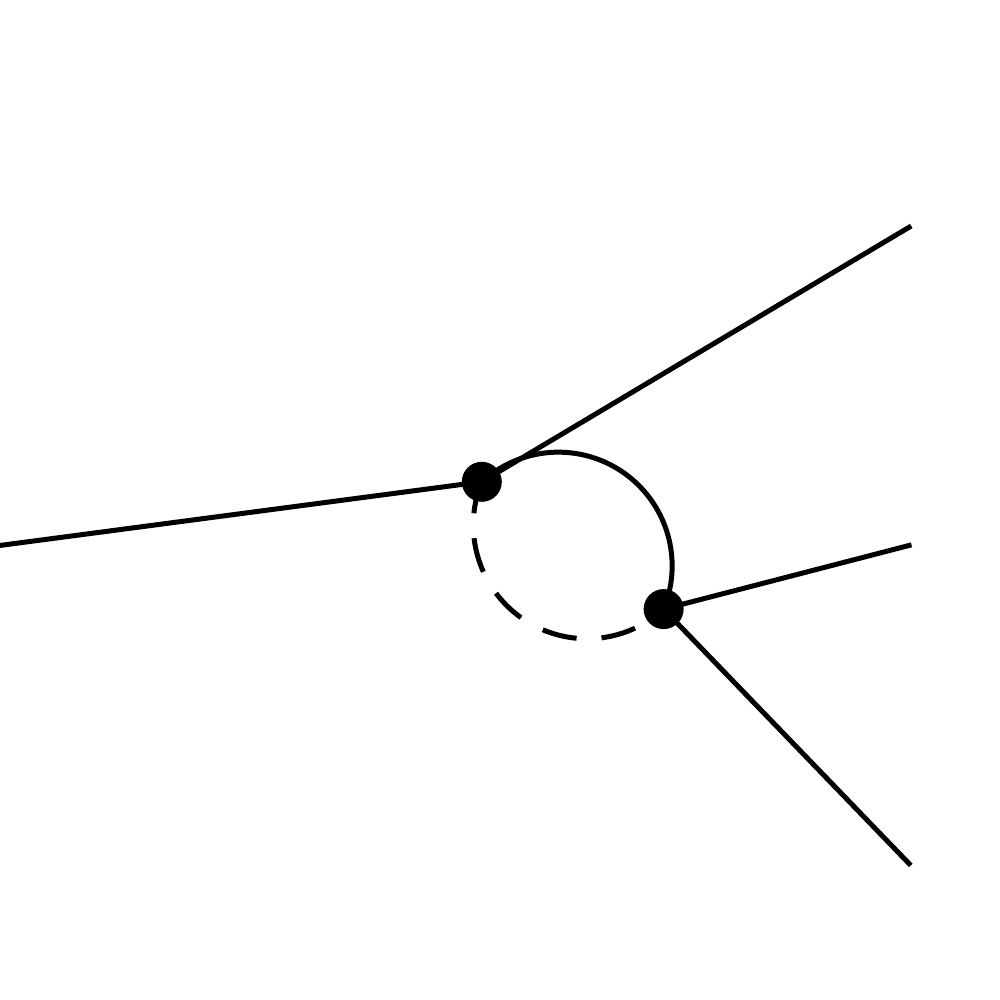}}  &   &$2 g_0+ 5 g_1, 0.5$\\
\hline
\caption{\label{gamma1311221l} One-loop contributions to  $\Gamma_{1,3}^{1122}(0)$.}\\
\end{longtable}
\end{center}
As the first two diagrams in Table~\ref{gamma1311221l} cancel each other there is no $g_2$ contribution to the renormalization constant $Z_1$ at the one-loop order. Table~\ref{g13bdiagrams} shows the two-loop diagrams contributing to $\Gamma_{1,3}^{1122}(0)$.

\begin{center}
\begin{longtable}{|c|}
\hline
\includegraphics[scale=1.75]{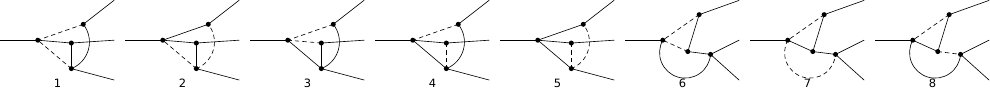}\\\\
\includegraphics[scale=1.75]{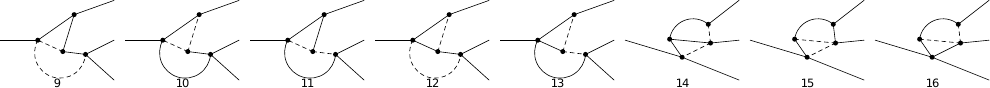}\\\\
\includegraphics[scale=1.75]{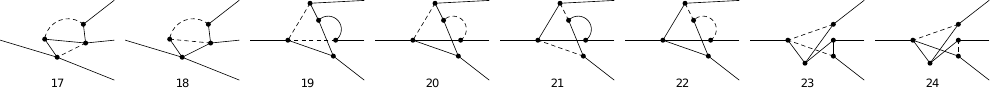}\\\\
\includegraphics[scale=1.75]{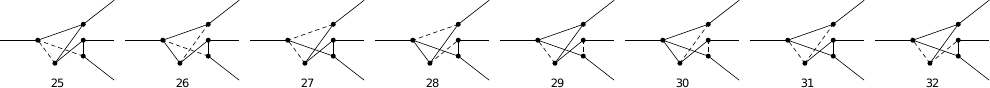}\\\\
\includegraphics[scale=1.75]{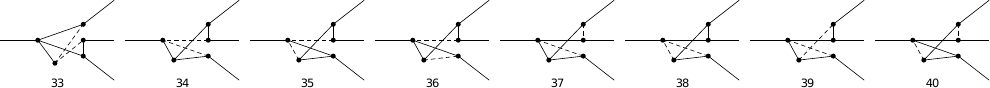}\\\\
\includegraphics[scale=1.75]{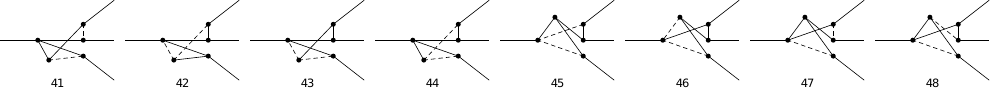}\\\\
\includegraphics[scale=1.75]{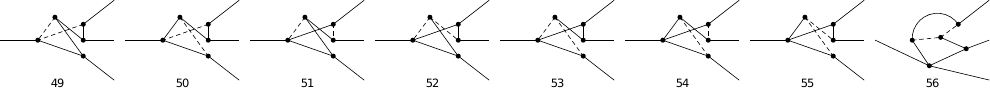}\\\\
\includegraphics[scale=1.75]{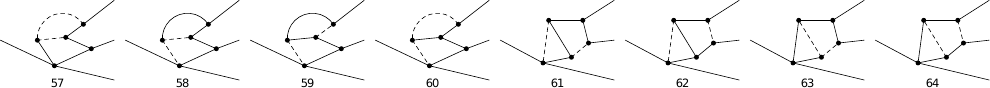}\\\\
\includegraphics[scale=1.75]{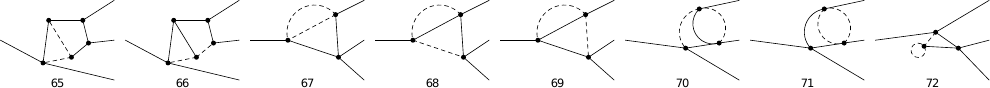}\\\\\includegraphics[scale=1.75]{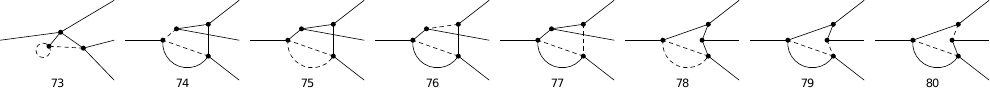}\\\\\includegraphics[scale=1.75]{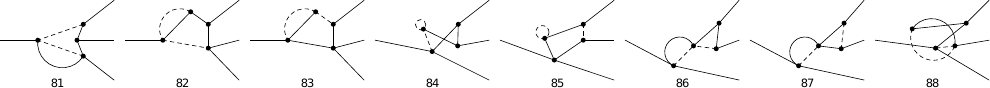}\\\\\includegraphics[scale=1.75]{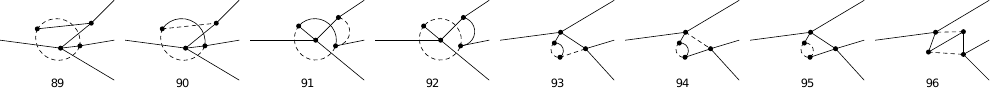}\\\\\includegraphics[scale=1.75]{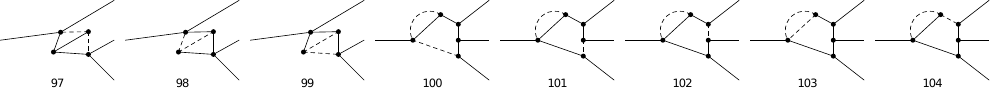}\\\\\includegraphics[scale=1.75]{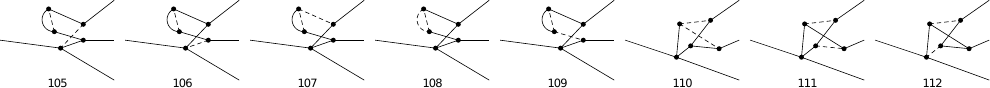}\\\\\includegraphics[scale=1.75]{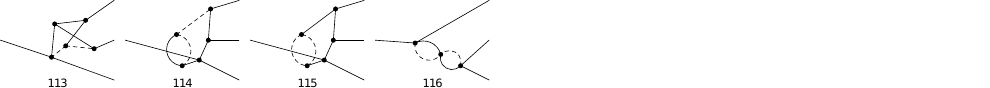}\\\\
\hline
\caption{\label{g13bdiagrams}Two-loop diagrams contributing to $\Gamma_{1,3}^{1122}(0)$}\\
\end{longtable}
\end{center}

The divergent contributions of the above diagrams to $\Gamma_{1,3}^{1122}(0)$ have the general form
$ r^{-\epsilon} u_1 \mathcal{A} \left( \frac{n}{\epsilon}+\frac{m}{\epsilon^2} \right)$. The parameters $\mathcal{A}$, $n$, and $m$ for each of the above diagrams are listed against their respective diagram numbers in the table below. 
\end{widetext}
\begin{longtable}{|L|L|L|L|}
\hline
No. & \mathcal{A} & n & m \\
\hline 
 1 & \left(g_0+4 g_1\right) g_2 & -0.0265397 & 0 \\
 2 & \left(g_0+4 g_1\right) g_2 & 0.0567936 & 0 \\
 3 & \left(g_0+4 g_1\right) g_2 & 0.00427981 & 0 \\
 4 & \left(g_0+4 g_1\right) g_2 & -0.0179801 & 0 \\
 5 & \left(g_0+4 g_1\right) g_2 & 0.0194067 & 0 \\
 6 & \left(3 g_0+7 g_1\right) g_2 & -0.00663493 & 0 \\
 7 & \left(3 g_0+7 g_1\right) g_2 & 0.0302668 & -0.046875 \\
 8 & \left(3 g_0+7 g_1\right) g_2 & 0.00345401 & -0.015625 \\
 9 & \left(3 g_0+7 g_1\right) g_2 & 0.0309125 & -0.046875 \\
 10 & \left(3 g_0+7 g_1\right) g_2 & 0.0141984 & 0 \\
 11 & \left(3 g_0+7 g_1\right) g_2 & -0.00482275 & -0.015625 \\
 12 & \left(3 g_0+7 g_1\right) g_2 & -0.0810842 & 0.09375 \\
 13 & \left(3 g_0+7 g_1\right) g_2 & -0.0128297 & 0.03125 \\
 14 & \left(3 g_0+7 g_1\right) g_2 & -0.00899006 & 0 \\
 15 & \left(3 g_0+7 g_1\right) g_2 & -0.0132699 & 0 \\
 16 & \left(3 g_0+7 g_1\right) g_2 & -0.0359603 & 0 \\
 17 & \left(3 g_0+7 g_1\right) g_2 & 0.00427981 & 0 \\
 18 & \left(3 g_0+7 g_1\right) g_2 & 0.0359603 & 0 \\
 19 & g_2^2 & -0.00706539 & 0 \\
 20 & g_2^2 & 0.0201538 & 0 \\
 21 & g_2^2 & 0.0161231 & 0\\
 22 & g_2^2 & -0.0208333 & 0 \\
 23 & \frac{\left(g_0+9 g_1\right) g_2^2}{g_1} & 0.00229571 & -0.00195313 \\
 24 & \frac{\left(g_0+9 g_1\right) g_2^2}{g_1} & -0.00400265 & 0.00390625 \\
 25 & \frac{\left(g_0+9 g_1\right) g_2^2}{g_1} & 0.00146635 & -0.00195313 \\
 26 & \frac{\left(g_0+9 g_1\right) g_2^2}{g_1} & -0.00542079 & 0.00390625 \\
 27 & g_2^2 & -0.010326 & 0 \\
 28 & g_2^2 & 0.00986803 & -0.0117188 \\
 29 & \frac{\left(g_0+9 g_1\right) g_2^2}{g_1} & -0.000328527 & 0.00390625 \\
 30 & \frac{\left(g_0+9 g_1\right) g_2^2}{g_1} & 0.00953951 & -0.0078125 \\
 31 & g_2^2 & 0.0261324 & 0 \\
 32 & g_2^2 & -0.0207763 & -0.0117188 \\
 33 & g_2^2 & -0.0234102 & 0.0234375 \\
 34 & \frac{\left(3 g_0+5 g_1\right) g_2^2}{g_1} & 0.00229571 & -0.00195313 \\
 35 & \frac{\left(3 g_0+5 g_1\right) g_2^2}{g_1} & 0.00146635 & -0.00195313 \\
 36 & \frac{\left(3 g_0+5 g_1\right) g_2^2}{g_1} & -0.00542079 & 0.00390625 \\
 37 & \frac{\left(3 g_0+5 g_1\right) g_2^2}{g_1} & -0.00400265 & 0.00390625 \\
 38 & \frac{\left(g_0+g_1\right) g_2^2}{g_1} & -0.00344199 & 0 \\
 39 & \frac{\left(g_0+g_1\right) g_2^2}{g_1} & 0.00328934 & -0.00390625 \\
 40 & \frac{\left(3 g_0+5 g_1\right) g_2^2}{g_1} & -0.000328527 & 0.00390625 \\
 41 & \frac{\left(3 g_0+5 g_1\right) g_2^2}{g_1} & 0.00953951 & -0.0078125 \\
 42 & \frac{\left(g_0+g_1\right) g_2^2}{g_1} & -0.00692543 & -0.00390625 \\
 43 & \frac{\left(g_0+g_1\right) g_2^2}{g_1} & 0.00871079 & 0 \\
 44 & \frac{\left(g_0+g_1\right) g_2^2}{g_1} & -0.00780339 & 0.0078125 \\
 45 & g_2^2 & 0.00459143 & -0.00390625 \\
 46 & g_2^2 & -0.00344199 & 0 \\
 47 & g_2^2 & -0.00800529 & 0.0078125 \\
 48 & g_2^2 & 0.00328934 & -0.00390625 \\
 49 & g_2^2 & 0.00293269 & -0.00390625 \\
 50 & g_2^2 & -0.0108416 & 0.0078125 \\
 51 & g_2^2 & -0.000657054 & 0.0078125 \\
 52 & g_2^2 & -0.00692543 & -0.00390625 \\
 53 & g_2^2 & 0.00871079 & 0 \\
 54 & g_2^2 & 0.019079 & -0.015625 \\
 55 & g_2^2 & -0.00780339 & 0.0078125 \\
 56 & \frac{\left(g_0+3 g_1\right) g_2^2}{g_1} & -0.0160893 & 0 \\
 57 & \frac{\left(g_0+3 g_1\right) g_2^2}{g_1} & 0.0123075 & 0 \\
 58 & \frac{\left(g_0+3 g_1\right) g_2^2}{g_1} & 0.00293269 & -0.00390625 \\
 59 & \frac{\left(g_0+3 g_1\right) g_2^2}{g_1} & 0.00459143 & -0.00390625 \\
 60 & \frac{\left(g_0+3 g_1\right) g_2^2}{g_1} & -0.0108416 & 0.0078125 \\
 61 & \frac{\left(g_0+5 g_1\right) g_2^2}{g_1} & 0.00459143 & -0.00390625 \\
 62 & \frac{\left(g_0+5 g_1\right) g_2^2}{g_1} & -0.00800529 & 0.0078125 \\
 63 & \frac{\left(g_0+5 g_1\right) g_2^2}{g_1} & -0.000192854 & 0.0078125 \\
 64 & \frac{\left(g_0+5 g_1\right) g_2^2}{g_1} & 0.00208369 & -0.0078125 \\
 65 & \frac{\left(g_0+5 g_1\right) g_2^2}{g_1} & -0.00344199 & 0 \\
 66 & \frac{\left(g_0+5 g_1\right) g_2^2}{g_1} & 0.00328934 & -0.00390625 \\
 67 & g_0^2+6 g_1 g_0+10 g_1^2 & -0.0359603 & 0 \\
 68 & g_0^2+6 g_1 g_0+10 g_1^2 & 0.127185 & -0.25 \\
 69 & g_0^2+6 g_1 g_0+10 g_1^2 & -0.00832817 & -0.125 \\
 70 & g_0^2+6 g_1 g_0+10 g_1^2 & 0.0635923 & -0.125 \\
 71 & g_0^2+6 g_1 g_0+10 g_1^2 & -0.107881 & 0 \\
 72 & 2 g_0^2+9 g_1 g_0+10 g_1^2 & 0.0625 & 0 \\
 73 & 2 g_0^2+9 g_1 g_0+10 g_1^2 & 0.0625 & 0 \\
 74 & \frac{\left(g_0^2+4 g_1 g_0+6 g_1^2\right) g_2}{g_1} & 0.0157615 & -0.015625 \\
 75 & \frac{\left(g_0^2+8 g_1 g_0+18 g_1^2\right) g_2}{g_1} & -0.00378173 & 0 \\
 76 & \frac{\left(g_0^2+4 g_1 g_0+6 g_1^2\right) g_2}{g_1} & -0.022533 & 0.03125 \\
 77 & \frac{\left(g_0^2+4 g_1 g_0+6 g_1^2\right) g_2}{g_1} & 0.00156315 & -0.015625 \\
 78 & \frac{\left(g_0^2+8 g_1^2\right) g_2}{g_1} & -0.00378173 & 0 \\
 79 & \frac{\left(g_0^2+8 g_1^2\right) g_2}{g_1} & 0.00156315 & -0.015625 \\
 80 & \frac{\left(g_0^2+8 g_1^2\right) g_2}{g_1} & -0.022533 & 0.03125 \\
 81 & \frac{\left(g_0^2+8 g_1^2\right) g_2}{g_1} & 0.0157615 & -0.015625 \\
 82 & \left(g_0+3 g_1\right) g_2 & 0.131256 & -0.140625 \\
 83 & \left(g_0+3 g_1\right) g_2 & -0.00208527 & 0.046875 \\
 84 & \frac{\left(g_0^2+5 g_1 g_0+6 g_1^2\right) g_2}{g_1} & 0.015625 & 0 \\
 85 & \frac{\left(g_0^2+5 g_1 g_0+6 g_1^2\right) g_2}{g_1} & -0.015625 & 0 \\
 86 & \frac{\left(g_0^2+2 g_1 g_0+7 g_1^2\right) g_2}{g_1} & 0.051701 & -0.0625 \\
 87 & \frac{\left(g_0^2+2 g_1 g_0+7 g_1^2\right) g_2}{g_1} & -0.051701 & 0.0625 \\
 88 & \left(g_0+3 g_1\right) g_2 & 0.126092 & -0.125 \\
 89 & \left(g_0+3 g_1\right) g_2 & -0.0719206 & 0 \\
 90 & \left(g_0+3 g_1\right) g_2 & -0.0609641 & 0.09375 \\
 91 & \left(g_0+3 g_1\right) g_2 & -0.0901321 & 0.125 \\
 92 & \left(g_0+3 g_1\right) g_2 & 0.0719206 & 0 \\
 93 & \left(2 g_0+5 g_1\right) g_2 & 0.0304821 & -0.046875 \\
 94 & \left(2 g_0+5 g_1\right) g_2 & 0.0437519 & -0.046875 \\
 95 & \left(2 g_0+5 g_1\right) g_2 & 0.0359602 & 0 \\
 96 & \frac{\left(g_0^2+2 g_1 g_0+7 g_1^2\right) g_2}{g_1} & 0.0100889 & -0.015625 \\
 97 & \frac{\left(g_0^2+2 g_1 g_0+7 g_1^2\right) g_2}{g_1} & 0.00345401 & -0.015625 \\
 98 & \frac{\left(g_0^2+2 g_1 g_0+7 g_1^2\right) g_2}{g_1} & -0.00663493 & 0 \\
 99 & \frac{\left(g_0^2+2 g_1 g_0+7 g_1^2\right) g_2}{g_1} & -0.00690807 & 0.03125 \\
 100 & \frac{\left(g_0+5 g_1\right) g_2^2}{g_1} & 0.0148442 & -0.0117188 \\
 101 & g_2^2 & -0.0757955 & 0.0703125 \\
 102 & g_2^2 & 0.0549622 & -0.0703125 \\
 103 & g_2^2 & -0.00702507 & 0 \\
 104 & g_2^2 & -0.0030681 & 0.0234375 \\
 105 & \frac{\left(g_0+3 g_1\right) g_2^2}{g_1} & 0.0148442 & -0.0117188 \\
 106 & \frac{\left(g_0+3 g_1\right) g_2^2}{g_1} & 0.0148442 & -0.0117188 \\
 107 & \frac{\left(g_0+3 g_1\right) g_2^2}{g_1} & -0.0378978 & 0.0351563 \\
 108 & \frac{\left(g_0+3 g_1\right) g_2^2}{g_1} & -0.0179801 & 0 \\
 109 & \frac{\left(g_0+3 g_1\right) g_2^2}{g_1} & -0.0378978 & 0.0351563 \\
 110 & g_2^2 & 0.0113452 & 0 \\
 111 & g_2^2 & -0.0113452 & 0 \\
 112 & g_2^2 & -0.0141308 & 0\\
 113 & g_2^2 & 0.0120923 & 0\\
 114 & \frac{\left(g_0+3 g_1\right) g_2^2}{g_1} & 0.0274811 & -0.0351563 \\
 115 & \frac{\left(g_0+3 g_1\right) g_2^2}{g_1} & 0.0217619 & 0 \\
 116 & 3 g_0^2+3 g_1 g_0+11 g_1^2 & 0.144304 & -0.25 \\
 \hline
\end{longtable}
\begin{widetext}
%Using Eq.~(\ref{rgconst}) and the renormalization condition, $ {\gammar^{1122}}_{1,3}(q_i=0)=u_{1R}$, we can now write down the renormalization constant $Z_1$ to two-loop order as

Collecting the divergences from all the above diagrams and applying the renormalization condition~(\ref{zu1cond}) yields,
\begin{align}\label{z1}
Z_1 & =1+\frac{1}{\epsilon}\left( \lambda _0+2.5\lambda _1-0.25 \lambda _0^2-2.5 \lambda
   _1^2-1.5 \lambda _0 \lambda _1+0.0676406 \lambda _2 \lambda _0+0.253861 \lambda _1 \lambda _2 +0.00136604 \lambda _2^2 \right.
\nonumber
\\   
 & \left. -\frac{0.0179802 \lambda _2 \lambda _0^2}{\lambda _1}-\frac{0.000747121 \lambda _2^2 \lambda _0}{\lambda _1}  \right)
+ \frac{1}{\epsilon^2} \left( 1.25 \lambda _0^2+7.75 \lambda _1^2+3.75 \lambda _1 \lambda _0+0.1875 \lambda _2 \lambda _0+0.46875 \lambda _1 \lambda _2 \right)
\end{align}

All the results obtained in Appendices~\ref{appg110} to~\ref{appg131122} using the computational method described in Appendix~\ref{appcomp} when truncated to one-loop order agree with the results obtained in Ref.~\citep{DuttaPark2011}, where calculations were performed only to this order. 
\end{widetext}
\bibliography{References} 
\end{document}